\documentclass[superscriptaddress,prx,twocolumn]{revtex4-2}
\usepackage{graphicx}
\usepackage{amsmath,amssymb,physics,dsfont}
\usepackage{mathrsfs}
\usepackage{color}
\usepackage{hyperref}
\usepackage{microtype}
\usepackage{braket}
\usepackage{stmaryrd}
\usepackage{academicons}
\usepackage{defs-private}
\usepackage{qcircuit}
\usepackage{booktabs}
\usepackage[dvipsnames]{xcolor}

\begin{document}

\title{Gutzwiller Quantum-Classical Embedding for Dynamics Simulations of Correlated Electron Systems}

\author{Joshua Aftergood}
\affiliation{Ames National Laboratory, U.S. Department of Energy, Ames, Iowa 50011, USA}

\author{Daniele Guerci}
\affiliation{Department of Physics, William \& Mary, Williamsburg, Virginia 23187, USA}

\author{Jacob Brunton}
\affiliation{Department of Physics and Astronomy, Iowa State University, Ames, Iowa 50011, USA}

\author{Feng Zhang}
\affiliation{Ames National Laboratory, U.S. Department of Energy, Ames, Iowa 50011, USA}
\affiliation{Department of Physics and Astronomy, Iowa State University, Ames, Iowa 50011, USA}

\author{Thomas Iadecola}
\affiliation{Ames National Laboratory, U.S. Department of Energy, Ames, Iowa 50011, USA}
\affiliation{Department of Physics and Astronomy, Iowa State University, Ames, Iowa 50011, USA}
\affiliation{Department of Physics, Pennsylvania State University, University Park, Pennsylvania 16802, USA}

\author{Peter P.~Orth}
\affiliation{Ames National Laboratory, U.S. Department of Energy, Ames, Iowa 50011, USA}
\affiliation{Department of Physics and Astronomy, Iowa State University, Ames, Iowa 50011, USA}
\affiliation{Department of Physics, Saarland University, 66123 Saarbr\"ucken, Germany}
\affiliation{Center for Quantum Technologies (QuTe), Saarland University, 66123 Saarbr\"ucken, Germany}

\author{Cai-Zhuang Wang}
\affiliation{Ames National Laboratory, U.S. Department of Energy, Ames, Iowa 50011, USA}
\affiliation{Department of Physics and Astronomy, Iowa State University, Ames, Iowa 50011, USA}

\author{Nicola Lanat\`a}
\affiliation{School of Physics and Astronomy, Rochester Institute of Technology, Rochester, New York 14623, USA}
\affiliation{Center for Computational Quantum Physics, Flatiron Institute, New York, New York 10010, USA}

\author{Yong-Xin Yao}
\email{ykent@iastate.edu}
\affiliation{Ames National Laboratory, U.S. Department of Energy, Ames, Iowa 50011, USA}
\affiliation{Department of Physics and Astronomy, Iowa State University, Ames, Iowa 50011, USA}

\begin{abstract}
Simulating the real-time dynamics of correlated electron systems requires a nonperturbative treatment of local correlations over long times, a challenge for classical methods. We present the time-dependent Gutzwiller quantum-classical embedding framework, t-GQCE, for generic multi-orbital lattice models with time-dependent Hamiltonians. Within the time-dependent ghost Gutzwiller approximation, a noninteracting quasiparticle lattice propagated classically is coupled to finite interacting embedding models propagated on quantum processors. The two parts exchange only embedding Hamiltonian parameters and single-particle density matrices, and the framework is agnostic to the quantum solver. We demonstrate it with adaptive variational quantum dynamics simulations of interaction quenches in the Hubbard model on the Bethe lattice. Without ghost orbitals, the embedding problem reduces to a single qubit, simulated on IBM quantum hardware with an ancilla qubit. With ghost orbitals, t-GQCE suppresses the artificial undamped oscillations of the conventional Gutzwiller approximation and approaches nonequilibrium dynamical mean-field theory results for weak and strong quenches. Statevector simulations with 2 to 6 ghost orbitals per spin, or 8 to 16 qubits, give adaptive circuits that follow the embedding states with infidelities below $5\times10^{-4}$. Under trapped-ion noise models, circuit compression, measurements in rotated orbital bases with post-selection, and noise mitigation reduce the errors two- to threefold. At one third of the error rates of a hardware-benchmarked noise model, they reach an absolute accuracy of $0.01$ for all single-particle density matrix elements and the double occupancy with $10^5$ shots per time step, and the Quantinuum H2-Emulator gives similar accuracy, suggesting that t-GQCE is a viable approach for early fault-tolerant quantum processors.
\end{abstract}

\maketitle

\section{Introduction}\label{sec: intro}
Ultrafast pump-probe experiments have made it possible to drive correlated electron materials far from equilibrium and to follow their electronic, magnetic, and structural response on femtosecond to picosecond time scales~\cite{ultrafast_approach, basov2017towards, de2021colloquium}. Light-induced transitions between metallic and insulating states, transient superconductivity, and the nonthermal control of competing orders all involve the real-time dynamics of strongly correlated electrons, which cannot be described by perturbative or independent-particle approaches. A quantitative theory of these phenomena requires methods that treat the local electronic correlations nonperturbatively and follow the coupled dynamics of the correlated orbitals and their environment over experimentally relevant times.

Quantum embedding theories provide such a nonperturbative treatment by mapping the lattice onto local interacting impurity problems that are coupled self-consistently to a noninteracting environment~\cite{sun2016quantum}. Among them, the Gutzwiller approximation and the equivalent rotationally invariant slave-boson theory have been applied in equilibrium to a wide range of correlated materials~\cite{ga_ce, ga_pu, ga_uo2, ga_tmo, ga_feas_bunemann, ga_smb6, ga_fe}. The Gutzwiller wave function has also been extended to a systematically improvable variational scheme for impurity models~\cite{grg}, and the relation of these approaches to other embedding theories has been explored~\cite{ga_dmet}. The ghost Gutzwiller approximation enlarges the variational freedom of this approach by introducing auxiliary ghost orbitals, so that each correlated orbital couples to $B$ bath orbitals, of which $B-1$ are ghosts~\cite{gga_lanata2017, Lanata2022operatorial, Frank2021QuantumED, Lee2023AccuracyGRISB, Lee2023multiorbGRISB, MejutoZaera2023screening, Lanata2023ghostDMET, mejuto2024quantum, Lee2024ChargeSC}. It becomes equivalent to the dynamical mean-field theory (DMFT) in the limit of infinitely many ghost orbitals~\cite{finiteTgGA2026}, and efficient implementations and machine-learning approaches have been developed to reduce its cost~\cite{Rogers2021bypassing, Frank2024activeLA, Giuli2026linear, GEM2026}.

To follow the dynamics in real time, several embedding theories have been extended out of equilibrium, including, among others, the nonequilibrium DMFT~\cite{RMPNonequilibriumDMFT}, the real-time extension of density matrix embedding theory~\cite{kretchmer18tddmet}, and the time-dependent Gutzwiller approximation~\cite{schiro10tdga, TDGutzwillerSC}. Nonequilibrium DMFT is exact in the limit of infinite dimensions and has established the main features of the quench dynamics of the Hubbard model~\cite{Eckstein09ThermalizationIQ}. Its real-time impurity problem with a continuous bath, however, remains demanding, especially for multi-orbital models and long times, which motivates complementary approaches. The time-dependent Gutzwiller approximation has a considerably lower computational cost, since its embedding problem contains only one bath orbital per correlated orbital, and it has nevertheless been shown to capture the short-time dynamics and the dynamical transition after an interaction quench~\cite{schiro10tdga}. With its compact variational form, it also produces coherent and undamped oscillations of local observables that are absent in the exact dynamics. Its ghost extension, the time-dependent ghost Gutzwiller approximation, suppresses these oscillations and approaches the nonequilibrium DMFT results with increasing $B$~\cite{Guerci2023TDgGA}. The price is an embedding Hamiltonian defined on $(B+1)\nu$ spin-orbitals for $\nu$ correlated spin-orbitals, whose Hilbert space grows exponentially with $B$ and $\nu$. The propagation of the interacting embedding states, rather than that of the noninteracting environment, then becomes the computational bottleneck.

Quantum computers can represent such many-body states with a number of qubits linear in the number of spin-orbitals, and real-time evolution is among their most natural applications~\cite{feynman82qc, lloyd1996, bauer2020quantum}. Systematically approaching the thermodynamic limit in direct simulations of the full finite-size lattice dynamics, however, remains challenging. Hybrid quantum-classical embedding approaches reduce the quantum part of the problem to the embedding models, while the noninteracting environment is treated on classical computers~\cite{hybrd_dmft, Kreula2016, rubin2016DMET, Vorwerk2022QuantumET}. In equilibrium, this strategy has been applied, among others, to DMFT~\cite{hybrd_dmft, hybrd_2sitedmft2, selisko2024DMFT, Hogan2026efficient}, density matrix embedding theory~\cite{rubin2016DMET, Mineh2022DMET}, and the Gutzwiller quantum-classical embedding (GQCE)~\cite{gqce, mukherjee2023comparative}, which was recently extended to ghost orbitals~\cite{chen2025quantum, Sriluckshmy2025QAgGA}. Out of equilibrium, Kreula \textit{et al.} proposed a nonlinear quantum-classical scheme to simulate nonequilibrium DMFT~\cite{Kreula2016}. The Gutzwiller embedding offers a convenient setting for this purpose, since its embedding Hamiltonian has a finite and small number of bath orbitals and couples to the classical part only through single-particle density matrices. For $N$ spin-orbitals, these contain $O(N^2)$ elements, which can all be obtained from a number of measurement circuits that grows only linearly with $N$ by rotating the orbital basis before measuring the occupation numbers~\cite{Arute2020HartreeFock, BonetMonroig2020}, or estimated with fermionic classical shadows~\cite{zhao2021fermionic, low2022classical}. On the quantum side, fault-tolerant algorithms simulate Hamiltonian dynamics, including that of time-dependent Hamiltonians~\cite{Berry2020TimeDependent}, with rigorous error bounds. They range from product formulas~\cite{lloyd1996, Childs2021TrotterError} to linear combinations of unitaries, truncated Taylor series, quantum signal processing and qubitization~\cite{Childs:2012, Berry:2015, Low2017QSP, Low2019hamiltonian}, and their circuit depths are suited to future fault-tolerant hardware. For near-term devices, variational quantum algorithms~\cite{cerezo2021variational, Bharti2022noisyiq} represent the evolving state by a parametrized circuit with a structure fixed in advance. Its parameters are propagated according to a variational principle~\cite{Li_Benjamin-PRX-2017, theory_vqs} or by projected updates~\cite{Barison2021efficientquantum}. Adaptive variational algorithms, such as the adaptive variational quantum dynamics simulation (AVQDS)~\cite{AVQDS, Zhang2025AdaptiveVQ} and the adaptive projected variational quantum dynamics~\cite{Linteau2024adaptiveprojectedVQD}, instead grow the ansatz during the evolution and thereby construct compact problem-specific circuits along the dynamical path. AVQDS and its imaginary-time counterpart have been applied to Green's functions and nonlinear susceptibilities~\cite{Gomes2023, mootz2024adaptive}, thermal averages of observables from minimally entangled typical thermal states~\cite{getelina2023adaptive} and classical transport equations~\cite{Alipanah2025advection}.

In this work, we develop the time-dependent GQCE framework, abbreviated t-GQCE, for the real-time dynamics of generic multi-orbital correlated electron systems with time-dependent Hamiltonians within the time-dependent ghost Gutzwiller approximation. The quasiparticle dynamics is propagated on classical processors, and the embedding states are propagated on quantum processors. The framework does not depend on the choice of the quantum dynamics solver: near-term variational algorithms and fault-tolerant Hamiltonian simulation algorithms can be used alike. For demonstration, we adopt AVQDS as the solver and take interaction quenches of the single-band Hubbard model on the Bethe lattice as an example. Without ghost orbitals, the embedding problem reduces to a single-qubit model. We simulate its dynamics on IBM quantum hardware using two physical qubits in total, since an ancilla qubit is needed for the Hadamard-test measurements of the quantum metric that enters the equations of motion. With $B=3$, $5$, and $7$, t-GQCE suppresses the artificial oscillations of the conventional Gutzwiller approximation and approaches the nonequilibrium DMFT results with increasing $B$ in both the weak and strong quench regimes. In statevector simulations, AVQDS follows the embedding states with infidelities below $5\times10^{-4}$ using circuits of a few hundred to about $3700$ CNOT gates. These simulations are driven by the embedding Hamiltonians from classical t-GQCE simulations, so that they assess AVQDS as the solver for the embedding states along the physical dynamical path. A fully self-consistent simulation with the quantum solver in the loop is left for future work. Finally, we assess how accurately the embedding observables can be measured on near-term quantum hardware. We use a noise model of a trapped-ion processor that was previously found to give results similar to those measured on the hardware itself~\cite{chen2025quantum}, as well as the Quantinuum H2-Emulator. Circuit compression, measurements in rotated orbital bases with post-selection, and noise mitigation reduce the errors two- to threefold. At the benchmarked error rates, the residual noise bias still exceeds the target accuracy of $0.01$ adopted in earlier GQCE work. Once the error rates are reduced threefold, this target is reached for the single-particle density matrix and the double occupancy with $10^5$ shots per time step. Runs on the Quantinuum H2-Emulator give errors of similar size.

The paper is organized as follows. Section~\ref{sec: framework} presents the t-GQCE framework and its implementation with AVQDS. Section~\ref{sec: hubbard} introduces the Hubbard model and the quench protocol used as the example. Sections~\ref{sec: b1} and~\ref{sec: bgt1} present the results without and with ghost orbitals, respectively. Section~\ref{sec: hardware} addresses the measurement of the embedding observables on quantum hardware, and Sec.~\ref{sec: conclusion} concludes with an outlook.

\begin{figure*}[t]
	\centering
	\includegraphics[width=\textwidth]{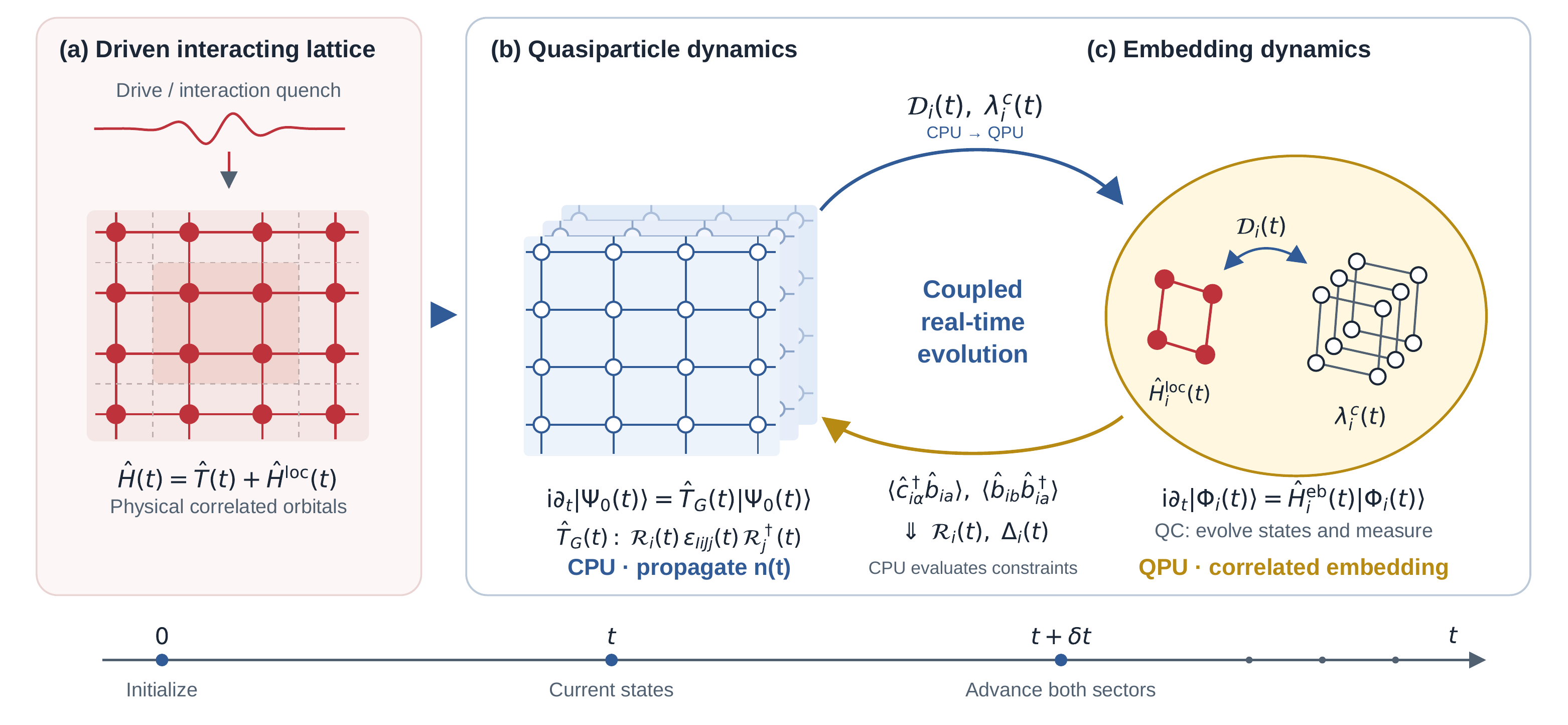}
	\caption{
	\textbf{Schematic illustration of the t-GQCE framework.} (a) An interacting fermionic lattice model with Hamiltonian $\h(t) = \hat{T}(t) + \h^\mathrm{loc}(t)$, driven out of equilibrium by a time-dependent field or an interaction quench. The unit cell, shaded, is treated as a single group $i$ of correlated orbitals. (b) Within t-GQCE, the dynamics of the lattice model is mapped to the coupled real-time evolution of a noninteracting quasiparticle lattice and of interacting embedding models. The quasiparticle state $\ket{\Psi_0(t)}$ evolves under $\hat{\mathcal{T}}_\mathrm{G}(t)$, whose inter-cell hoppings are renormalized to $\R_i(t)\, \varepsilon_{IiJj}(t)\, \R_j^\dag(t)$ and whose unit cell contains $B$ times as many auxiliary orbitals. Its density matrix $n(t)$ is propagated on CPUs. (c) The embedding state $\ket{\Phi_i(t)}$ evolves under $\h^\mathrm{eb}_i(t)$, composed of $\h^\mathrm{loc}_i$ on the physical orbitals, shown in red, and the bath orbitals described by $\lambda^\mathrm{c}_i(t)$, coupled with strength $\D_i(t)$. It is propagated and measured on QPUs. At each time step, the CPU passes $\D_i(t)$ and $\lambda^\mathrm{c}_i(t)$ to the QPU. The single-particle density matrix elements $\langle \cc_{i\alpha}\ba_{ia}\rangle$ and $\langle \ba_{ib}\bc_{ia}\rangle$ measured on the QPU return to the CPU, which evaluates $\R_i(t)$ and $\Delta_i(t)$. Both sectors then advance together from $t$ to $t+\delta t$ along the same physical time axis, starting from the initial states at $t=0$.
	}
	\label{fig: gqceg}
\end{figure*}

\section{General exposition of the time-dependent GQCE framework}\label{sec: framework}
We consider the real-time evolution of a generic multi-orbital interacting fermionic system on a periodic lattice, described by a time-dependent Hamiltonian:
\begin{equation}
\begin{aligned}
\h(t) = &\sum_{I,J=1}^{\mathcal{N}}\sum_{i,j=0}^{\mathcal{C}}\sum_{\alpha=1}^{\nu_i}\sum_{\beta=1}^{\nu_j} 
\varepsilon_{IiJj}^{\alpha\beta}(t)\, \cc_{Ii\alpha} \ca_{Jj\beta} \\ 
&+ \sum_{I=1}^{\mathcal{N}}\sum_{i=1}^{\mathcal{C}} \h^\mathrm{loc}_{Ii}(t).
\label{eq: model}
\end{aligned}  
\end{equation}
Here $I,J$ index the $\mathcal{N}$ unit cells, and each unit cell contains $\mathcal{C}+1$ groups of orbitals labeled by $i=0,\dots,\mathcal{C}$. The $i=0$ group collects ``uncorrelated'' orbitals, often extended $s$- or $p$-states, while the groups $i\geq 1$ contain ``correlated'' orbitals, typically $d$- or $f$-states, with screened Coulomb interactions. The indices $\alpha$ and $\beta$ run over the $\nu_i$ and $\nu_j$ spin-orbitals within groups $i$ and $j$, respectively. 

The first term in Eq.~\eqref{eq: model}, the hopping term $\hat{T}(t)$, contains hopping amplitudes $\varepsilon_{IiJj}^{\alpha\beta}(t)$ between creation ($\cc$) and annihilation ($\ca$) operators. The second term $\h^\mathrm{loc}_{Ii}(t)$ contains all local one-body terms, such as the crystal field and spin-orbit coupling, and the two-body interaction terms defined on the correlated orbitals in group $i$. For $i\geq 1$, on-site hoppings $\varepsilon_{IiIi}^{\alpha\beta}(t)$ are set to zero so that all local physics is carried by $\h^\mathrm{loc}_{Ii}(t)$. Both hopping and local terms can, in principle, be time-dependent, for example, due to lattice vibrations, when modeling electric or magnetic field pulses by Peierls substitution with minimal coupling, or under interaction quenches.

To approximate the many-body state at time $t$, we employ the time-dependent ghost Gutzwiller ansatz~\cite{Guerci2023TDgGA}:
\be
\ket{\Psi_\mathrm{G}(t)} = \hat{\mathcal{P}}_\mathrm{G}(t)\,\ket{\Psi_0(t)},
\label{eq:gwf}
\ee
where $\ket{\Psi_0(t)}$ is an uncorrelated single-particle reference state, a Slater determinant, and $\hat{\mathcal{P}}_\mathrm{G}(t)$ is a product of local projectors acting on the correlated subspaces:
\be
\hat{\mathcal{P}}_\mathrm{G}(t) = \prod_{I=1}^{\mathcal{N}}\prod_{i=1}^{\mathcal{C}} \hat{\mathcal{P}}_{Ii}(t).
\ee
Each $\hat{\mathcal{P}}_{Ii}(t)$ maps between the physical Fock space of the $\nu_i$ correlated $c$-modes and an enlarged auxiliary space with $B\nu_i$ $f$-modes including the $(B-1)\nu_i$ ghost orbitals:
\be
\hat{\mathcal{P}}_{Ii} (t) = \sum_{\Gamma=0}^{2^{\nu_i}-1}\sum_{n=0}^{2^{B\nu_i}-1} [\Lambda_i(t)]_{\Gamma n}\, \ket{\Gamma, Ii}\bra{n, Ii}.
\ee
Here $\ket{\Gamma, Ii} = \prod_{\alpha=1}^{\nu_i} (\cc_{Ii\alpha})^{q_\alpha(\Gamma)}\ket{0}$ denotes a correlated-site Fock state specified by the occupation pattern of the $c$-modes, and $\ket{n, Ii} = \prod_{a=1}^{B\nu_i} (\fc_{Iia})^{q_a(n)}\ket{0}$ an $f$-mode Fock state. The binary digits of the integers $\Gamma$ and $n$ give the respective occupations $q_\alpha(\Gamma)$ and $q_a(n)$. The bath orbital number $B$ controls the size of the ghost sector: $B=1$ recovers the conventional Gutzwiller wave function without ghost orbitals, while $B>1$ introduces $(B-1)\nu_i$ auxiliary modes per correlated group, enabling a richer variational description. For simplicity of notation, we hereafter drop the parameter $t$ by assuming implicit time dependence, and omit the unit cell index $I$ for quantities that are translationally invariant.

The matrices $\Lambda_i(t)$ contain the time-dependent variational parameters. Symmetry constraints can restrict their structure: for example, in a $\mathrm{U}(1)$-symmetric, particle-number-conserving setting, one must have
\be
\sum_{a=1}^{B\nu_i} q_a(n) - \sum_{\alpha=1}^{\nu_i} q_\alpha(\Gamma) = m_i
\quad \text{whenever} \quad [\Lambda_i(t)]_{\Gamma n} \neq 0,
\ee
where we fix $m_i=(B-1)\nu_i/2$ in practice.

The dynamics can be studied by the Dirac--Frenkel variational principle, which yields the following Lagrange function within the time-dependent ghost Gutzwiller approximation~\cite{Guerci2023TDgGA}:
{\allowdisplaybreaks
\begin{align}
	\mathcal{L} =& \frac{1}{\mathcal{N}} \avihi{\Psi_0(t)}{i\partial_t-\hat{\mathcal{T}}_\textrm{G}(t)} \nonumber \\
	& + \sum_i \avihi{\Phi_i(t)}{i\partial_t-\h^\mathrm{eb}_{i}(t)} \nonumber \\
	& + \sum_i \sum_{a,b=1}^{B\nu_i}[\lambda_i^\textrm{c}(t)]_{ab}[\Delta_i (t)]_{ab} \nonumber \\
	& + \sum_i \sum_{a,b=1}^{B\nu_i}\sum_{\alpha=1}^{\nu_i} \Big[[\D_i(t)]_{a\alpha}[\R_i(t)]_{b\alpha} \nonumber \\
	& \quad \times \left[\Delta_i(t)(1-\Delta_i(t))\right]_{ba}^{1/2} + \textrm{c.c.} \Big].
\end{align}}
Here $\ket{\Phi_i(t)}$ is the embedding state of group $i$, defined in the Fock space of the $\nu_i$ physical modes $\ca_{i\alpha}$ and the $B\nu_i$ bath modes $\ba_{ia}$, whose coefficients encode the variational matrix $\Lambda_i(t)$~\cite{Guerci2023TDgGA}. The renormalized kinetic operator is defined as:
\begin{equation}
    \hat{\mathcal{T}}_\textrm{G}(t) = \sum_{I,J=1}^{\mathcal{N}}\sum_{i,j=0}^{\mathcal{C}}\sum_{a=1}^{B\nu_i}\sum_{b=1}^{B\nu_j}\left[\R_i\varepsilon_{IiJj}\R_j^\dag\right]_{ab}\fc_{Iia}\fa_{Jjb}\,,
\end{equation}
with $\R_0\equiv 1$ for the uncorrelated group $i=0$, which carries no ghost orbitals, so that $B\nu_0$ is replaced by $\nu_0$, and the embedding Hamiltonian is defined as:
\begin{equation}
	\begin{split}
		\h^\mathrm{eb}_{i}(t) ={}& \h_i^\textrm{loc}[\cc_{i\alpha}, \ca_{i\alpha}] + \sum_{a,b=1}^{B\nu_i}[\lambda_i^\textrm{c}]_{ab}\ba_{ib}\bc_{ia} \\
		& + \sum_{a=1}^{B\nu_i}\sum_{\alpha=1}^{\nu_i}\left([\D_i]_{a\alpha} \cc_{i\alpha}\ba_{ia} + \textrm{H.c.} \right).
	\end{split} \label{eq: heb}
\end{equation}
Here $a,b$ label the bath orbitals including the spin index $\sigma$ unless $\sigma$ is explicitly present. The local quasiparticle density matrix $\Delta_i$ and the Lagrange multipliers $\{\lambda^\textrm{c}_i, \R_i, \D_i\}$ are introduced to linearize the problem. The extremization of the Lagrange function with respect to these quantities and the state functions $\{\Psi_0(t), \Phi_i(t)\}$ yields the following set of equations of motion:
{\allowdisplaybreaks
\begin{gather}
	\begin{split}
	i\partial_t n^{I'i'a'}_{J'j'b'} &- \sum_{J=1}^{\mathcal{N}}\sum_{j=0}^{\mathcal{C}}\sum_{b=1}^{B\nu_j} \left[\R_{j'}\varepsilon_{J'j'Jj}\R_j^\dag\right]_{b'b} n^{I'i'a'}_{Jjb} \\
	&+ \sum_{J=1}^{\mathcal{N}}\sum_{j=0}^{\mathcal{C}}\sum_{b=1}^{B\nu_j} \left[\R_{j}\varepsilon_{JjI'i'}\R_{i'}^\dag\right]_{ba'} n^{Jjb}_{J'j'b'} = 0,
	\end{split} \label{eq: eom_n} \\
	i\partial_t \ket{\Phi_i(t)} - \h^\mathrm{eb}_{i}(t)\ket{\Phi_i(t)} = 0, \label{eq: eom_phi} \\
	\begin{split}
	\sum_{J=1}^{\mathcal{N}}\sum_{j=0}^{\mathcal{C}}\sum_{b=1}^{B\nu_j} \sum_{\beta=1}^{\nu_j} &\varepsilon_{1iJj}^{\alpha\beta}[\R_j^\dag]_{\beta b}\, n^{1ia}_{Jjb} \\
	&- \left[\sqrt{\Delta_i(1-\Delta_i)} \D_i \right]_{a \alpha} = 0,
	\end{split} \label{eq: eom_d} \\
	\frac{\partial \left[\Tr{\R_i^\text{T}\sqrt{\Delta_i(1-\Delta_i)}\D_i} + \textrm{c.c.}\right]}{\partial \left[\Delta_i\right]_{ab}} + \left[\lambda_i^\mathrm{c}\right]_{ab} = 0, \label{eq: eom_l} \\
	\avihi{\Phi_i}{\ba_{i b}\bc_{i a}} - [\Delta_i]_{ab} = 0, \label{eq: eom_delta} \\
	\avihi{\Phi_i}{\cc_{i\alpha}\ba_{i a}} - \left[\R_i^\text{T}\sqrt{\Delta_i(1-\Delta_i)} \right]_{\alpha a} = 0, \label{eq: eom_r}
\end{gather}}
where Eq.~\eqref{eq: eom_n} is derived according to the time-dependent Schr\"odinger equation, $ i\partial_t \ket{\Psi_0(t)} = \hat{\mathcal{T}}_\textrm{G}(t)\ket{\Psi_0(t)}$, for the quasiparticle density matrix $n^{Iia}_{Jjb} \equiv \avihi{\Psi_0}{\fc_{Iia}\fa_{Jjb}}$. In Eq.~\eqref{eq: eom_d}, the unit cell index $1$ denotes a reference cell, which is equivalent to any other cell by translational invariance. We set $\hbar$ to unity.
The coupled time evolution of $n^{Iia}_{Jjb}$ and the many-body embedding states $\ket{\Phi_i(t)}$ is governed by the nonlinear first-order differential equations, Eqs.~\eqref{eq: eom_n} and~\eqref{eq: eom_phi}. The associated parameters, $\{\Delta_i, \lambda^\textrm{c}_i, \R_i, \D_i\}$, are completely determined by $n^{1ia}_{Jjb}$ and $\Phi_i(t)$ via Eqs.~\eqref{eq: eom_d}--\eqref{eq: eom_r}. Therefore, the dynamics simulations can be performed by numerically integrating the equations of motion using standard approaches such as the Runge--Kutta method.

The classical computational bottleneck for the above dynamics simulations arises from the need to evolve the many-body states $\ket{\Phi_i(t)}$, which can be computationally expensive, especially for large embedding system sizes and long time evolution. This challenge can be addressed by leveraging quantum processing units (QPUs) to propagate $\ket{\Phi_i(t)}$ under the embedding Hamiltonian $\h^\mathrm{eb}_{i}(t)$, while the dynamics of the uncorrelated reference state $\ket{\Psi_0(t)}$ can be efficiently handled on classical processing units (CPUs). This hybrid quantum-classical approach forms the basis of the t-GQCE framework, as illustrated in Fig.~\ref{fig: gqceg}. To be processed on a QPU, the $(B+1)\nu_i$ fermionic modes of the embedding model are mapped onto qubits by a fermion-to-qubit encoding. With the Jordan--Wigner encoding used in this work, each mode is represented by one qubit, so that $\ket{\Phi_i(t)}$ is a state of $(B+1)\nu_i$ qubits, or $2(B+1)$ qubits for a single-band model with $\nu_i=2$, and $\h^\mathrm{eb}_i(t)$ becomes a linear combination of Pauli strings.

Recent advances in fault-tolerant quantum algorithms, such as quantum signal processing and qubitization~\cite{Low2017QSP, Low2019hamiltonian}, have enabled the simulation of time-independent Hamiltonians with asymptotically optimal query complexity, scaling additively in the evolution time and logarithmically in the inverse target error. Time-dependent Hamiltonians such as $\h^\mathrm{eb}_i(t)$ can be simulated with a cost that scales with the $L^1$ norm of the Hamiltonian over time~\cite{Berry2020TimeDependent}. The circuit depths of these algorithms, however, are suited to fault-tolerant hardware rather than to current devices, whose qubit counts and gate fidelities continue to improve. The t-GQCE framework itself is agnostic to the choice of quantum dynamics solver. For the present demonstration, we adopt the adaptive variational quantum dynamics simulation (AVQDS) algorithm~\cite{AVQDS} as one of several near-term approaches to propagate the embedding states $\ket{\Phi_i(t)}$. AVQDS adaptively constructs problem-specific, compact quantum circuits that evolve $\ket{\Phi_i(t)}$ under $\h^\mathrm{eb}_{i}(t)$ with controllable accuracy.

Within AVQDS, the embedding state is represented by a pseudo-Trotterized ansatz:
\begin{equation}
\ket{\Phi_i(\bth[t])} = \prod_{\mu=1}^{N_{\bth}} e^{-i\theta_\mu (t) \hat{G}_\mu/2} \ket{\phi_0}\,, \label{eq: Phi}
\end{equation}
where the product is ordered such that the $\mu=1$ factor acts first on the reference state $\ket{\phi_0}$, a product state of the $(B+1)\nu_i$ qubits such as a computational basis state, and the factor $1/2$ in the exponent follows the convention of single-qubit rotation gates. This convention differs from the original formulation of AVQDS~\cite{AVQDS}, which uses $e^{-i\theta_\mu \hat{G}_\mu}$, so that the angles here are twice those of Ref.~\cite{AVQDS} and the expressions for $M$ and $V$ below are rescaled accordingly. At the initial time $t=0$, the ansatz represents the embedding ground state, as discussed below. The $N_{\bth}$ time-dependent variational parameters $\{\theta_\mu(t)\}$ are associated with Hermitian generators $\{\hat{G}_\mu\}$, typically Pauli strings. The equation of motion for $\bth$ follows from the McLachlan variational principle, which is formulated at the level of the density matrix $\varrho_i = \ket{\Phi_i}\bra{\Phi_i}$ to eliminate the need to track the global phase of $\ket{\Phi_i}$~\cite{AVQDS}. The McLachlan distance between the variationally evolving state and the exactly propagating state, governed by the von Neumann equation $\dot{\varrho}_i = -i[\h_i^\mathrm{eb}, \varrho_i]$, is defined as
\begin{equation}
	L^2 = \frac{1}{2}\Big\| \sum_{\mu} \frac{\partial \varrho_i[\bth]}{\partial \theta_\mu}\dot{\theta}_\mu + i\big[\h_i^\mathrm{eb}, \varrho_i[\bth]\big] \Big\|_\mathrm{F}^2\,, \label{eq: mclachlan}
\end{equation}
where $\|A\|_\mathrm{F} = \sqrt{\Tr[A^\dag A]}$ denotes the Frobenius norm; with the prefactor $1/2$, $L^2$ equals the energy variance of the state when the parameters are held fixed. Minimizing $L^2$ with respect to $\dot{\bth}$ yields
\begin{equation}
	\sum_{\nu} M_{\mu\nu} \dot{\theta}_\nu = V_\mu\,, \label{eq: eom_theta}
\end{equation}
where the symmetric matrix $M_{\mu\nu} = \frac{1}{2}\Tr\big[\frac{\partial \varrho_i}{\partial \theta_\mu}\frac{\partial \varrho_i}{\partial \theta_\nu}\big]$, related to the quantum Fisher information matrix, is given by the real part of the quantum geometric tensor, that is, the quantum metric:
\begin{equation}
	M_{\mu\nu} = \Re\left[\ovij{\frac{\partial \Phi_i}{\partial \theta_\mu}} {\frac{\partial \Phi_i}{\partial \theta_\nu}} + \ovij{\Phi_i}{\frac{\partial \Phi_i}{\partial \theta_\mu}} \ovij{\Phi_i}{\frac{\partial \Phi_i}{\partial \theta_\nu}} \right], \label{eq: M}
\end{equation}
and the vector $V_\mu$ is given by:
\begin{equation}
	V_\mu = \Im\left[\avihj{\frac{\partial \Phi_i}{\partial \theta_\mu}}{\h_i^\mathrm{eb}}{\Phi_i} + \ovij{\Phi_i}{\frac{\partial \Phi_i}{\partial \theta_\mu}} \avihi{\Phi_i}{\h_i^\mathrm{eb}}\right]. \label{eq: V}
\end{equation}
Both $M$ and $V$ are expectation values that can be measured on QPUs, e.g., using Hadamard-test circuits. The second terms in Eqs.~\eqref{eq: M} and~\eqref{eq: V} originate from the global phase contribution, which is eliminated in the density-matrix formulation; they should be dropped if the ansatz tracks the global phase explicitly through a dedicated variational parameter. The accuracy of the ansatz at each time step is quantified by the residual McLachlan distance, i.e., $L^2$ evaluated at the optimal $\dot{\bth}$ solving Eq.~\eqref{eq: eom_theta}:
\begin{equation}
	L^2_\mathrm{min} = \mathrm{Var}_{\bth}\big[\h_i^\mathrm{eb}\big] - \sum_{\mu} V_\mu \dot{\theta}_\mu, \label{eq: mclachlan_min}
\end{equation}
where $\mathrm{Var}_{\bth}[\h_i^\mathrm{eb}] = \avihi{\Phi_i}{(\h_i^\mathrm{eb})^2} - \avihi{\Phi_i}{\h_i^\mathrm{eb}}^2$ is the energy variance of the variational state. The residual $L^2_\mathrm{min}$ vanishes only if the ansatz can exactly follow the time evolution. Along the dynamical path, whenever $L^2_\mathrm{min}$ exceeds a preset threshold, the ansatz is expanded by appending the generator from a predefined operator pool that most reduces $L^2_\mathrm{min}$. Each new parameter is initialized to zero, so that the state stays continuous in time, and the expansion is repeated until $L^2_\mathrm{min}$ falls below the threshold. In practice, we find it beneficial to cap the number of generators appended per time step, as implemented in our AVQDS code~\cite{CyQC}. If $L^2_\mathrm{min}$ still exceeds the threshold when the cap is reached, the step is taken with the enlarged ansatz and $L^2_\mathrm{min}$ slightly above the threshold, and the expansion continues at the next step. Compared with expanding the ansatz until the threshold is met at every step, this can yield shallower circuits with comparable final accuracy.

The initial embedding state before the dynamics starts, $\ket{\Phi_i(\bth[0])}$, is the ground state of $\h^\mathrm{eb}_i(t=0)$ in the self-consistent equilibrium GQCE solution~\cite{gqce, chen2025quantum}. Within GQCE, this ground state is prepared on QPUs using the adaptive variational quantum imaginary-time evolution (AVQITE) algorithm~\cite{AVQITE}, the imaginary-time counterpart of AVQDS. AVQITE adaptively constructs a compact circuit of the same pseudo-Trotterized form as Eq.~\eqref{eq: Phi} acting on the product state $\ket{\phi_0}$. The resulting AVQITE generators and optimized angles define the initial ansatz of AVQDS at $t=0$, so that $N_{\bth}>0$ in general. Importantly, these angles are not held fixed. They are propagated according to Eq.~\eqref{eq: eom_theta} together with the angles of the generators subsequently appended by AVQDS, which gives the ansatz more flexibility to follow the dynamics without additional circuit depth~\cite{mootz2024adaptive}.

In summary, each time step of the t-GQCE dynamics with AVQDS proceeds as follows. (i) The single-particle density matrices of the embedding states in Eqs.~\eqref{eq: eom_delta} and~\eqref{eq: eom_r} are measured on QPUs, which determine $\Delta_i$ and $\R_i$. (ii) Together with the quasiparticle density matrix $n^{Iia}_{Jjb}$, they determine $\D_i$ and $\lambda_i^\mathrm{c}$ via Eqs.~\eqref{eq: eom_d} and~\eqref{eq: eom_l} on CPUs, which updates the embedding Hamiltonians $\h^\mathrm{eb}_i(t)$. (iii) The matrix $M$ and vector $V$ for each embedding Hamiltonian are measured on QPUs. (iv) The quasiparticle density matrix and variational parameters $\bth$ are advanced on CPUs by integrating Eqs.~\eqref{eq: eom_n} and~\eqref{eq: eom_theta}.

In simple cases, this workflow can be greatly simplified. For instance, for the single-band Hubbard model with $B=1$ presented below, the quasiparticle density matrix remains time-independent, and the embedding Hamiltonian depends only on a single expectation value of the embedding state. The full t-GQCE dynamics then reduces to a single-qubit problem with an explicitly known ansatz. In general, however, the cost of t-GQCE is dominated by the quantum resources needed to propagate the embedding states. Since this work aims to present the t-GQCE framework with an emphasis on the associated circuit complexity, for the demonstrations with $B=3$, $5$, and $7$ below we decouple the AVQDS simulations from the self-consistent feedback loop. Specifically, we first perform t-GQCE simulations on CPUs, where the embedding states are propagated by exact diagonalization (ED), and extract the resulting time-dependent embedding Hamiltonians $\h^\mathrm{eb}_i(t)$. AVQDS simulations are then performed with the so-obtained $\h^\mathrm{eb}_i(t)$. This allows a direct assessment of the accuracy and circuit complexity of AVQDS in propagating the embedding states along the physical t-GQCE dynamical path.

\section{Hubbard model under interaction quenches}\label{sec: hubbard}
To illustrate the application of the t-GQCE framework, we consider the single-band Hubbard model on a Bethe lattice, defined by:
\be
\h = \sum_{\braket{I,J}}\sum_{\sigma}\varepsilon\, \cc_{I\sigma}\ca_{J\sigma} + \sum_{I} \frac{U(t)}{2} (\hat{n}_I - 1)^2\,. \label{eq: sbhm}
\ee
In this model, $U(t)$ denotes the time-dependent onsite interaction strength, and $\hat{n}_I = \sum_{\sigma=\uparrow, \downarrow} \cc_{I\sigma}\ca_{I\sigma}$ is the electron number operator at site $I$. Unlike the general multi-orbital case in Eq.~\eqref{eq: model}, this Hamiltonian contains only one correlated orbital, so the summations over orbital groups are omitted and the spin-orbital indices reduce to the spin index $\sigma$. The notation $\braket{I,J}$ indicates summation over nearest-neighbor sites. We consider the Bethe lattice in the limit of infinite coordination. With the hopping amplitude $\varepsilon=W/(2\sqrt{z})$ scaled with the coordination number $z\to\infty$, the kinetic term has a semicircular density of states, $\rho_0(\omega) = \frac{2}{\pi W}\sqrt{1-(\omega/W)^2}$, where the half-bandwidth $W$ is set to unity, $W=1$, and serves as the energy unit. With the chemical potential set to zero in the following simulations, the interaction term ensures particle-hole symmetry and $\braket{\hat{n}_I} = 1$ in the ground state.

The embedding Hamiltonian~\eqref{eq: heb} can now be simplified as follows:
\begin{equation}
	\begin{aligned}
		\h^\mathrm{eb} =& \frac{U(t)}{2} (\hat{n} - 1)^2 + \sum_{a=1}^{B}\sum_{\sigma=\uparrow, \downarrow}(\D_a\cc_\sigma\ba_{a\sigma} + \text{H.c.}) \\
		&+\sum_{a,b=1}^{B}\sum_{\sigma=\uparrow, \downarrow}\lambda^\text{c}_{ab} \ba_{b\sigma}\bc_{a\sigma},
	\end{aligned}
\end{equation}
where we omit the site index since there is only one unique embedding Hamiltonian. Note that the spin index $\sigma$ is explicitly included here, so that $a,b$ run only over the $B$ bath orbitals. Due to spin degeneracy in the paramagnetic phase, $\D$ and $\lambda^\textrm{c}$ do not carry a spin index. 

Equations~\eqref{eq: eom_n} and~\eqref{eq: eom_d} associated with the quasiparticle density matrix reduce to the following forms:
\begin{align}
	i\partial_t n_{ab}(\omega) - \omega\sum_{b'=1}^{B} \left[\R_{b} \R^\dag_{b'} n_{ab'}(\omega) - \R_{b'} \R^\dag_{a} n_{b'b}(\omega)\right] &= 0, \label{eq: eom_n_sbhm} \\
	\sum_{b=1}^{B} \int_{-W}^{W} d\omega \rho_0(\omega)\omega \R^\dag_b n_{ab}(\omega) - \left[\sqrt{\Delta(1-\Delta)} \D \right]_{a} &= 0\,, \label{eq: eom_d_sbhm}
\end{align}
where $n_{ab}(\omega) \equiv \braket{\Psi_0(t) | \fc_{\omega a} \fa_{\omega b} | \Psi_0(t)}$ is the quasiparticle density matrix in the energy eigenbasis of the hopping part of the Hubbard model Hamiltonian~\eqref{eq: sbhm}, whose density of states is $\rho_0(\omega)$.

We consider the time evolution of the Hubbard model under interaction quenching, where the onsite interaction strength $U(t)$ is initially set to $U(t<0)=0$, and then quenched to a finite value $U(t\ge 0) = U_\textrm{f}$ at time $t=0$. Some characteristic behaviors for this protocol have been established~\cite{Moeckel08InteractionQH, Eckstein09ThermalizationIQ}. In particular, two qualitatively different dynamical regimes are observed, with a crossover at $U_\textrm{c}\approx1.6$~\cite{Eckstein09ThermalizationIQ}. For weak quenches with $U_\textrm{f} < U_\textrm{c}$, the system rapidly approaches a ``prethermalized'' plateau, which is governed by quasi-conserved quantities and can be captured by low-order perturbation theory. For strong quenches with $U_\textrm{f} > U_\textrm{c}$, the dynamics is dominated by pronounced collapse-and-revival oscillations with a frequency set by $U_\textrm{f}$ and a slow damping arising from the kinetic terms. Thermalization generally requires a long-time simulation. The exception is the special quench to $U_\textrm{f} = U_\textrm{c}$, where the system rapidly relaxes and observables, such as the double occupancy, $\braket{\uparrow\downarrow} = \avihi{\Phi}{\cc_{\uparrow}\ca_{\uparrow}\cc_{\downarrow}\ca_{\downarrow}}$, take on their thermal values.

\begin{figure}[ht]
	\centering
	\includegraphics[width=\linewidth]{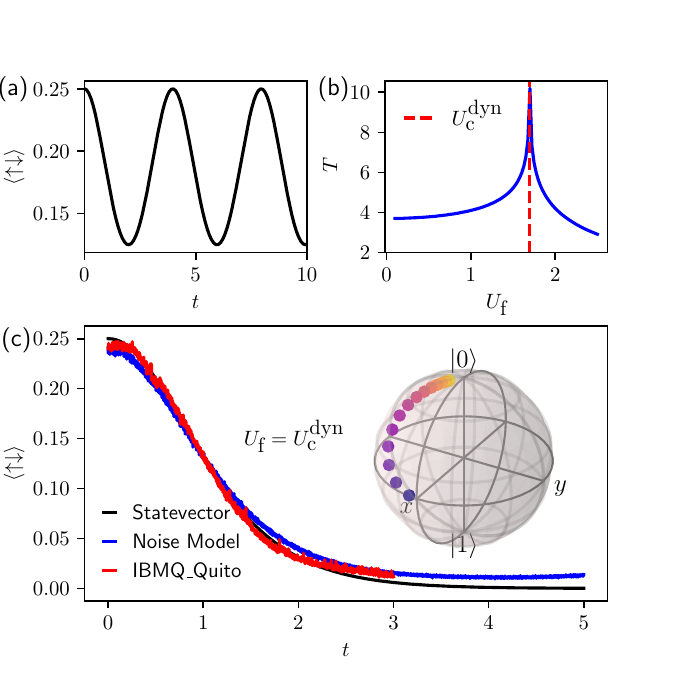}
	\caption{
	\textbf{Quench dynamics simulations of the Hubbard model using t-GQCE with $B=1$.} 
	(a) A representative periodic oscillation of the double occupancy $\braket{\uparrow\downarrow}$ as a function of time $t$ under interaction quenching from $U(t<0)=0$ to $U(t\ge 0)=U_\textrm{c}^\textrm{dyn}/2$. (b) The oscillation period $T$ as a function of the quench interaction strength $U_\textrm{f}$, showing a divergence at $U_\textrm{c}^\textrm{dyn}$. (c) The time evolution of the double occupancy $\braket{\uparrow\downarrow}$ under interaction quenching to $U_\textrm{f}=U_\textrm{c}^\textrm{dyn}$. The results obtained on the \texttt{IBMQ\_Quito} device and with its derived noise model are shown as red and blue lines, respectively, and the noiseless statevector simulation results as a black line for reference. Inset: real-time trace of the embedding state $\ket{\Phi}$ on the Bloch sphere. The calculations involve in total two qubits, one for the system and one ancilla, with which $M$~\eqref{eq: M} and $V$~\eqref{eq: V} are measured by Hadamard tests. At each time step, $2\times10^4$ shots are applied to each of $24$ circuits: $10$ Hadamard tests for the independent elements of the symmetric $4\times4$ matrix $M$, $12$ for the four elements of $V$, three per element for the three Pauli terms of the Hamiltonian~\eqref{eq: heb_b1q}, one circuit for $\R$, which determines $\D=-4\R/(3\pi)$ and thus the Hamiltonian, and one for $\braket{\uparrow\downarrow}$.
	}
	\label{fig: ibmq_quito_device}
\end{figure}

\section{Results with \texorpdfstring{$B=1$}{B=1}}\label{sec: b1}
In this section, we present t-GQCE results with $B=1$, corresponding to the conventional time-dependent Gutzwiller approximation without ghost orbitals. Following a sudden quench of the interaction strength from 0 to a finite $U_\textrm{f}$, the double occupancy $\braket{\uparrow\downarrow}$ exhibits undamped periodic oscillations, as shown in Fig.~\figref{fig: ibmq_quito_device}{(a)}. The oscillation period diverges at an intermediate dynamical critical interaction strength $U_\textrm{c}^\textrm{dyn} = U_\textrm{c}^\textrm{BR}/2\approx U_\textrm{c}$ as illustrated in Fig.~\figref{fig: ibmq_quito_device}{(b)}, where $U_\textrm{c}^\textrm{BR}=\frac{32}{3\pi}$ is the Brinkman--Rice critical interaction strength. The coherent oscillation of the double occupancy largely reflects the restricted treatment of quantum fluctuations in the Gutzwiller approximation with $B=1$. The short-time dynamics and the dynamical critical behavior at $U_\textrm{c}^\textrm{dyn}$, however, are qualitatively consistent with accurate nonequilibrium DMFT results~\cite{Eckstein09ThermalizationIQ}. Therefore, we focus on the quench dynamics at the critical interaction strength as a benchmark example for t-GQCE.

The embedding Hamiltonian with $B=1$ reduces to a two-site Anderson impurity model:
\begin{equation}
	\h^\mathrm{eb} = \frac{U(t)}{2} (\hat{n} - 1)^2 + \sum_{\sigma=\uparrow, \downarrow}\left[\D(t) \cc_\sigma\ba_{\sigma} + \text{H.c.}\right] \,. \label{eq: heb_b1}
\end{equation}
The embedding state holds two electrons in the impurity and bath orbitals, and the ground state of $\h^\mathrm{eb}$, corresponding to the paramagnetic solution of the Hubbard model, is a spin singlet. Since $\h^\mathrm{eb}(t)$ conserves the particle number and the total spin, the embedding state remains in the space of the three two-electron states with $S=0$. These are the singlet with one electron on each site and the two states with both electrons on the same site, $\ket{\uparrow\downarrow,0}$ and $\ket{0,\uparrow\downarrow}$, written with the impurity occupation first and the bath occupation second. Particle-hole symmetry exchanges the latter two, so that only their symmetric combination enters, and the embedding state takes the form:
\begin{equation}
	\ket{\Phi} = \frac{c_1(t)}{\sqrt{2}}(-\ket{\uparrow,\downarrow} + \ket{\downarrow,\uparrow}) + \frac{c_2(t)}{\sqrt{2}}(\ket{\uparrow\downarrow,0} + \ket{0,\uparrow\downarrow})\,.
\end{equation}
Note that the solution satisfies the particle-hole symmetry of the model, which also mandates $\lambda^\text{c}=0$ in the embedding Hamiltonian~\eqref{eq: heb_b1}. 
As expected, Eq.~\eqref{eq: eom_delta} reduces to a constant $\Delta(t) = 1/2$ due to normalization of $\ket{\Phi}$, and Eq.~\eqref{eq: eom_r} leads to $\R(t) = -2\Re{c_1(t)c_2^*(t)}$. The quasiparticle density matrix $n_\sigma (\omega) = \braket{\Psi_0(t) | \fc_{\omega \sigma} \fa_{\omega \sigma} | \Psi_0(t)} = f(\omega)$ is time-independent, with Eq.~\eqref{eq: eom_n_sbhm} simplified as $\partial_t n_\sigma(\omega) = 0$. Here $f(\omega)$ is the Fermi--Dirac distribution function at zero temperature. Equation~\eqref{eq: eom_d_sbhm} yields $\D(t) = -4\R (t) / (3\pi)$. The dynamics is fully determined by the time evolution of the embedding wave function via two coefficients $c_1(t)$ and $c_2(t)$, which can be cast into the following form by referencing Eq.~\eqref{eq: eom_phi} and the embedding Hamiltonian~\eqref{eq: heb_b1}:
\begin{equation}
	i\partial_t \ket{\Phi(t)} = \left[\frac{U(t)}{4} (\sigma^0 - \sigma^z) + \frac{U_\textrm{c}^\textrm{BR}}{4}\R(t)\sigma^x\right] \ket{\Phi(t)}\,. \label{eq: heb_b1q}
\end{equation}
Here $\ket{\Phi(t)}=c_1(t)\ket{0} + c_2(t)\ket{1}$, $\sigma^x$ and $\sigma^z$ are the Pauli matrices, and $\sigma^0$ is the identity matrix. The renormalization factor can be measured as $\R(t) = -\avihi{\Phi(t)}{\sigma^x}$.

We adopt the following ansatz for the dynamics simulations of the above single-qubit model~\eqref{eq: heb_b1q} to maintain a simple circuit along the dynamical path:
\begin{equation}
	\ket{\Phi(\bth[t])} = e^{-i\theta_4\sigma^0/2} e^{-i\theta_3\sigma^z/2} e^{-i\theta_2\sigma^x/2} e^{-i\theta_1\sigma^z/2} \ket{+}\,, \label{eq: phi_b1}
\end{equation}
where three angles $\theta_1$, $\theta_2$, and $\theta_3$ fully specify the single-qubit unitary. We introduce $\theta_4$ to explicitly track the global phase. The state $\ket{+} = \frac{\ket{0} + \ket{1}}{\sqrt{2}}$ corresponds to the ground state of the embedding Hamiltonian for the noninteracting model, $U(t<0)=0$. Since the ansatz is completely general, we can simply follow the time evolution of the angles $\theta_\mu(t)$ according to Eq.~\eqref{eq: eom_theta} to capture the dynamics of the system without incurring the adaptive ansatz growth overhead in AVQDS. In the circuit implementation, we introduce an ancilla qubit to measure the matrix $M$~\eqref{eq: M} and the vector $V$~\eqref{eq: V} via Hadamard tests, so that the hardware simulations of this single-qubit problem use two physical qubits in total. Note that the second terms in $M$~\eqref{eq: M} and $V$~\eqref{eq: V} originate from the contribution of the global phase and will be omitted here since the ansatz explicitly encodes the global phase via $\theta_4$.

In the calculations, the angles $\bth(t)$ are initialized to zero and evolve by numerically integrating Eq.~\eqref{eq: eom_theta} using the Euler method with a time step of $\delta t = 0.001$. To solve for $\dot{\bth}$ in Eq.~\eqref{eq: eom_theta} at each time step, we adopt the optimization method with bound constraints on the variables as implemented in the SciPy function \texttt{scipy.optimize.lsq\_linear}, with the bounds set to $[-2, 2]$.

Figure~\figref{fig: ibmq_quito_device}{(c)} illustrates the time evolution of the double occupancy, $\braket{\uparrow\downarrow} = \avihi{\Phi}{\frac{\sigma^0-\sigma^z}{4}}$, under interaction quenching to the critical value $U_\textrm{f} = U_\textrm{c}^\textrm{dyn}$, simulated using the t-GQCE framework with $B=1$. The results from the \texttt{IBMQ\_Quito} device, shown as a red line, are compared with those from its noise model, shown as a blue line, and with the noiseless statevector simulation, shown as a black line.

The noiseless statevector simulation reveals a monotonic decrease in the double occupancy from its initial value of $0.25$ at $U=0$, consistent with the diverging oscillation period characteristic of quenching at the critical interaction strength $U_\textrm{c}^\textrm{dyn}$, as shown in Fig.~\figref{fig: ibmq_quito_device}{(b)}. The results obtained using the noise model closely align with the noiseless simulation, reaching a slightly higher double occupancy at $t=5$. The experimental results on the \texttt{IBMQ\_Quito} device also closely resemble the noiseless simulation, benefiting from the high fidelity of single-qubit gates and the limited use of two-qubit gates, at most two, in the Hadamard-test circuits.

The noise model simulations incorporate a readout error mitigation technique, which is based on the inversion of the confusion matrix characterizing the probabilities of measuring each output state given the true quantum state. This technique is not applied to the experimental results obtained on the \texttt{IBMQ\_Quito} device. The close agreement between the noise model simulations and the noiseless statevector results indicates that the impact of noise is minimal for the single-qubit model during the simulated time period. The inset of Fig.~\figref{fig: ibmq_quito_device}{(c)} illustrates the real-time evolution of the embedding state $\ket{\Phi}$ on the Bloch sphere. The state transitions from the initial $\ket{+}$ state toward $\ket{0}$, corresponding to a decay of the double occupancy.

\section{Results with \texorpdfstring{$B>1$}{B>1}}\label{sec: bgt1}
The t-GQCE framework with $B>1$ introduces ghost orbitals, which enable a richer variational description of the embedding state and can capture quantum fluctuations beyond the conventional Gutzwiller approximation. This improvement is expected to yield more accurate dynamics simulations, which approach the nonequilibrium DMFT results with increasing $B$, as found in previous benchmarks~\cite{Guerci2023TDgGA}. In this section, we present benchmark calculations of t-GQCE with $B=3$, $5$, and $7$. They show significant improvements in capturing the quench dynamics of the Hubbard model, especially the suppression of the artificial oscillations of the double occupancy observed with $B=1$ in both the weak and strong quench regimes.

With $B>1$, the dynamics of the embedding state $\ket{\Phi(t)}$~\eqref{eq: eom_phi} becomes more complex due to the enlarged Hilbert space from the ghost orbitals and cannot be mapped to a single-qubit problem. Therefore, we adopt the AVQDS algorithm to simulate the embedding state dynamics, where the ansatz circuit is adaptively constructed along the dynamical path. 
Additionally, there is a technical difficulty in obtaining the initial equilibrium GQCE solution at $U=0$ with $B>1$ due to a degeneracy of the energy landscape. To address this issue, similar to Ref.~\cite{Guerci2023TDgGA}, we instead start from the converged GQCE solution at a small interaction strength $U(t<0)=0.05$, which is close to the noninteracting limit and can be easily obtained by numerical calculations. The dynamics simulations are then performed by quenching the interaction strength from $U=0.05$ to the target value $U_\textrm{f}$.

\begin{figure*}[ht]
	\centering
	\includegraphics[width=\textwidth]{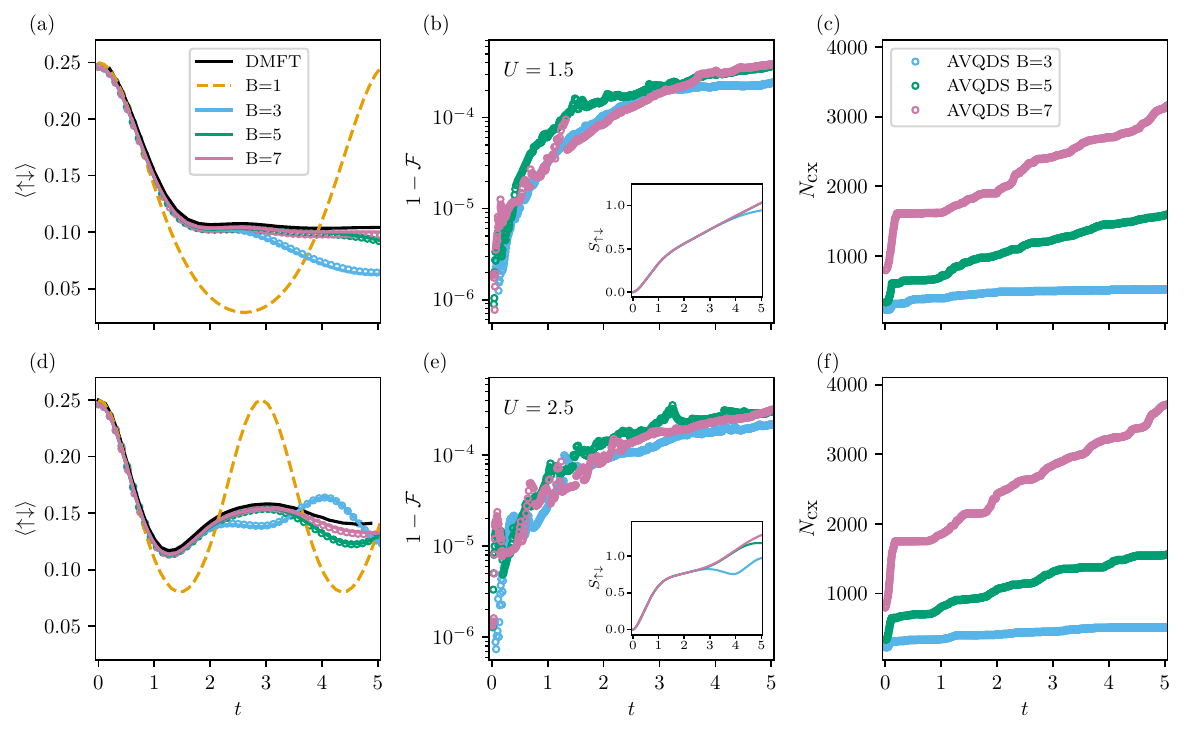}
	\caption{
	\textbf{Quench dynamics simulations of the Hubbard model using t-GQCE with different numbers of bath orbitals, $B=3,5,7$.}
	The upper (a--c) and lower (d--f) rows correspond to interaction quenches from $U(t<0)=0.05$ to the weak quench $U_\textrm{f}=1.5<U_\textrm{c}\approx1.6$ and the strong quench $U_\textrm{f}=2.5>U_\textrm{c}$, respectively.
	(a, d) Time evolution of the double occupancy $\braket{\uparrow\downarrow}$. The t-GQCE results with $B=3$, $5$, and $7$ are shown in blue, green, and pink, with the embedding states propagated by ED as solid lines and by AVQDS as open circles. AVQDS is driven by the time-dependent embedding Hamiltonians extracted from the ED-based t-GQCE simulations. The results are compared with the conventional time-dependent Gutzwiller results with $B=1$, shown as orange dashed lines, and with the nonequilibrium DMFT benchmark~\cite{Eckstein09ThermalizationIQ}, shown as black lines, which starts from $U=0$ exactly.
	(b, e) Infidelity $1-\mathcal{F}$ of the AVQDS embedding state $\ket{\Phi(\bth[t])}$ with respect to the ED-propagated embedding state $\ket{\Phi_\mathrm{ED}(t)}$ along the same dynamical path, with $\mathcal{F}=|\braket{\Phi_\mathrm{ED}(t)|\Phi(\bth[t])}|^2$. Insets: entanglement entropy $S_{\uparrow\downarrow}$ between the spin-up and spin-down spin-orbitals of the ED-propagated embedding state, i.e., the von Neumann entropy of the reduced density matrix of the spin-up spin-orbitals, which vanishes for an uncorrelated embedding state.
	(c, f) Number of CNOT gates, $N_\textrm{cx}$, in the adaptively constructed AVQDS ansatz circuit as a function of time $t$.
	}
	\label{fig: B35}
\end{figure*}

Figure~\ref{fig: B35} summarizes the t-GQCE results with $B=3$, $5$, and $7$ for two representative quenches: $U_\textrm{f}=1.5$ in the weak quench regime and $U_\textrm{f}=2.5$ in the strong quench regime. The embedding Hamiltonian is now defined on $2(B+1)$ spin-orbitals, which amounts to $8$, $12$, and $16$ qubits for $B=3$, $5$, and $7$, respectively. Following the protocol described in Sec.~\ref{sec: framework}, the time-dependent embedding Hamiltonian $\h^\mathrm{eb}(t)$ is extracted from t-GQCE simulations with ED for the embedding state evolution, and is then used to drive the AVQDS simulations. The choice of the reference state $\ket{\phi_0}$ for AVQITE becomes important with increasing $B$. For $B=3$ and $5$, we use the half-filled product state with the impurity orbital doubly occupied and a contiguous block of bath orbitals filled. From this state, AVQITE reaches fidelities above $0.9999$ with $48$ Pauli-rotation generators for $B=3$ and $98$ for $B=5$. For $B=7$, this state has a negligible squared overlap of about $10^{-13}$ with the embedding ground state, and AVQITE starting from it relaxes into a low-lying excited state. We therefore take as $\ket{\phi_0}$ the computational basis state with the lowest energy expectation value. It keeps the impurity orbital doubly occupied but places one electron in each particle-hole pair of bath orbitals, and has a squared overlap of $0.18$ with the ground state. From this state, AVQITE reaches a fidelity of $0.9999$ with $241$ generators. Since both quenches start from the same equilibrium state at $U=0.05$, the same AVQITE circuit serves as the initial state of AVQDS for both. In all AVQDS simulations, the operator pool consists of the Pauli strings of the Jordan--Wigner-encoded embedding Hamiltonian, the threshold for $L^2_\mathrm{min}$ is $10^{-4}$, and at most five generators are appended per time step. Equation~\eqref{eq: eom_theta} is solved with the regularized matrix $M+10^{-6}\,\mathds{1}$, which keeps the solution stable when $M$ is ill-conditioned. The time step is chosen adaptively as $\delta t=\min(0.05,\,0.01/\max_\mu|\dot{\theta}_\mu|)$. For each $B$, we compare the double occupancy evaluated with the AVQDS embedding state against the ED-based t-GQCE result, together with the $B=1$ results and the nonequilibrium DMFT benchmark~\cite{Eckstein09ThermalizationIQ}.

For the weak quench to $U_\textrm{f}=1.5$ shown in Fig.~\figref{fig: B35}{(a)}, the DMFT double occupancy rapidly decreases from $0.25$ and settles at a prethermalized plateau of $\braket{\uparrow\downarrow}\approx 0.10$ for $t\gtrsim 1.5$. In contrast, the $B=1$ result exhibits a large-amplitude oscillation, with $\braket{\uparrow\downarrow}$ dropping to about $0.03$ at $t\approx 2.7$ and returning close to its initial value at $t\approx 5$. Introducing ghost orbitals qualitatively changes this picture. The results with $B=3$, $5$, and $7$ all reproduce the initial decay of the DMFT result and develop a plateau instead of the coherent oscillation. The $B=3$ result starts to deviate from the plateau at $t\approx 2.5$ and decreases to about $0.06$ at $t=5$, whereas the $B=5$ result remains close to the DMFT plateau over a substantially longer time window, with a much weaker drift to about $0.09$ at $t=5$. With $B=7$, the plateau persists over the entire time window, with $\braket{\uparrow\downarrow}\approx 0.100$ at $t=5$ compared with $0.104$ from DMFT. This systematic improvement with increasing $B$ is consistent with the expected convergence of t-GQCE toward the DMFT limit as $B\to\infty$.

For the strong quench to $U_\textrm{f}=2.5$ shown in Fig.~\figref{fig: B35}{(d)}, the DMFT double occupancy displays damped collapse-and-revival oscillations, with a first minimum of $\braket{\uparrow\downarrow}\approx 0.12$ at $t\approx 1.3$ and a revival peak of about $0.158$ at $t\approx 3$. The $B=1$ result captures the qualitative oscillatory behavior, but strongly overestimates the oscillation amplitude, with $\braket{\uparrow\downarrow}$ oscillating between $0.08$ and $0.25$ without damping. With $B=3$, $5$, and $7$, the first minimum is accurately reproduced in both position and depth. The $B=5$ result further captures the subsequent revival with a peak of about $0.154$ at $t\approx 3$, in good agreement with DMFT. The $B=3$ result, on the other hand, shows a flattened revival followed by a delayed peak at $t\approx 4$, indicating that more ghost orbitals are needed to describe the dynamics at longer times. The $B=7$ result coincides with the $B=5$ result up to the revival peak, and follows the subsequent decay of the DMFT result more closely, with $\braket{\uparrow\downarrow}\approx 0.141$ at $t=4$ compared with $0.133$ for $B=5$ and $0.146$ for DMFT. Overall, the ghost orbitals strongly suppress the artificial undamped oscillations of the $B=1$ approximation in both the weak and strong quench regimes.

In all cases, the AVQDS results, shown as open circles, are nearly indistinguishable from the ED-based t-GQCE results, shown as solid lines. Since the AVQDS simulations are driven by the embedding Hamiltonian extracted from the ED-based t-GQCE simulations, rather than being fed back self-consistently, this agreement directly reflects the accuracy of AVQDS in propagating the embedding states. In a fully self-consistent t-GQCE simulation, AVQDS errors would additionally feed back into $\h^\mathrm{eb}(t)$ through Eqs.~\eqref{eq: eom_d}--\eqref{eq: eom_r}. This feedback is not included here, and its effect remains to be examined. Figures~\figref{fig: B35}{(b)} and~\figref{fig: B35}{(e)} show the infidelity $1-\mathcal{F}$, with the fidelity $\mathcal{F}=|\braket{\Phi_\mathrm{ED}(t)|\Phi(\bth[t])}|^2$ between the AVQDS and the ED-propagated embedding states. It grows gradually from about $10^{-6}$ at early times but remains below $5\times10^{-4}$ up to $t=5$ for both quenches and all three values of $B$. The adaptively constructed ansatz thus faithfully follows the many-body dynamics of the embedding state. The corresponding circuit complexity, measured by the number of CNOT gates $N_\textrm{cx}$, is shown in Figs.~\figref{fig: B35}{(c)} and~\figref{fig: B35}{(f)}. For $B=3$, $N_\textrm{cx}$ grows from $224$ to $516$ and $510$ at $t=5$ for $U_\textrm{f}=1.5$ and $2.5$, respectively. For $B=5$, the circuits are considerably deeper, with $N_\textrm{cx}$ growing from $332$ to about $1600$ and $1550$ at $t=5$ for $U_\textrm{f}=1.5$ and $2.5$, respectively. For $B=7$, $N_\textrm{cx}$ starts from about $800$ for the AVQITE ground-state circuit and reaches about $3100$ and $3700$ at $t=5$ for $U_\textrm{f}=1.5$ and $2.5$, respectively, two to two and a half times the $B=5$ values.

The rapid increase of $N_\textrm{cx}$ right after the quench, for $t\lesssim 0.2$, reflects the limited variational freedom of the ansatz at early times. Taking $B=5$ as an example, the residual $L^2_\mathrm{min}$ remains slightly above the threshold of $10^{-4}$, at $1$--$2.5\times10^{-4}$, for $0.04\lesssim t\lesssim 0.11$, because the preset maximum of five generators appended per time step is reached at every step. In the median, each appended generator reduces it by only $0.6\%$ for $U_\textrm{f}=1.5$ and $0.9\%$ for $U_\textrm{f}=2.5$, compared with $3$--$8\%$ at later times. Many generators are therefore needed before the ansatz can follow the dynamics: $64$ and $81$ generators are appended for $t\lesssim 0.2$ in the two quenches, while $\braket{\uparrow\downarrow}$ changes by at most $0.01$. The subsequent decay of $\braket{\uparrow\downarrow}$ for $0.2\lesssim t\lesssim 1$, during which the embedding state changes most, requires only $21$ and $34$ additional generators, respectively. The quench itself, in contrast, is accommodated at once. At the first step after the quench, $L^2_\mathrm{min}=0.13$ and $0.38$ for the two quenches, in agreement with the variance $(\delta U)^2\,\mathrm{Var}[(\hat{n}-1)^2/2]=(\delta U)^2 d(\frac{1}{2}-d)\approx(\delta U)^2/16$ of the quenched interaction term in the initial state, with $\delta U=U_\textrm{f}-0.05$ and $d=\braket{\uparrow\downarrow}\approx\frac{1}{4}$. This deviation is removed by a single generator, the rotation generated by the product $Z_\uparrow Z_\downarrow$ of the Pauli-$Z$ operators on the two impurity qubits, which corresponds to the interaction term itself.

The growth of the circuits can be compared with that of the correlations in the embedding state, which we quantify by the entanglement entropy $S_{\uparrow\downarrow}$ between the spin-up and spin-down spin-orbitals, shown in the insets of Figs.~\figref{fig: B35}{(b)} and~\figref{fig: B35}{(e)}. With the embedding state expanded in the Fock states $\ket{n_\uparrow}$ and $\ket{n_\downarrow}$ of the $B+1$ spin-up and the $B+1$ spin-down spin-orbitals, with all spin-up modes ordered first, $\ket{\Phi}=\sum_{n_\uparrow n_\downarrow} C_{n_\uparrow n_\downarrow}\ket{n_\uparrow}\otimes\ket{n_\downarrow}$, the entropy is given by
\begin{equation}
	S_{\uparrow\downarrow} = -\Tr\left[\varrho_\uparrow\ln\varrho_\uparrow\right] = -\sum_k s_k^2\ln s_k^2,
	\label{eq: S_updn}
\end{equation}
where $\varrho_\uparrow = \Tr_\downarrow\left[\ket{\Phi}\bra{\Phi}\right] = CC^\dag$ is the reduced density matrix of the spin-up spin-orbitals and $s_k$ are the singular values of the coefficient matrix $C$. Since the two spin species are coupled only by the interaction on the impurity orbital, $S_{\uparrow\downarrow}$ vanishes for an uncorrelated embedding state, a product of Slater determinants for the two spins, and its growth measures the interaction-induced correlations. The more familiar entanglement entropy between the impurity and bath orbitals is not an independent measure here. By particle-number, spin, and particle-hole symmetry, the reduced density matrix of the impurity orbital in the basis $\{\ket{0},\ket{\uparrow},\ket{\downarrow},\ket{\uparrow\downarrow}\}$ is $\mathrm{diag}(d,\,\frac{1}{2}-d,\,\frac{1}{2}-d,\,d)$ with $d=\braket{\uparrow\downarrow}$. This entropy, $-2d\ln d-(1-2d)\ln(\frac{1}{2}-d)$, is therefore fixed by the double occupancy and decreases from $\ln 4$ at $d=\frac{1}{4}$. For $B=5$, $S_{\uparrow\downarrow}$ reaches only $0.035$ for $U_\textrm{f}=1.5$ and $0.080$ for $U_\textrm{f}=2.5$ by $t=0.2$, compared with $0.33$ and $0.59$ at $t=1$ and about $1.0$ and $1.2$ at $t=5$. This confirms that the early burst of generators occurs while the correlations are still small. Afterwards, $N_\textrm{cx}$ grows slowly and steadily, in parallel with the steady growth of $S_{\uparrow\downarrow}$, which continues after $\braket{\uparrow\downarrow}$ has settled at the plateau for $U_\textrm{f}=1.5$. The circuit depth grows mainly with $B$, reflecting the enlarged Hilbert space of the embedding problem, while the difference between the two quenches is smaller. These results highlight the trade-off between the accuracy of t-GQCE, which improves systematically with $B$, and the quantum resources required to simulate the embedding dynamics.

\section{Toward quantum hardware: measuring the embedding observables}\label{sec: hardware}

The statevector results above establish that AVQDS follows the embedding state accurately, with circuits that grow over $0\le t\le5$ from a few hundred CNOT gates for $B=3$ to about $3700$ for $B=7$. Circuits of this size remain challenging for current quantum hardware. In a self-consistent t-GQCE simulation, the quantum processor must moreover supply at every time step the single-particle density matrix of the embedding state, which enters Eqs.~\eqref{eq: eom_delta} and~\eqref{eq: eom_r}. The double occupancy $\braket{\uparrow\downarrow}$, in contrast, does not feed back into the equations of motion: it is needed only for the analysis of the dynamics and can therefore be measured on a much sparser time mesh. In this section, we take $B=5$, the weak quench to $U_\textrm{f}=1.5$, and ten times $t=0.5,1,\dots,5$ to assess the practical requirements of measuring these observables on real quantum hardware, taking a trapped-ion device as an example. The embedding problem then has $12$ qubits, and the AVQDS circuits contain between about $650$ and $1600$ CNOT gates. Following Ref.~\cite{chen2025quantum}, we take as the target an absolute accuracy of $0.01$ for $\braket{\uparrow\downarrow}$ and for each element of the single-particle density matrix $\rho_{ab}=\avihi{\Phi}{\cc_a\ca_b}$. Here $a,b$ run over the impurity and bath spin-orbitals of one spin block, and $\ca_a$ denotes the annihilation operator of spin-orbital $a$, impurity or bath.

\begin{figure*}[ht]
	\centering
	\includegraphics[width=\textwidth]{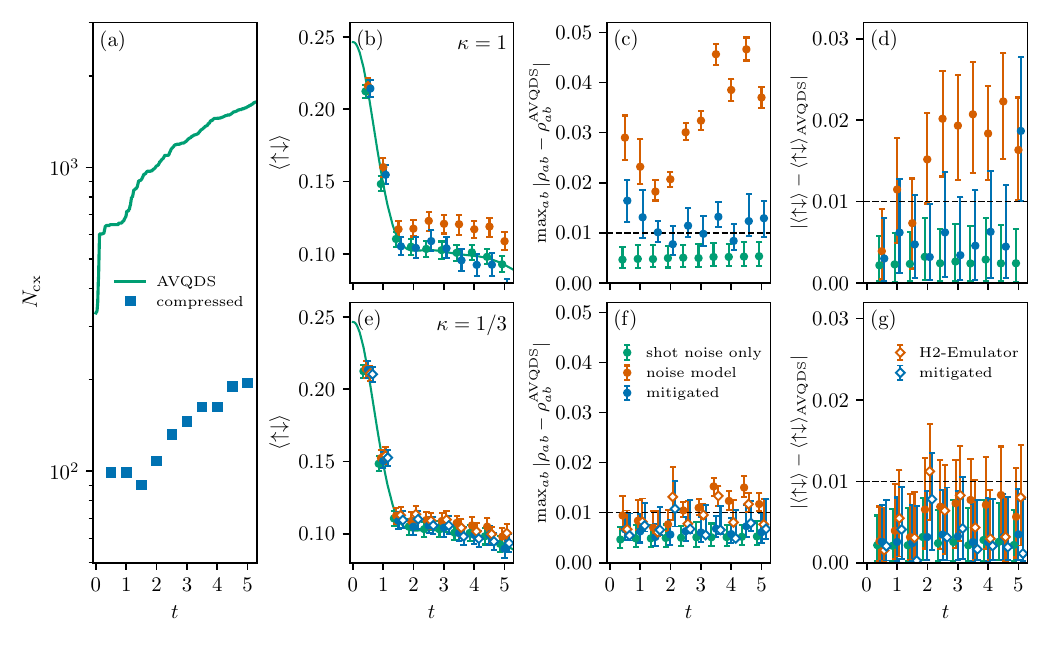}
	\caption{
	\textbf{Measuring the embedding observables for $B=5$, $U_\textrm{f}=1.5$ under trapped-ion noise models.}
	(a) $N_\textrm{cx}$ of the AVQDS circuit, shown as a line, and of the compressed circuits, shown as squares.
	(b, e) Double occupancy estimated with $10^5$ shots per time, with shot noise only in green, with the noise model in orange, and with the noise model and mirror-circuit mitigation in blue, compared with the AVQDS result shown as a line.
	(c, f) Maximum error of the single-particle density matrix and (d, g) error of the double occupancy relative to AVQDS. The upper row uses the noise model of Ref.~\cite{chen2025quantum} with error rates scaled by $\kappa=1$, the lower row by $\kappa=1/3$; the shot-noise-only results are the same in both rows. Symbols: median, bars: $5$--$95\%$ of $500$ repetitions. The lower row also shows the results of the Quantinuum H2-Emulator, post-selected in orange and mitigated in blue, as hollow diamonds, from a single measurement of $10^5$ shots per time with bars from $500$ bootstrap resamplings. Dashed lines: target accuracy $0.01$.
	}
	\label{fig: hw}
\end{figure*}

AVQDS is designed to generate compact circuits that represent the state evolution along the entire dynamical path: generators are appended only when the ansatz can no longer follow the dynamics, and the same circuit structure serves all later times with continuously evolving parameters. The adaptive construction is, however, not guaranteed to yield the most compact circuit. The state at a particular time may be represented by a considerably shallower circuit, especially if a small additional error is tolerated, and possibly even exactly. We therefore compress the AVQDS state at each time into a brickwork circuit of general two-qubit gates on a linear chain of qubits, each gate compiled into at most three CNOTs~\cite{Vatan2004optimal, Shende2004minimal}. The gates are optimized in sweeps over the circuit. In each sweep, every gate in turn is replaced by the two-qubit unitary that maximizes the overlap with the AVQDS state while the other gates are kept fixed, which is obtained from the polar decomposition of its $4\times4$ environment matrix. This matrix is the outer product of the reference state propagated forward through the preceding gates and the AVQDS state propagated backward through the subsequent gates, traced over all qubits except the two on which the gate acts. The circuit is deepened by one layer at a time. We keep the shallowest circuit whose $\rho$ and $\braket{\uparrow\downarrow}$ deviate from the AVQDS values by at most $2\times10^{-3}$ and $10^{-3}$, respectively. As shown in Fig.~\figref{fig: hw}{(a)}, this reduces $N_\textrm{cx}$ by a factor of $6$--$8$, to $90$--$195$ CNOTs in $6$ to $12$ brickwork layers, with infidelities $1-\mathcal{F}\le3\times10^{-3}$ relative to the AVQDS state. Because this compression uses the classically computed state, it serves here as a compilation step to assess the measurement. Whether comparable compression can be achieved more generally, for instance with tensor-network methods for the moderately entangled embedding states shown in the insets of Fig.~\figref{fig: B35}{(b)}, remains to be explored~\cite{lin2021real, Barratt2021, Martin2024combining}.

We simulate the compressed circuits after compilation to the native gate set of Quantinuum H-series trapped-ion processors, which consists of single-qubit rotations and two-qubit $ZZ$ rotations with arbitrary angles on all-to-all connected qubits. After compilation, the circuits retain a similar number of two-qubit gates, $88$--$189$. Since a $ZZ$ rotation by a negative angle equals the rotation by the positive angle conjugated with a single-qubit $X$ gate, which merges into the neighboring single-qubit rotations, we compile all $ZZ$ angles to be positive. For the noisy simulations, we adopt the noise model of Refs.~\cite{self2024protecting, chen2025quantum}, which is based on randomized benchmarking of the Quantinuum H1-2 processor. A depolarizing channel acts after every single- and two-qubit gate, with error probabilities $p_\mathrm{1q}=4\times10^{-4}$ and $p_\mathrm{2q}=3\times10^{-3}$. A bit-flip channel acts after the initialization and before the measurement of every qubit, with error probabilities $4\times10^{-4}$ and $3\times10^{-3}$. In Ref.~\cite{chen2025quantum}, noisy simulations with this model gave density-matrix errors and a spectral function similar to those measured on the Quantinuum H1-1 processor for a circuit with $55$ two-qubit gates. To assess how the results depend on the hardware quality, we scale all error rates by a common factor $\kappa$ and consider $\kappa=1$, the error rates of Ref.~\cite{chen2025quantum}, and $\kappa=1/3$, which represents hardware with error rates three times lower. As an independent check with a more detailed noise model, we also run the same circuits on the Quantinuum H2-Emulator. It models the H2 processor with a single-qubit error of $2.9\times10^{-5}$, a two-qubit error of $1.28\times10^{-3}$ that grows with the rotation angle, state-dependent measurement errors, and dephasing of idle qubits, among other error sources. Its two-qubit error is about $0.4$ times the $p_\mathrm{2q}$ of the noise model above, between $\kappa=1/3$ and $\kappa=1$. The emulator assigns a $ZZ$ rotation by a negative angle the error of the equivalent large positive angle. With the angles of an unconstrained compilation, most of which are negative, the errors on the emulator are about twice as large, which motivates the choice of positive angles. Whether the hardware shows the same dependence on the sign of the angle remains to be tested.

We order the $12$ spin-orbitals as the $B=5$ spin-up bath orbitals, the spin-up and spin-down impurity orbitals, and the $5$ spin-down bath orbitals, and map them onto a linear chain of qubits in this order with the Jordan--Wigner encoding, so that the two impurity qubits are adjacent. With the occupation operator $\hat{n}_a=(1-Z_a)/2$, a diagonal element is then $\rho_{aa}=(1-\braket{Z_a})/2$, and the double occupancy is $\braket{\uparrow\downarrow}=(1-\braket{Z_\uparrow}-\braket{Z_\downarrow}+\braket{Z_\uparrow Z_\downarrow})/4$, where $Z_\uparrow$ and $Z_\downarrow$ act on the two impurity qubits. An off-diagonal element $\rho_{ab}$ becomes a combination of the Pauli strings $X_aZ\cdots ZX_b$ and $Y_aZ\cdots ZY_b$ for its real part and $X_aZ\cdots ZY_b$ and $Y_aZ\cdots ZX_b$ for its imaginary part, whose weight grows with the distance between the qubits. Together, $\rho$ of both spin blocks and $\braket{\uparrow\downarrow}$ involve $133$ distinct Pauli strings, and the strings of different elements do not commute qubit by qubit. We compared four strategies to measure $\rho$ and $\braket{\uparrow\downarrow}$ under the noise model: direct measurement of these Pauli strings sorted into $36$ qubit-wise commuting groups, classical shadows in random Pauli bases~\cite{huang2020predicting}, particle-number-conserving fermionic shadows~\cite{zhao2021fermionic, low2022classical}, and a small set of fixed orbital rotations, as used to measure the single-particle density matrix in Ref.~\cite{Arute2020HartreeFock}. For the present problem, the orbital rotations turn out to be the most efficient, for three reasons. First, a rotation $\hat{O}$ of the spin-orbitals maps every element of $\rho$ onto occupation numbers, which are measured simultaneously in the computational basis, so that a few settings suffice. Second, $\hat{O}$ conserves the particle number of each spin block, so that every shot can be post-selected on the correct particle numbers, which removes a large part of the noise. Third, the rotation circuits are shallow and add at most $30$ two-qubit gates.

A measurement setting consists of the compressed circuit, a rotation circuit $\hat{O}_s$, and a measurement of all qubits in the computational basis, so that each setting is one distinct circuit. The first setting has no rotation: it gives the occupations, which are the diagonal elements $\rho_{aa}$, and, from the joint outcomes of the two impurity qubits, the double occupancy $\braket{\uparrow\downarrow}$. The off-diagonal elements are obtained with $50{:}50$ beam-splitter rotations $\exp[-i\frac{\pi}{4}(e^{i\phi}\cc_a\ca_b+\mathrm{H.c.})]$ on pairs of spin-orbitals, which turn the occupations of $a$ and $b$ into $\frac{1}{2}(\rho_{aa}+\rho_{bb})\pm\mathrm{Im}\,\rho_{ab}$ for $\phi=0$ and $\frac{1}{2}(\rho_{aa}+\rho_{bb})\pm\mathrm{Re}\,\rho_{ab}$ for $\phi=\pi/2$. Since the $B+1=6$ spin-orbitals of a spin block split into $3$ disjoint pairs, one setting rotates $3$ pairs at once, and the $15$ pairs are covered by the $B=5$ rounds of a round-robin schedule. With the two phases, this gives $2B=10$ rotated settings, applied identically to both spin blocks, and $2B+1=11$ settings in total. For a spin block of $N$ spin-orbitals with $N$ even, the same construction gives $2N-1$ settings, so that their number grows linearly with $N$, as in other schemes for the partial tomography of fermionic reduced density matrices~\cite{Arute2020HartreeFock, BonetMonroig2020}. Each rotation $\hat{O}_s$ is an orbital rotation and can be compiled into at most $N(N-1)/2$ Givens rotations with a depth linear in $N$ on a linear chain of qubits~\cite{BabbushLinearDepth}. The density matrix $\rho$ follows from a linear least-squares fit to all measured occupations. Post-selection keeps $62$--$81\%$ of the shots for $\kappa=1$, $85$--$93\%$ for $\kappa=1/3$, and $86$--$93\%$ on the H2-Emulator. All shot numbers quoted in this section count the shots before post-selection.

For this problem, the orbital-rotation settings give the smallest errors of the four strategies at both noise strengths, while Pauli-string measurements and classical shadows do not rely on particle-number conservation and apply to more general observables. We compare the strategies by the error of $\rho$, which must be measured at every time step, with $10^5$ shots per time divided among the circuits of each strategy and without the symmetrization described below. The error is the maximum element-wise deviation of $\rho$ from the AVQDS value, and we quote the range over $0.5\le t\le5$ of its median over $20$ repetitions. With the orbital-rotation settings, it is $0.013$--$0.018$ at $\kappa=1/3$ and $0.023$--$0.052$ at $\kappa=1$. With the commuting groups, which cannot be post-selected, it is $0.020$--$0.033$ and $0.047$--$0.090$. With Pauli shadows, whose variance grows as $3^w$ with the weight $w$ of the Pauli strings, it is $0.049$--$0.057$ and $0.062$--$0.091$. Fermionic shadows have the smallest statistical error per shot, but they require a random orbital rotation with about $60$ additional two-qubit gates. This raises the noise bias and gives errors of $0.014$--$0.023$ at $\kappa=1/3$ and $0.030$--$0.058$ at $\kappa=1$, above those of the orbital-rotation settings. Since each shot requires a distinct circuit, fermionic shadows also add a per-circuit overhead in compilation and execution on quantum hardware.

We further exploit the spin and particle-hole symmetries of the paramagnetic half-filled problem, $\rho_\downarrow=\rho_\uparrow$ and $\rho=K(1-Q\rho^\mathrm{T}Q)K$, where $Q$ exchanges the particle-hole partner bath orbitals and $K$ changes the sign of the impurity orbital. Averaging the four equivalent estimates reduces the statistical error by a factor of about $2.6$.

After post-selection, the residual noise acts approximately as depolarizing noise within the particle-number sector, $\rho\to \eta\rho+(1-\eta)\mathds{1}/2$. We estimate $\eta$ with a noise-estimation circuit~\cite{Urbnek2021MitigatingDN}: the mirror circuit $\hat{\mathcal{U}}^\dagger\hat{\mathcal{U}}$ applied to the reference configuration from which the compressed circuit $\hat{\mathcal{U}}$ prepares the state. Ideally, it returns that configuration with occupations $0$ or $1$. Since the mirror circuit contains twice the two-qubit gates of $\hat{\mathcal{U}}$, its measured occupations deviate from $0$ or $1$ by $(1-\eta^2)/2$ on average, which gives $\eta$ for each spin block. The value for each setting is extrapolated from its number of two-qubit gates, assuming the same factor per gate, and the measured occupations are rescaled as $n\to[n-(1-\eta)/2]/\eta$. Without shot noise and with the symmetrization, this rescaling reduces the maximum error of $\rho$ from $0.006$--$0.015$ to $0.002$--$0.006$ for $\kappa=1/3$, and from $0.018$--$0.047$ to $0.005$--$0.016$ for $\kappa=1$. At $\kappa=1$, the residual noise deviates more strongly from the depolarizing form: after the rescaling, the off-diagonal elements of $\rho$ remain about $3\%$ too small, and for the deepest circuit at $t=5$ the double occupancy is overcorrected by $0.019$. Even a single factor $\eta$ per spin block fitted to the exact answer leaves errors above $0.01$ at several times, so that this limitation lies in the depolarizing model rather than in the estimate of $\eta$.

Figure~\ref{fig: hw} summarizes the resulting accuracy at $10^5$ shots per time. The shots are distributed evenly over the $11$ settings and the mirror circuit, except that the computational-basis setting, the only setting that yields $\braket{\uparrow\downarrow}$, receives three times as many. We compare three cases: shot noise only, obtained with the noiseless compressed circuits, which is the same for both noise strengths; the noise model with post-selection; and the noise model with post-selection and the mirror-circuit mitigation. To capture the statistical spread, we repeat the complete measurement of $10^5$ shots $500$ times for each case, time and noise strength. The symbols in Figs.~\figref{fig: hw}{(b)}--\figref{fig: hw}{(g)} mark the median over these repetitions, and the bars span the $5\%$ to $95\%$ quantiles.

Figures~\figref{fig: hw}{(b)} and~\figref{fig: hw}{(e)} show that the estimated double occupancy follows the AVQDS curve in all cases. Without mitigation, the noise shifts the estimate upward, by up to about $0.02$ for $\kappa=1$ and $0.008$ for $\kappa=1/3$. This shift reflects the depolarizing character of the noise, since the fully mixed state within the particle-number sector has $\braket{\uparrow\downarrow}=1/4$, above the values of $0.09$--$0.21$ along the dynamics. The mitigation removes most of this shift. With shot noise only, the maximum error of $\rho$ in Figs.~\figref{fig: hw}{(c)} and~\figref{fig: hw}{(f)} is about $0.005$ in the median, set by the number of shots, and the compression contributes at most $0.0014$. For $\kappa=1$, the mitigated error of $\rho$ remains above the target accuracy at seven of the ten times. The mitigated error of $\braket{\uparrow\downarrow}$, shown in Fig.~\figref{fig: hw}{(d)}, stays below $0.015$ in $95\%$ of the repetitions, except at $t=5$, where the deepest circuit leads to an overcorrection and errors of up to $0.028$. For $\kappa=1/3$, the mitigation brings the errors of $\rho$ and $\braket{\uparrow\downarrow}$ close to the shot-noise level. The $95\%$ quantiles stay at or below $0.012$ for $\rho$ and below the target accuracy at all times for $\braket{\uparrow\downarrow}$, as shown in Figs.~\figref{fig: hw}{(f)} and~\figref{fig: hw}{(g)}. The H2-Emulator results, shown as hollow diamonds in the lower row, lie close to those for $\kappa=1/3$. After mitigation, the error of $\rho$ is below the target accuracy at nine of the ten times, with $0.011$ at $t=2$, and that of $\braket{\uparrow\downarrow}$ stays below $0.008$.

The main message of Fig.~\ref{fig: hw} is that, after compression, the accuracy of the measurement is limited by the combination of shot noise and the residual noise bias left by the mitigation, and that the latter depends sensitively on the hardware error rates. With the error rates of Ref.~\cite{chen2025quantum}, the combination of orbital-rotation measurements, post-selection, symmetry averaging, and mirror-circuit mitigation reduces the errors by a factor of two to three. The residual bias of the depolarizing rescaling, however, keeps the error of $\rho$ above the target accuracy of $0.01$ at most times, independently of the number of shots. Even at $10^6$ shots per time, the median error is still $0.006$--$0.017$. With error rates three times lower, the same protocol reaches the target accuracy for $\rho$ and $\braket{\uparrow\downarrow}$ in the median at all ten times with $10^5$ shots per time. Tripling the number of shots to $3\times10^5$ brings the $95\%$ quantiles to at most $0.008$ for $\rho$ and $0.007$ for $\braket{\uparrow\downarrow}$. In this case, the mitigated errors do not grow along the dynamics, although the circuit depth doubles between $t=0.5$ and $t=5$. The H2-Emulator, with its model of the H2 processor, gives errors similar to those for $\kappa=1/3$. Mitigation schemes that account for the structure of the residual noise beyond the depolarizing form may extend this accuracy to higher error rates. Since the density matrix must be measured at every time step of a self-consistent t-GQCE simulation, whereas $\braket{\uparrow\downarrow}$ can be sampled on a sparser mesh, the measurement of $\rho$ sets the shot budget. In implementations with a variational solver such as AVQDS, the measurements of $M$ and $V$ add to this budget. This overhead can be reduced by measurement-efficient variants of AVQDS~\cite{Zhang2025AdaptiveVQ} and by distributing the shots according to the sensitivity of the dynamics to each measured quantity, as explored for imaginary-time dynamics simulations~\cite{Zhang2026sampling}. With a fault-tolerant Hamiltonian simulation algorithm as the solver, the embedding state is propagated without variational parameters, so that this overhead is removed and only the density matrix needs to be measured.

\section{Conclusion and outlook}\label{sec: conclusion}
We have presented the t-GQCE framework for simulating the real-time dynamics of generic multi-orbital correlated electron systems on hybrid quantum-classical computers. Within the time-dependent ghost Gutzwiller approximation, the lattice dynamics is mapped onto the coupled evolution of a noninteracting quasiparticle lattice, propagated on classical processors, and of finite interacting embedding models, propagated on quantum processors. The two parts exchange only the parameters of the embedding Hamiltonian and the single-particle density matrices of the embedding states at each time step, and the framework is agnostic to the quantum dynamics solver. For demonstration, we have used the adaptive variational algorithm AVQDS.

Taking interaction quenches of the Hubbard model on the Bethe lattice as an example, the results illustrate both the accuracy of the approximation and the cost of its quantum part. Without ghost orbitals, that is, with a single bath orbital per spin, $B=1$, the dynamics reduces to a single-qubit problem. We have simulated it on IBM quantum hardware, using a second qubit as the ancilla for measuring the quantum metric, in close agreement with the noiseless result. With $B=3$, $5$, and $7$ bath orbitals per spin, t-GQCE suppresses the artificial undamped oscillations of the conventional Gutzwiller approximation and approaches the nonequilibrium DMFT results systematically. For the weak quench, the double occupancy with $B=7$ stays on the prethermalized plateau over the entire time window, at $0.100$ compared with $0.104$ from DMFT at $t=5$. For the strong quench, $B=5$ and $7$ capture the first collapse and revival. AVQDS follows the embedding states with infidelities below $5\times10^{-4}$, but its circuits grow with time and with $B$, to about $1550$--$1600$ CNOT gates for $B=5$ and $3100$--$3700$ for $B=7$ at $t=5$. The early growth reflects the limited variational freedom of the ansatz right after the quench rather than a fast build-up of correlations, whereas the later steady growth follows the growth of the entanglement between the two spin species.

For the measurements that a quantum processor must supply in a self-consistent simulation, our analysis of the $B=5$ weak quench under trapped-ion noise models identifies, besides the circuit depth, the measurement accuracy as a key requirement and shows how it can be improved. After the AVQDS states at individual times are compressed by a factor of $6$--$8$ in CNOT count, the single-particle density matrix and the double occupancy are obtained from $11$ measurement settings in rotated orbital bases. These settings give smaller errors than measuring commuting groups of Pauli strings or classical shadows, and are combined with post-selection on the particle numbers, symmetry averaging, and a mirror-circuit estimate of the residual depolarizing noise. With the noise model of Ref.~\cite{chen2025quantum}, whose results were found to be similar to those measured on trapped-ion hardware, this protocol reduces the errors by a factor of two to three. The residual bias of the depolarizing rescaling, however, keeps the error of the density matrix above the target accuracy of $0.01$ at most times. With error rates three times lower, the target accuracy is reached in the median at all times studied with $10^5$ shots per time, and the remaining errors are set by shot noise and residual noise bias of comparable size. The Quantinuum H2-Emulator, with its model of the H2 processor, gives errors of similar size, below the target at nine of the ten times.

Several directions follow from this work. The AVQDS simulations with $B>1$ were driven by embedding Hamiltonians from classical t-GQCE simulations. In a fully self-consistent simulation, the statistical and noise-induced errors of the measured density matrices feed back into the embedding Hamiltonian at every time step. How these errors accumulate over long times, and whether filtering or a modified integration scheme is beneficial, remains to be examined. Recent work on how sampling noise affects variational imaginary-time evolution, and on how it can be controlled by regularization and an optimized distribution of measurements~\cite{Zhang2026sampling}, provides a starting point. On the hardware side, mitigation schemes that capture the structure of the residual noise beyond the depolarizing form may allow the target accuracy to be reached at the error rates of the noise model considered here. Compression schemes that operate directly on the quantum processor, or that exploit the moderate entanglement of the embedding states, would reduce the circuit depth without relying on the classically computed state. The framework itself applies directly to multi-orbital models and to general electric or magnetic field pulses, where the embedding problems are larger and quantum processors may offer an advantage. As fault-tolerant hardware becomes available, Hamiltonian simulation algorithms with provable error bounds, including those for time-dependent Hamiltonians~\cite{Childs:2012, Berry:2015, Low2017QSP, Low2019hamiltonian, Berry2020TimeDependent}, can replace the variational solver for the embedding states, while the classical part of t-GQCE remains unchanged.

\section*{Data availability}
The data shown in all figures and the scripts that generate the figures are openly available on Zenodo at \url{https://doi.org/10.5281/zenodo.23887709}.

\section*{Acknowledgments}
This work was supported by the U.S. Department of Energy (DOE), Office of Science, Basic Energy Sciences, Materials Science and Engineering Division, including the grant of computer time at the National Energy Research Scientific Computing Center (NERSC) in Berkeley, California. The research was performed at the Ames National Laboratory, which is operated for the U.S. DOE by Iowa State University under Contract No. DE-AC02-07CH11358. N.L. gratefully acknowledges funding from the National Science Foundation under Award No. DMR-2532771 and from the Simons Foundation (Grant No. 00010910).
\bibliography{bib/master}

\begin{thebibliography}{83}%
\makeatletter
\providecommand \@ifxundefined [1]{%
 \@ifx{#1\undefined}
}%
\providecommand \@ifnum [1]{%
 \ifnum #1\expandafter \@firstoftwo
 \else \expandafter \@secondoftwo
 \fi
}%
\providecommand \@ifx [1]{%
 \ifx #1\expandafter \@firstoftwo
 \else \expandafter \@secondoftwo
 \fi
}%
\providecommand \natexlab [1]{#1}%
\providecommand \enquote  [1]{``#1''}%
\providecommand \bibnamefont  [1]{#1}%
\providecommand \bibfnamefont [1]{#1}%
\providecommand \citenamefont [1]{#1}%
\providecommand \href@noop [0]{\@secondoftwo}%
\providecommand \href [0]{\begingroup \@sanitize@url \@href}%
\providecommand \@href[1]{\@@startlink{#1}\@@href}%
\providecommand \@@href[1]{\endgroup#1\@@endlink}%
\providecommand \@sanitize@url [0]{\catcode `\\12\catcode `\$12\catcode
  `\&12\catcode `\#12\catcode `\^12\catcode `\_12\catcode `\%12\relax}%
\providecommand \@@startlink[1]{}%
\providecommand \@@endlink[0]{}%
\providecommand \url  [0]{\begingroup\@sanitize@url \@url }%
\providecommand \@url [1]{\endgroup\@href {#1}{\urlprefix }}%
\providecommand \urlprefix  [0]{URL }%
\providecommand \Eprint [0]{\href }%
\providecommand \doibase [0]{https://doi.org/}%
\providecommand \selectlanguage [0]{\@gobble}%
\providecommand \bibinfo  [0]{\@secondoftwo}%
\providecommand \bibfield  [0]{\@secondoftwo}%
\providecommand \translation [1]{[#1]}%
\providecommand \BibitemOpen [0]{}%
\providecommand \bibitemStop [0]{}%
\providecommand \bibitemNoStop [0]{.\EOS\space}%
\providecommand \EOS [0]{\spacefactor3000\relax}%
\providecommand \BibitemShut  [1]{\csname bibitem#1\endcsname}%
\let\auto@bib@innerbib\@empty
\bibitem [{\citenamefont {Giannetti}\ \emph {et~al.}(2016)\citenamefont
  {Giannetti}, \citenamefont {Capone}, \citenamefont {Fausti}, \citenamefont
  {Fabrizio}, \citenamefont {Parmigiani},\ and\ \citenamefont
  {Mihailovic}}]{ultrafast_approach}%
  \BibitemOpen
  \bibfield  {author} {\bibinfo {author} {\bibfnamefont {C.}~\bibnamefont
  {Giannetti}}, \bibinfo {author} {\bibfnamefont {M.}~\bibnamefont {Capone}},
  \bibinfo {author} {\bibfnamefont {D.}~\bibnamefont {Fausti}}, \bibinfo
  {author} {\bibfnamefont {M.}~\bibnamefont {Fabrizio}}, \bibinfo {author}
  {\bibfnamefont {F.}~\bibnamefont {Parmigiani}},\ and\ \bibinfo {author}
  {\bibfnamefont {D.}~\bibnamefont {Mihailovic}},\ }\bibfield  {title}
  {\bibinfo {title} {Ultrafast optical spectroscopy of strongly correlated
  materials and high-temperature superconductors: a non-equilibrium approach},\
  }\href {https://doi.org/10.1080/00018732.2016.1194044} {\bibfield  {journal}
  {\bibinfo  {journal} {Adv. Phys.}\ }\textbf {\bibinfo {volume} {65}},\
  \bibinfo {pages} {58} (\bibinfo {year} {2016})}\BibitemShut {NoStop}%
\bibitem [{\citenamefont {Basov}\ \emph {et~al.}(2017)\citenamefont {Basov},
  \citenamefont {Averitt},\ and\ \citenamefont {Hsieh}}]{basov2017towards}%
  \BibitemOpen
  \bibfield  {author} {\bibinfo {author} {\bibfnamefont {D.}~\bibnamefont
  {Basov}}, \bibinfo {author} {\bibfnamefont {R.}~\bibnamefont {Averitt}},\
  and\ \bibinfo {author} {\bibfnamefont {D.}~\bibnamefont {Hsieh}},\ }\bibfield
   {title} {\bibinfo {title} {Towards properties on demand in quantum
  materials},\ }\href {https://doi.org/10.1038/nmat5017} {\bibfield  {journal}
  {\bibinfo  {journal} {Nat. Mater.}\ }\textbf {\bibinfo {volume} {16}},\
  \bibinfo {pages} {1077} (\bibinfo {year} {2017})}\BibitemShut {NoStop}%
\bibitem [{\citenamefont {de~la Torre}\ \emph {et~al.}(2021)\citenamefont
  {de~la Torre}, \citenamefont {Kennes}, \citenamefont {Claassen},
  \citenamefont {Gerber}, \citenamefont {McIver},\ and\ \citenamefont
  {Sentef}}]{de2021colloquium}%
  \BibitemOpen
  \bibfield  {author} {\bibinfo {author} {\bibfnamefont {A.}~\bibnamefont
  {de~la Torre}}, \bibinfo {author} {\bibfnamefont {D.~M.}\ \bibnamefont
  {Kennes}}, \bibinfo {author} {\bibfnamefont {M.}~\bibnamefont {Claassen}},
  \bibinfo {author} {\bibfnamefont {S.}~\bibnamefont {Gerber}}, \bibinfo
  {author} {\bibfnamefont {J.~W.}\ \bibnamefont {McIver}},\ and\ \bibinfo
  {author} {\bibfnamefont {M.~A.}\ \bibnamefont {Sentef}},\ }\bibfield  {title}
  {\bibinfo {title} {Colloquium: Nonthermal pathways to ultrafast control in
  quantum materials},\ }\href {https://doi.org/10.1103/RevModPhys.93.041002}
  {\bibfield  {journal} {\bibinfo  {journal} {Rev. Mod. Phys.}\ }\textbf
  {\bibinfo {volume} {93}},\ \bibinfo {pages} {041002} (\bibinfo {year}
  {2021})}\BibitemShut {NoStop}%
\bibitem [{\citenamefont {Sun}\ and\ \citenamefont
  {Chan}(2016)}]{sun2016quantum}%
  \BibitemOpen
  \bibfield  {author} {\bibinfo {author} {\bibfnamefont {Q.}~\bibnamefont
  {Sun}}\ and\ \bibinfo {author} {\bibfnamefont {G.~K.-L.}\ \bibnamefont
  {Chan}},\ }\bibfield  {title} {\bibinfo {title} {Quantum embedding
  theories},\ }\href {https://doi.org/10.1021/acs.accounts.6b00356} {\bibfield
  {journal} {\bibinfo  {journal} {Acc. Chem. Res.}\ }\textbf {\bibinfo {volume}
  {49}},\ \bibinfo {pages} {2705} (\bibinfo {year} {2016})}\BibitemShut
  {NoStop}%
\bibitem [{\citenamefont {Lanata}\ \emph {et~al.}(2013)\citenamefont {Lanata},
  \citenamefont {Yao}, \citenamefont {Wang}, \citenamefont {Ho}, \citenamefont
  {Schmalian}, \citenamefont {Haule},\ and\ \citenamefont {Kotliar}}]{ga_ce}%
  \BibitemOpen
  \bibfield  {author} {\bibinfo {author} {\bibfnamefont {N.}~\bibnamefont
  {Lanata}}, \bibinfo {author} {\bibfnamefont {Y.-X.}\ \bibnamefont {Yao}},
  \bibinfo {author} {\bibfnamefont {C.-Z.}\ \bibnamefont {Wang}}, \bibinfo
  {author} {\bibfnamefont {K.-M.}\ \bibnamefont {Ho}}, \bibinfo {author}
  {\bibfnamefont {J.}~\bibnamefont {Schmalian}}, \bibinfo {author}
  {\bibfnamefont {K.}~\bibnamefont {Haule}},\ and\ \bibinfo {author}
  {\bibfnamefont {G.}~\bibnamefont {Kotliar}},\ }\bibfield  {title} {\bibinfo
  {title} {$\gamma$- $\alpha$ isostructural transition in cerium},\ }\href
  {https://doi.org/10.1103/PhysRevLett.111.196801} {\bibfield  {journal}
  {\bibinfo  {journal} {Phys. Rev. Lett.}\ }\textbf {\bibinfo {volume} {111}},\
  \bibinfo {pages} {196801} (\bibinfo {year} {2013})}\BibitemShut {NoStop}%
\bibitem [{\citenamefont {Lanat{\`a}}\ \emph {et~al.}(2015)\citenamefont
  {Lanat{\`a}}, \citenamefont {Yao}, \citenamefont {Wang}, \citenamefont {Ho},\
  and\ \citenamefont {Kotliar}}]{ga_pu}%
  \BibitemOpen
  \bibfield  {author} {\bibinfo {author} {\bibfnamefont {N.}~\bibnamefont
  {Lanat{\`a}}}, \bibinfo {author} {\bibfnamefont {Y.-X.}\ \bibnamefont {Yao}},
  \bibinfo {author} {\bibfnamefont {C.-Z.}\ \bibnamefont {Wang}}, \bibinfo
  {author} {\bibfnamefont {K.-M.}\ \bibnamefont {Ho}},\ and\ \bibinfo {author}
  {\bibfnamefont {G.}~\bibnamefont {Kotliar}},\ }\bibfield  {title} {\bibinfo
  {title} {Phase diagram and electronic structure of praseodymium and
  plutonium},\ }\href {https://doi.org/10.1103/PhysRevX.5.011008} {\bibfield
  {journal} {\bibinfo  {journal} {Phys. Rev. X}\ }\textbf {\bibinfo {volume}
  {5}},\ \bibinfo {pages} {011008} (\bibinfo {year} {2015})}\BibitemShut
  {NoStop}%
\bibitem [{\citenamefont {Lanat\`a}\ \emph {et~al.}(2017)\citenamefont
  {Lanat\`a}, \citenamefont {Yao}, \citenamefont {Deng}, \citenamefont
  {Dobrosavljevi\ifmmode~\acute{c}\else \'{c}\fi{}},\ and\ \citenamefont
  {Kotliar}}]{ga_uo2}%
  \BibitemOpen
  \bibfield  {author} {\bibinfo {author} {\bibfnamefont {N.}~\bibnamefont
  {Lanat\`a}}, \bibinfo {author} {\bibfnamefont {Y.-X.}\ \bibnamefont {Yao}},
  \bibinfo {author} {\bibfnamefont {X.}~\bibnamefont {Deng}}, \bibinfo {author}
  {\bibfnamefont {V.}~\bibnamefont {Dobrosavljevi\ifmmode~\acute{c}\else
  \'{c}\fi{}}},\ and\ \bibinfo {author} {\bibfnamefont {G.}~\bibnamefont
  {Kotliar}},\ }\bibfield  {title} {\bibinfo {title} {Slave boson theory of
  orbital differentiation with crystal field effects: Application to
  {{${\mathrm{UO}}_{2}$}}},\ }\href
  {https://doi.org/10.1103/PhysRevLett.118.126401} {\bibfield  {journal}
  {\bibinfo  {journal} {Phys. Rev. Lett.}\ }\textbf {\bibinfo {volume} {118}},\
  \bibinfo {pages} {126401} (\bibinfo {year} {2017})}\BibitemShut {NoStop}%
\bibitem [{\citenamefont {Lanat{\`a}}\ \emph {et~al.}(2019)\citenamefont
  {Lanat{\`a}}, \citenamefont {Lee}, \citenamefont {Yao}, \citenamefont
  {Stevanovi{\'c}},\ and\ \citenamefont {Dobrosavljevi{\'c}}}]{ga_tmo}%
  \BibitemOpen
  \bibfield  {author} {\bibinfo {author} {\bibfnamefont {N.}~\bibnamefont
  {Lanat{\`a}}}, \bibinfo {author} {\bibfnamefont {T.-H.}\ \bibnamefont {Lee}},
  \bibinfo {author} {\bibfnamefont {Y.-X.}\ \bibnamefont {Yao}}, \bibinfo
  {author} {\bibfnamefont {V.}~\bibnamefont {Stevanovi{\'c}}},\ and\ \bibinfo
  {author} {\bibfnamefont {V.}~\bibnamefont {Dobrosavljevi{\'c}}},\ }\bibfield
  {title} {\bibinfo {title} {Connection between {Mott} physics and crystal
  structure in a series of transition metal binary compounds},\ }\href
  {https://doi.org/10.1038/s41524-019-0169-0} {\bibfield  {journal} {\bibinfo
  {journal} {npj Comput. Mater.}\ }\textbf {\bibinfo {volume} {5}},\ \bibinfo
  {pages} {30} (\bibinfo {year} {2019})}\BibitemShut {NoStop}%
\bibitem [{\citenamefont {Schickling}\ \emph {et~al.}(2012)\citenamefont
  {Schickling}, \citenamefont {Gebhard}, \citenamefont {B{\"u}nemann},
  \citenamefont {Boeri}, \citenamefont {Andersen},\ and\ \citenamefont
  {Weber}}]{ga_feas_bunemann}%
  \BibitemOpen
  \bibfield  {author} {\bibinfo {author} {\bibfnamefont {T.}~\bibnamefont
  {Schickling}}, \bibinfo {author} {\bibfnamefont {F.}~\bibnamefont {Gebhard}},
  \bibinfo {author} {\bibfnamefont {J.}~\bibnamefont {B{\"u}nemann}}, \bibinfo
  {author} {\bibfnamefont {L.}~\bibnamefont {Boeri}}, \bibinfo {author}
  {\bibfnamefont {O.~K.}\ \bibnamefont {Andersen}},\ and\ \bibinfo {author}
  {\bibfnamefont {W.}~\bibnamefont {Weber}},\ }\bibfield  {title} {\bibinfo
  {title} {Gutzwiller theory of band magnetism in {LaOFeAs}},\ }\href
  {https://doi.org/10.1103/PhysRevLett.108.036406} {\bibfield  {journal}
  {\bibinfo  {journal} {Phys. Rev. Lett.}\ }\textbf {\bibinfo {volume} {108}},\
  \bibinfo {pages} {036406} (\bibinfo {year} {2012})}\BibitemShut {NoStop}%
\bibitem [{\citenamefont {Lu}\ \emph {et~al.}(2013)\citenamefont {Lu},
  \citenamefont {Zhao}, \citenamefont {Weng}, \citenamefont {Fang},\ and\
  \citenamefont {Dai}}]{ga_smb6}%
  \BibitemOpen
  \bibfield  {author} {\bibinfo {author} {\bibfnamefont {F.}~\bibnamefont
  {Lu}}, \bibinfo {author} {\bibfnamefont {J.}~\bibnamefont {Zhao}}, \bibinfo
  {author} {\bibfnamefont {H.}~\bibnamefont {Weng}}, \bibinfo {author}
  {\bibfnamefont {Z.}~\bibnamefont {Fang}},\ and\ \bibinfo {author}
  {\bibfnamefont {X.}~\bibnamefont {Dai}},\ }\bibfield  {title} {\bibinfo
  {title} {Correlated topological insulators with mixed valence},\ }\href
  {https://doi.org/10.1103/PhysRevLett.110.096401} {\bibfield  {journal}
  {\bibinfo  {journal} {Phys. Rev. Lett.}\ }\textbf {\bibinfo {volume} {110}},\
  \bibinfo {pages} {096401} (\bibinfo {year} {2013})}\BibitemShut {NoStop}%
\bibitem [{\citenamefont {Borghi}\ \emph {et~al.}(2014)\citenamefont {Borghi},
  \citenamefont {Fabrizio},\ and\ \citenamefont {Tosatti}}]{ga_fe}%
  \BibitemOpen
  \bibfield  {author} {\bibinfo {author} {\bibfnamefont {G.}~\bibnamefont
  {Borghi}}, \bibinfo {author} {\bibfnamefont {M.}~\bibnamefont {Fabrizio}},\
  and\ \bibinfo {author} {\bibfnamefont {E.}~\bibnamefont {Tosatti}},\
  }\bibfield  {title} {\bibinfo {title} {{Gutzwiller} electronic structure
  calculations applied to transition metals: Kinetic energy gain with
  ferromagnetic order in bcc {Fe}},\ }\href
  {https://doi.org/10.1103/PhysRevB.90.125102} {\bibfield  {journal} {\bibinfo
  {journal} {Phys. Rev. B}\ }\textbf {\bibinfo {volume} {90}},\ \bibinfo
  {pages} {125102} (\bibinfo {year} {2014})}\BibitemShut {NoStop}%
\bibitem [{\citenamefont {Lanata}\ \emph {et~al.}(2016)\citenamefont {Lanata},
  \citenamefont {Yao}, \citenamefont {Deng}, \citenamefont {Wang},
  \citenamefont {Ho},\ and\ \citenamefont {Kotliar}}]{grg}%
  \BibitemOpen
  \bibfield  {author} {\bibinfo {author} {\bibfnamefont {N.}~\bibnamefont
  {Lanata}}, \bibinfo {author} {\bibfnamefont {Y.-X.}\ \bibnamefont {Yao}},
  \bibinfo {author} {\bibfnamefont {X.}~\bibnamefont {Deng}}, \bibinfo {author}
  {\bibfnamefont {C.-Z.}\ \bibnamefont {Wang}}, \bibinfo {author}
  {\bibfnamefont {K.-M.}\ \bibnamefont {Ho}},\ and\ \bibinfo {author}
  {\bibfnamefont {G.}~\bibnamefont {Kotliar}},\ }\bibfield  {title} {\bibinfo
  {title} {Gutzwiller renormalization group},\ }\href
  {https://doi.org/10.1103/PhysRevB.93.045103} {\bibfield  {journal} {\bibinfo
  {journal} {Phys. Rev. B}\ }\textbf {\bibinfo {volume} {93}},\ \bibinfo
  {pages} {045103} (\bibinfo {year} {2016})}\BibitemShut {NoStop}%
\bibitem [{\citenamefont {Lee}\ \emph {et~al.}(2019)\citenamefont {Lee},
  \citenamefont {Ayral}, \citenamefont {Yao}, \citenamefont {Lanata},\ and\
  \citenamefont {Kotliar}}]{ga_dmet}%
  \BibitemOpen
  \bibfield  {author} {\bibinfo {author} {\bibfnamefont {T.-H.}\ \bibnamefont
  {Lee}}, \bibinfo {author} {\bibfnamefont {T.}~\bibnamefont {Ayral}}, \bibinfo
  {author} {\bibfnamefont {Y.-X.}\ \bibnamefont {Yao}}, \bibinfo {author}
  {\bibfnamefont {N.}~\bibnamefont {Lanata}},\ and\ \bibinfo {author}
  {\bibfnamefont {G.}~\bibnamefont {Kotliar}},\ }\bibfield  {title} {\bibinfo
  {title} {Rotationally invariant slave-boson and density matrix embedding
  theory: Unified framework and comparative study on the one-dimensional and
  two-dimensional {Hubbard} model},\ }\href
  {https://doi.org/10.1103/PhysRevB.99.115129} {\bibfield  {journal} {\bibinfo
  {journal} {Phys. Rev. B}\ }\textbf {\bibinfo {volume} {99}},\ \bibinfo
  {pages} {115129} (\bibinfo {year} {2019})}\BibitemShut {NoStop}%
\bibitem [{\citenamefont {Lanata}\ \emph {et~al.}(2017)\citenamefont {Lanata},
  \citenamefont {Lee}, \citenamefont {Yao},\ and\ \citenamefont
  {Dobrosavljevi{\'c}}}]{gga_lanata2017}%
  \BibitemOpen
  \bibfield  {author} {\bibinfo {author} {\bibfnamefont {N.}~\bibnamefont
  {Lanata}}, \bibinfo {author} {\bibfnamefont {T.-H.}\ \bibnamefont {Lee}},
  \bibinfo {author} {\bibfnamefont {Y.-X.}\ \bibnamefont {Yao}},\ and\ \bibinfo
  {author} {\bibfnamefont {V.}~\bibnamefont {Dobrosavljevi{\'c}}},\ }\bibfield
  {title} {\bibinfo {title} {Emergent {Bloch} excitations in {Mott} matter},\
  }\href {https://doi.org/10.1103/PhysRevB.96.195126} {\bibfield  {journal}
  {\bibinfo  {journal} {Phys. Rev. B}\ }\textbf {\bibinfo {volume} {96}},\
  \bibinfo {pages} {195126} (\bibinfo {year} {2017})}\BibitemShut {NoStop}%
\bibitem [{\citenamefont {Lanat{\`a}}(2022)}]{Lanata2022operatorial}%
  \BibitemOpen
  \bibfield  {author} {\bibinfo {author} {\bibfnamefont {N.}~\bibnamefont
  {Lanat{\`a}}},\ }\bibfield  {title} {\bibinfo {title} {Operatorial
  formulation of the ghost rotationally invariant slave-boson theory},\ }\href
  {https://doi.org/10.1103/PhysRevB.105.045111} {\bibfield  {journal} {\bibinfo
   {journal} {Phys. Rev. B}\ }\textbf {\bibinfo {volume} {105}},\ \bibinfo
  {pages} {045111} (\bibinfo {year} {2022})}\BibitemShut {NoStop}%
\bibitem [{\citenamefont {Frank}\ \emph {et~al.}(2021)\citenamefont {Frank},
  \citenamefont {Lee}, \citenamefont {Bhattacharyya}, \citenamefont {Tsang},
  \citenamefont {Quito}, \citenamefont {Dobrosavljevi\ifmmode~\acute{c}\else
  \'{c}\fi{}}, \citenamefont {Christiansen},\ and\ \citenamefont
  {Lanat\`a}}]{Frank2021QuantumED}%
  \BibitemOpen
  \bibfield  {author} {\bibinfo {author} {\bibfnamefont {M.~S.}\ \bibnamefont
  {Frank}}, \bibinfo {author} {\bibfnamefont {T.-H.}\ \bibnamefont {Lee}},
  \bibinfo {author} {\bibfnamefont {G.}~\bibnamefont {Bhattacharyya}}, \bibinfo
  {author} {\bibfnamefont {P.~K.~H.}\ \bibnamefont {Tsang}}, \bibinfo {author}
  {\bibfnamefont {V.~L.}\ \bibnamefont {Quito}}, \bibinfo {author}
  {\bibfnamefont {V.}~\bibnamefont {Dobrosavljevi\ifmmode~\acute{c}\else
  \'{c}\fi{}}}, \bibinfo {author} {\bibfnamefont {O.}~\bibnamefont
  {Christiansen}},\ and\ \bibinfo {author} {\bibfnamefont {N.}~\bibnamefont
  {Lanat\`a}},\ }\bibfield  {title} {\bibinfo {title} {Quantum embedding
  description of the {Anderson} lattice model with the ghost {Gutzwiller}
  approximation},\ }\href {https://doi.org/10.1103/PhysRevB.104.L081103}
  {\bibfield  {journal} {\bibinfo  {journal} {Phys. Rev. B}\ }\textbf {\bibinfo
  {volume} {104}},\ \bibinfo {pages} {L081103} (\bibinfo {year}
  {2021})}\BibitemShut {NoStop}%
\bibitem [{\citenamefont {Lee}\ \emph {et~al.}(2023{\natexlab{a}})\citenamefont
  {Lee}, \citenamefont {Lanat\`a},\ and\ \citenamefont
  {Kotliar}}]{Lee2023AccuracyGRISB}%
  \BibitemOpen
  \bibfield  {author} {\bibinfo {author} {\bibfnamefont {T.-H.}\ \bibnamefont
  {Lee}}, \bibinfo {author} {\bibfnamefont {N.}~\bibnamefont {Lanat\`a}},\ and\
  \bibinfo {author} {\bibfnamefont {G.}~\bibnamefont {Kotliar}},\ }\bibfield
  {title} {\bibinfo {title} {Accuracy of ghost rotationally invariant
  slave-boson and dynamical mean field theory as a function of the
  impurity-model bath size},\ }\href
  {https://doi.org/10.1103/PhysRevB.107.L121104} {\bibfield  {journal}
  {\bibinfo  {journal} {Phys. Rev. B}\ }\textbf {\bibinfo {volume} {107}},\
  \bibinfo {pages} {L121104} (\bibinfo {year}
  {2023}{\natexlab{a}})}\BibitemShut {NoStop}%
\bibitem [{\citenamefont {Lee}\ \emph {et~al.}(2023{\natexlab{b}})\citenamefont
  {Lee}, \citenamefont {Melnick}, \citenamefont {Adler}, \citenamefont
  {Lanat{\`a}},\ and\ \citenamefont {Kotliar}}]{Lee2023multiorbGRISB}%
  \BibitemOpen
  \bibfield  {author} {\bibinfo {author} {\bibfnamefont {T.-H.}\ \bibnamefont
  {Lee}}, \bibinfo {author} {\bibfnamefont {C.}~\bibnamefont {Melnick}},
  \bibinfo {author} {\bibfnamefont {R.}~\bibnamefont {Adler}}, \bibinfo
  {author} {\bibfnamefont {N.}~\bibnamefont {Lanat{\`a}}},\ and\ \bibinfo
  {author} {\bibfnamefont {G.}~\bibnamefont {Kotliar}},\ }\bibfield  {title}
  {\bibinfo {title} {Accuracy of ghost-rotationally-invariant slave-boson
  theory for multiorbital {Hubbard} models and realistic materials},\ }\href
  {https://doi.org/10.1103/PhysRevB.108.245147} {\bibfield  {journal} {\bibinfo
   {journal} {Phys. Rev. B}\ }\textbf {\bibinfo {volume} {108}},\ \bibinfo
  {pages} {245147} (\bibinfo {year} {2023}{\natexlab{b}})}\BibitemShut
  {NoStop}%
\bibitem [{\citenamefont {Mejuto-Zaera}\ and\ \citenamefont
  {Fabrizio}(2023)}]{MejutoZaera2023screening}%
  \BibitemOpen
  \bibfield  {author} {\bibinfo {author} {\bibfnamefont {C.}~\bibnamefont
  {Mejuto-Zaera}}\ and\ \bibinfo {author} {\bibfnamefont {M.}~\bibnamefont
  {Fabrizio}},\ }\bibfield  {title} {\bibinfo {title} {Efficient computational
  screening of strongly correlated materials: Multiorbital phenomenology within
  the ghost {Gutzwiller} approximation},\ }\href
  {https://doi.org/10.1103/PhysRevB.107.235150} {\bibfield  {journal} {\bibinfo
   {journal} {Phys. Rev. B}\ }\textbf {\bibinfo {volume} {107}},\ \bibinfo
  {pages} {235150} (\bibinfo {year} {2023})}\BibitemShut {NoStop}%
\bibitem [{\citenamefont {Lanat\`a}(2023)}]{Lanata2023ghostDMET}%
  \BibitemOpen
  \bibfield  {author} {\bibinfo {author} {\bibfnamefont {N.}~\bibnamefont
  {Lanat\`a}},\ }\bibfield  {title} {\bibinfo {title} {Derivation of the ghost
  {Gutzwiller} approximation from quantum embedding principles: Ghost density
  matrix embedding theory},\ }\href
  {https://doi.org/10.1103/PhysRevB.108.235112} {\bibfield  {journal} {\bibinfo
   {journal} {Phys. Rev. B}\ }\textbf {\bibinfo {volume} {108}},\ \bibinfo
  {pages} {235112} (\bibinfo {year} {2023})}\BibitemShut {NoStop}%
\bibitem [{\citenamefont {Mejuto-Zaera}(2024)}]{mejuto2024quantum}%
  \BibitemOpen
  \bibfield  {author} {\bibinfo {author} {\bibfnamefont {C.}~\bibnamefont
  {Mejuto-Zaera}},\ }\bibfield  {title} {\bibinfo {title} {Quantum embedding
  for molecules using auxiliary particles--the ghost {Gutzwiller} ansatz},\
  }\href {https://doi.org/10.1039/D4FD00053F} {\bibfield  {journal} {\bibinfo
  {journal} {Faraday Discussions}\ }\textbf {\bibinfo {volume} {254}},\
  \bibinfo {pages} {653} (\bibinfo {year} {2024})}\BibitemShut {NoStop}%
\bibitem [{\citenamefont {Lee}\ \emph {et~al.}(2024)\citenamefont {Lee},
  \citenamefont {Melnick}, \citenamefont {Adler}, \citenamefont {Sun},
  \citenamefont {Yao}, \citenamefont {Lanat\`a},\ and\ \citenamefont
  {Kotliar}}]{Lee2024ChargeSC}%
  \BibitemOpen
  \bibfield  {author} {\bibinfo {author} {\bibfnamefont {T.-H.}\ \bibnamefont
  {Lee}}, \bibinfo {author} {\bibfnamefont {C.}~\bibnamefont {Melnick}},
  \bibinfo {author} {\bibfnamefont {R.}~\bibnamefont {Adler}}, \bibinfo
  {author} {\bibfnamefont {X.}~\bibnamefont {Sun}}, \bibinfo {author}
  {\bibfnamefont {Y.-X.}\ \bibnamefont {Yao}}, \bibinfo {author} {\bibfnamefont
  {N.}~\bibnamefont {Lanat\`a}},\ and\ \bibinfo {author} {\bibfnamefont
  {G.}~\bibnamefont {Kotliar}},\ }\bibfield  {title} {\bibinfo {title} {Charge
  self-consistent density functional theory plus ghost rotationally invariant
  slave-boson theory for correlated materials},\ }\href
  {https://doi.org/10.1103/PhysRevB.110.115126} {\bibfield  {journal} {\bibinfo
   {journal} {Phys. Rev. B}\ }\textbf {\bibinfo {volume} {110}},\ \bibinfo
  {pages} {115126} (\bibinfo {year} {2024})}\BibitemShut {NoStop}%
\bibitem [{\citenamefont {Giuli}\ \emph
  {et~al.}(2026{\natexlab{a}})\citenamefont {Giuli}, \citenamefont {Lee},
  \citenamefont {Yao}, \citenamefont {Kotliar}, \citenamefont {Ruckenstein},
  \citenamefont {Gingras},\ and\ \citenamefont {Lanat{\`a}}}]{finiteTgGA2026}%
  \BibitemOpen
  \bibfield  {author} {\bibinfo {author} {\bibfnamefont {S.}~\bibnamefont
  {Giuli}}, \bibinfo {author} {\bibfnamefont {T.-H.}\ \bibnamefont {Lee}},
  \bibinfo {author} {\bibfnamefont {Y.-X.}\ \bibnamefont {Yao}}, \bibinfo
  {author} {\bibfnamefont {G.}~\bibnamefont {Kotliar}}, \bibinfo {author}
  {\bibfnamefont {A.~E.}\ \bibnamefont {Ruckenstein}}, \bibinfo {author}
  {\bibfnamefont {O.}~\bibnamefont {Gingras}},\ and\ \bibinfo {author}
  {\bibfnamefont {N.}~\bibnamefont {Lanat{\`a}}},\ }\href
  {https://doi.org/10.48550/arXiv.2603.20559} {\bibinfo {title} {Unifying
  variational and dynamical quantum embedding: From ghost {Gutzwiller}
  approximation to dynamical mean-field theory}} (\bibinfo {year}
  {2026}{\natexlab{a}}),\ \Eprint {https://arxiv.org/abs/2603.20559}
  {arXiv:2603.20559 [cond-mat.str-el]} \BibitemShut {NoStop}%
\bibitem [{\citenamefont {Rogers}\ \emph {et~al.}(2021)\citenamefont {Rogers},
  \citenamefont {Lee}, \citenamefont {Pakdel}, \citenamefont {Xu},
  \citenamefont {Dobrosavljevi\'c}, \citenamefont {Yao}, \citenamefont
  {Christiansen},\ and\ \citenamefont {Lanat\`a}}]{Rogers2021bypassing}%
  \BibitemOpen
  \bibfield  {author} {\bibinfo {author} {\bibfnamefont {J.}~\bibnamefont
  {Rogers}}, \bibinfo {author} {\bibfnamefont {T.-H.}\ \bibnamefont {Lee}},
  \bibinfo {author} {\bibfnamefont {S.}~\bibnamefont {Pakdel}}, \bibinfo
  {author} {\bibfnamefont {W.}~\bibnamefont {Xu}}, \bibinfo {author}
  {\bibfnamefont {V.}~\bibnamefont {Dobrosavljevi\'c}}, \bibinfo {author}
  {\bibfnamefont {Y.-X.}\ \bibnamefont {Yao}}, \bibinfo {author} {\bibfnamefont
  {O.}~\bibnamefont {Christiansen}},\ and\ \bibinfo {author} {\bibfnamefont
  {N.}~\bibnamefont {Lanat\`a}},\ }\bibfield  {title} {\bibinfo {title}
  {Bypassing the computational bottleneck of quantum-embedding theories for
  strong electron correlations with machine learning},\ }\href
  {https://doi.org/10.1103/PhysRevResearch.3.013101} {\bibfield  {journal}
  {\bibinfo  {journal} {Phys. Rev. Research}\ }\textbf {\bibinfo {volume}
  {3}},\ \bibinfo {pages} {013101} (\bibinfo {year} {2021})}\BibitemShut
  {NoStop}%
\bibitem [{\citenamefont {Frank}\ \emph {et~al.}(2024)\citenamefont {Frank},
  \citenamefont {Artiukhin}, \citenamefont {Lee}, \citenamefont {Yao},
  \citenamefont {Barros}, \citenamefont {Christiansen},\ and\ \citenamefont
  {Lanat{\`a}}}]{Frank2024activeLA}%
  \BibitemOpen
  \bibfield  {author} {\bibinfo {author} {\bibfnamefont {M.~S.}\ \bibnamefont
  {Frank}}, \bibinfo {author} {\bibfnamefont {D.~G.}\ \bibnamefont
  {Artiukhin}}, \bibinfo {author} {\bibfnamefont {T.-H.}\ \bibnamefont {Lee}},
  \bibinfo {author} {\bibfnamefont {Y.-X.}\ \bibnamefont {Yao}}, \bibinfo
  {author} {\bibfnamefont {K.}~\bibnamefont {Barros}}, \bibinfo {author}
  {\bibfnamefont {O.}~\bibnamefont {Christiansen}},\ and\ \bibinfo {author}
  {\bibfnamefont {N.}~\bibnamefont {Lanat{\`a}}},\ }\bibfield  {title}
  {\bibinfo {title} {Active learning approach to simulations of strongly
  correlated matter with the ghost {Gutzwiller} approximation},\ }\href
  {https://doi.org/10.1103/PhysRevResearch.6.013242} {\bibfield  {journal}
  {\bibinfo  {journal} {Physical Review Research}\ }\textbf {\bibinfo {volume}
  {6}},\ \bibinfo {pages} {013242} (\bibinfo {year} {2024})}\BibitemShut
  {NoStop}%
\bibitem [{\citenamefont {Giuli}\ \emph
  {et~al.}(2026{\natexlab{b}})\citenamefont {Giuli}, \citenamefont {Hasan},
  \citenamefont {Kloss}, \citenamefont {Frank}, \citenamefont {Lee},
  \citenamefont {Gingras}, \citenamefont {Yao},\ and\ \citenamefont
  {Lanat\`a}}]{Giuli2026linear}%
  \BibitemOpen
  \bibfield  {author} {\bibinfo {author} {\bibfnamefont {S.}~\bibnamefont
  {Giuli}}, \bibinfo {author} {\bibfnamefont {H.}~\bibnamefont {Hasan}},
  \bibinfo {author} {\bibfnamefont {B.}~\bibnamefont {Kloss}}, \bibinfo
  {author} {\bibfnamefont {M.~S.}\ \bibnamefont {Frank}}, \bibinfo {author}
  {\bibfnamefont {T.-H.}\ \bibnamefont {Lee}}, \bibinfo {author} {\bibfnamefont
  {O.}~\bibnamefont {Gingras}}, \bibinfo {author} {\bibfnamefont {Y.-X.}\
  \bibnamefont {Yao}},\ and\ \bibinfo {author} {\bibfnamefont {N.}~\bibnamefont
  {Lanat\`a}},\ }\bibfield  {title} {\bibinfo {title} {Linear foundation model
  for quantum embedding: Data-driven compression of the ghost {Gutzwiller}
  variational space},\ }\href {https://doi.org/10.1103/phk3-pvhm} {\bibfield
  {journal} {\bibinfo  {journal} {PRX Intelligence}\ }\textbf {\bibinfo
  {volume} {1}},\ \bibinfo {pages} {013001} (\bibinfo {year}
  {2026}{\natexlab{b}})}\BibitemShut {NoStop}%
\bibitem [{\citenamefont {Giuli}\ \emph
  {et~al.}(2026{\natexlab{c}})\citenamefont {Giuli}, \citenamefont {Lee},
  \citenamefont {Yao}, \citenamefont {Park}, \citenamefont {LaBollita},
  \citenamefont {Pasqua}, \citenamefont {Wentzell}, \citenamefont
  {Lanat{\`a}},\ and\ \citenamefont {Gingras}}]{GEM2026}%
  \BibitemOpen
  \bibfield  {author} {\bibinfo {author} {\bibfnamefont {S.}~\bibnamefont
  {Giuli}}, \bibinfo {author} {\bibfnamefont {T.-H.}\ \bibnamefont {Lee}},
  \bibinfo {author} {\bibfnamefont {Y.-X.}\ \bibnamefont {Yao}}, \bibinfo
  {author} {\bibfnamefont {I.}~\bibnamefont {Park}}, \bibinfo {author}
  {\bibfnamefont {H.}~\bibnamefont {LaBollita}}, \bibinfo {author}
  {\bibfnamefont {I.}~\bibnamefont {Pasqua}}, \bibinfo {author} {\bibfnamefont
  {N.}~\bibnamefont {Wentzell}}, \bibinfo {author} {\bibfnamefont
  {N.}~\bibnamefont {Lanat{\`a}}},\ and\ \bibinfo {author} {\bibfnamefont
  {O.}~\bibnamefont {Gingras}},\ }\href
  {https://doi.org/10.48550/arXiv.2609.32633} {\bibinfo {title} {{GEM}: An
  implementation of the ghost-{Gutzwiller} approximation for simulating
  interacting quantum systems}} (\bibinfo {year} {2026}{\natexlab{c}}),\
  \Eprint {https://arxiv.org/abs/2609.32633} {arXiv:2609.32633
  [cond-mat.str-el]} \BibitemShut {NoStop}%
\bibitem [{\citenamefont {Aoki}\ \emph {et~al.}(2014)\citenamefont {Aoki},
  \citenamefont {Tsuji}, \citenamefont {Eckstein}, \citenamefont {Kollar},
  \citenamefont {Oka},\ and\ \citenamefont {Werner}}]{RMPNonequilibriumDMFT}%
  \BibitemOpen
  \bibfield  {author} {\bibinfo {author} {\bibfnamefont {H.}~\bibnamefont
  {Aoki}}, \bibinfo {author} {\bibfnamefont {N.}~\bibnamefont {Tsuji}},
  \bibinfo {author} {\bibfnamefont {M.}~\bibnamefont {Eckstein}}, \bibinfo
  {author} {\bibfnamefont {M.}~\bibnamefont {Kollar}}, \bibinfo {author}
  {\bibfnamefont {T.}~\bibnamefont {Oka}},\ and\ \bibinfo {author}
  {\bibfnamefont {P.}~\bibnamefont {Werner}},\ }\bibfield  {title} {\bibinfo
  {title} {Nonequilibrium dynamical mean-field theory and its applications},\
  }\href {https://doi.org/10.1103/RevModPhys.86.779} {\bibfield  {journal}
  {\bibinfo  {journal} {Rev. Mod. Phys.}\ }\textbf {\bibinfo {volume} {86}},\
  \bibinfo {pages} {779} (\bibinfo {year} {2014})}\BibitemShut {NoStop}%
\bibitem [{\citenamefont {Kretchmer}\ and\ \citenamefont
  {Chan}(2018)}]{kretchmer18tddmet}%
  \BibitemOpen
  \bibfield  {author} {\bibinfo {author} {\bibfnamefont {J.~S.}\ \bibnamefont
  {Kretchmer}}\ and\ \bibinfo {author} {\bibfnamefont {G.~K.-L.}\ \bibnamefont
  {Chan}},\ }\bibfield  {title} {\bibinfo {title} {A real-time extension of
  density matrix embedding theory for non-equilibrium electron dynamics},\
  }\href {https://doi.org/10.1063/1.5012766} {\bibfield  {journal} {\bibinfo
  {journal} {J. Chem. Phys.}\ }\textbf {\bibinfo {volume} {148}},\ \bibinfo
  {pages} {054108} (\bibinfo {year} {2018})}\BibitemShut {NoStop}%
\bibitem [{\citenamefont {Schir{\'o}}\ and\ \citenamefont
  {Fabrizio}(2010)}]{schiro10tdga}%
  \BibitemOpen
  \bibfield  {author} {\bibinfo {author} {\bibfnamefont {M.}~\bibnamefont
  {Schir{\'o}}}\ and\ \bibinfo {author} {\bibfnamefont {M.}~\bibnamefont
  {Fabrizio}},\ }\bibfield  {title} {\bibinfo {title} {Time-dependent mean
  field theory for quench dynamics in correlated electron systems},\ }\href
  {https://doi.org/10.1103/PhysRevLett.105.076401} {\bibfield  {journal}
  {\bibinfo  {journal} {Phys. Rev. Lett.}\ }\textbf {\bibinfo {volume} {105}},\
  \bibinfo {pages} {076401} (\bibinfo {year} {2010})}\BibitemShut {NoStop}%
\bibitem [{\citenamefont {Mazza}\ and\ \citenamefont
  {Georges}(2017)}]{TDGutzwillerSC}%
  \BibitemOpen
  \bibfield  {author} {\bibinfo {author} {\bibfnamefont {G.}~\bibnamefont
  {Mazza}}\ and\ \bibinfo {author} {\bibfnamefont {A.}~\bibnamefont
  {Georges}},\ }\bibfield  {title} {\bibinfo {title} {Nonequilibrium
  superconductivity in driven alkali-doped fullerides},\ }\href
  {https://doi.org/10.1103/PhysRevB.96.064515} {\bibfield  {journal} {\bibinfo
  {journal} {Phys. Rev. B}\ }\textbf {\bibinfo {volume} {96}},\ \bibinfo
  {pages} {064515} (\bibinfo {year} {2017})}\BibitemShut {NoStop}%
\bibitem [{\citenamefont {Eckstein}\ \emph {et~al.}(2009)\citenamefont
  {Eckstein}, \citenamefont {Kollar},\ and\ \citenamefont
  {Werner}}]{Eckstein09ThermalizationIQ}%
  \BibitemOpen
  \bibfield  {author} {\bibinfo {author} {\bibfnamefont {M.}~\bibnamefont
  {Eckstein}}, \bibinfo {author} {\bibfnamefont {M.}~\bibnamefont {Kollar}},\
  and\ \bibinfo {author} {\bibfnamefont {P.}~\bibnamefont {Werner}},\
  }\bibfield  {title} {\bibinfo {title} {Thermalization after an interaction
  quench in the {Hubbard} model},\ }\href
  {https://doi.org/10.1103/PhysRevLett.103.056403} {\bibfield  {journal}
  {\bibinfo  {journal} {Phys. Rev. Lett.}\ }\textbf {\bibinfo {volume} {103}},\
  \bibinfo {pages} {056403} (\bibinfo {year} {2009})}\BibitemShut {NoStop}%
\bibitem [{\citenamefont {Guerci}\ \emph {et~al.}(2023)\citenamefont {Guerci},
  \citenamefont {Capone},\ and\ \citenamefont {Lanat\`a}}]{Guerci2023TDgGA}%
  \BibitemOpen
  \bibfield  {author} {\bibinfo {author} {\bibfnamefont {D.}~\bibnamefont
  {Guerci}}, \bibinfo {author} {\bibfnamefont {M.}~\bibnamefont {Capone}},\
  and\ \bibinfo {author} {\bibfnamefont {N.}~\bibnamefont {Lanat\`a}},\
  }\bibfield  {title} {\bibinfo {title} {Time-dependent ghost {Gutzwiller}
  nonequilibrium dynamics},\ }\href
  {https://doi.org/10.1103/PhysRevResearch.5.L032023} {\bibfield  {journal}
  {\bibinfo  {journal} {Phys. Rev. Res.}\ }\textbf {\bibinfo {volume} {5}},\
  \bibinfo {pages} {L032023} (\bibinfo {year} {2023})}\BibitemShut {NoStop}%
\bibitem [{\citenamefont {Feynman}(1982)}]{feynman82qc}%
  \BibitemOpen
  \bibfield  {author} {\bibinfo {author} {\bibfnamefont {R.~P.}\ \bibnamefont
  {Feynman}},\ }\bibfield  {title} {\bibinfo {title} {Simulating physics with
  computers},\ }\href {https://doi.org/10.1007/BF02650179} {\bibfield
  {journal} {\bibinfo  {journal} {Int. J. Theor. Phys.}\ }\textbf {\bibinfo
  {volume} {21}},\ \bibinfo {pages} {467} (\bibinfo {year} {1982})}\BibitemShut
  {NoStop}%
\bibitem [{\citenamefont {Lloyd}(1996)}]{lloyd1996}%
  \BibitemOpen
  \bibfield  {author} {\bibinfo {author} {\bibfnamefont {S.}~\bibnamefont
  {Lloyd}},\ }\bibfield  {title} {\bibinfo {title} {Universal quantum
  simulators},\ }\href {https://doi.org/10.1126/science.273.5278.1073}
  {\bibfield  {journal} {\bibinfo  {journal} {Science}\ }\textbf {\bibinfo
  {volume} {273}},\ \bibinfo {pages} {1073} (\bibinfo {year}
  {1996})}\BibitemShut {NoStop}%
\bibitem [{\citenamefont {Bauer}\ \emph {et~al.}(2020)\citenamefont {Bauer},
  \citenamefont {Bravyi}, \citenamefont {Motta},\ and\ \citenamefont
  {Chan}}]{bauer2020quantum}%
  \BibitemOpen
  \bibfield  {author} {\bibinfo {author} {\bibfnamefont {B.}~\bibnamefont
  {Bauer}}, \bibinfo {author} {\bibfnamefont {S.}~\bibnamefont {Bravyi}},
  \bibinfo {author} {\bibfnamefont {M.}~\bibnamefont {Motta}},\ and\ \bibinfo
  {author} {\bibfnamefont {G.~K.-L.}\ \bibnamefont {Chan}},\ }\bibfield
  {title} {\bibinfo {title} {Quantum algorithms for quantum chemistry and
  quantum materials science},\ }\href
  {https://doi.org/10.1021/acs.chemrev.9b00829} {\bibfield  {journal} {\bibinfo
   {journal} {Chem. Rev.}\ }\textbf {\bibinfo {volume} {120}},\ \bibinfo
  {pages} {12685} (\bibinfo {year} {2020})}\BibitemShut {NoStop}%
\bibitem [{\citenamefont {Bauer}\ \emph {et~al.}(2016)\citenamefont {Bauer},
  \citenamefont {Wecker}, \citenamefont {Millis}, \citenamefont {Hastings},\
  and\ \citenamefont {Troyer}}]{hybrd_dmft}%
  \BibitemOpen
  \bibfield  {author} {\bibinfo {author} {\bibfnamefont {B.}~\bibnamefont
  {Bauer}}, \bibinfo {author} {\bibfnamefont {D.}~\bibnamefont {Wecker}},
  \bibinfo {author} {\bibfnamefont {A.~J.}\ \bibnamefont {Millis}}, \bibinfo
  {author} {\bibfnamefont {M.~B.}\ \bibnamefont {Hastings}},\ and\ \bibinfo
  {author} {\bibfnamefont {M.}~\bibnamefont {Troyer}},\ }\bibfield  {title}
  {\bibinfo {title} {Hybrid quantum-classical approach to correlated
  materials},\ }\href {https://doi.org/10.1103/PhysRevX.6.031045} {\bibfield
  {journal} {\bibinfo  {journal} {Phys. Rev. X}\ }\textbf {\bibinfo {volume}
  {6}},\ \bibinfo {pages} {031045} (\bibinfo {year} {2016})}\BibitemShut
  {NoStop}%
\bibitem [{\citenamefont {Kreula}\ \emph {et~al.}(2016)\citenamefont {Kreula},
  \citenamefont {Clark},\ and\ \citenamefont {Jaksch}}]{Kreula2016}%
  \BibitemOpen
  \bibfield  {author} {\bibinfo {author} {\bibfnamefont {J.~M.}\ \bibnamefont
  {Kreula}}, \bibinfo {author} {\bibfnamefont {S.~R.}\ \bibnamefont {Clark}},\
  and\ \bibinfo {author} {\bibfnamefont {D.}~\bibnamefont {Jaksch}},\
  }\bibfield  {title} {\bibinfo {title} {Non-linear quantum-classical scheme to
  simulate non-equilibrium strongly correlated fermionic many-body dynamics},\
  }\href {https://doi.org/10.1038/srep32940} {\bibfield  {journal} {\bibinfo
  {journal} {Sci. Rep.}\ }\textbf {\bibinfo {volume} {6}},\ \bibinfo {pages}
  {32940} (\bibinfo {year} {2016})}\BibitemShut {NoStop}%
\bibitem [{\citenamefont {Rubin}(2016)}]{rubin2016DMET}%
  \BibitemOpen
  \bibfield  {author} {\bibinfo {author} {\bibfnamefont {N.~C.}\ \bibnamefont
  {Rubin}},\ }\bibfield  {title} {\bibinfo {title} {A hybrid classical/quantum
  approach for large-scale studies of quantum systems with density matrix
  embedding theory},\ }\bibfield  {journal} {\bibinfo  {journal} {arXiv
  preprint arXiv:1610.06910}\ }\href
  {https://doi.org/10.48550/arXiv.1610.06910} {10.48550/arXiv.1610.06910}
  (\bibinfo {year} {2016})\BibitemShut {NoStop}%
\bibitem [{\citenamefont {Vorwerk}\ \emph {et~al.}(2022)\citenamefont
  {Vorwerk}, \citenamefont {Sheng}, \citenamefont {Govoni}, \citenamefont
  {Huang},\ and\ \citenamefont {Galli}}]{Vorwerk2022QuantumET}%
  \BibitemOpen
  \bibfield  {author} {\bibinfo {author} {\bibfnamefont {C.}~\bibnamefont
  {Vorwerk}}, \bibinfo {author} {\bibfnamefont {N.}~\bibnamefont {Sheng}},
  \bibinfo {author} {\bibfnamefont {M.}~\bibnamefont {Govoni}}, \bibinfo
  {author} {\bibfnamefont {B.}~\bibnamefont {Huang}},\ and\ \bibinfo {author}
  {\bibfnamefont {G.}~\bibnamefont {Galli}},\ }\bibfield  {title} {\bibinfo
  {title} {Quantum embedding theories to simulate condensed systems on quantum
  computers},\ }\href {https://doi.org/10.1038/s43588-022-00279-0} {\bibfield
  {journal} {\bibinfo  {journal} {Nat. Comput. Sci.}\ }\textbf {\bibinfo
  {volume} {2}},\ \bibinfo {pages} {424} (\bibinfo {year} {2022})}\BibitemShut
  {NoStop}%
\bibitem [{\citenamefont {Rungger}\ \emph {et~al.}(2019)\citenamefont
  {Rungger}, \citenamefont {Fitzpatrick}, \citenamefont {Chen}, \citenamefont
  {Alderete}, \citenamefont {Apel}, \citenamefont {Cowtan}, \citenamefont
  {Patterson}, \citenamefont {Ramo}, \citenamefont {Zhu}, \citenamefont
  {Nguyen} \emph {et~al.}}]{hybrd_2sitedmft2}%
  \BibitemOpen
  \bibfield  {author} {\bibinfo {author} {\bibfnamefont {I.}~\bibnamefont
  {Rungger}}, \bibinfo {author} {\bibfnamefont {N.}~\bibnamefont
  {Fitzpatrick}}, \bibinfo {author} {\bibfnamefont {H.}~\bibnamefont {Chen}},
  \bibinfo {author} {\bibfnamefont {C.~H.}\ \bibnamefont {Alderete}}, \bibinfo
  {author} {\bibfnamefont {H.}~\bibnamefont {Apel}}, \bibinfo {author}
  {\bibfnamefont {A.}~\bibnamefont {Cowtan}}, \bibinfo {author} {\bibfnamefont
  {A.}~\bibnamefont {Patterson}}, \bibinfo {author} {\bibfnamefont {D.~M.}\
  \bibnamefont {Ramo}}, \bibinfo {author} {\bibfnamefont {Y.}~\bibnamefont
  {Zhu}}, \bibinfo {author} {\bibfnamefont {N.~H.}\ \bibnamefont {Nguyen}},
  \emph {et~al.},\ }\bibfield  {title} {\bibinfo {title} {Dynamical mean field
  theory algorithm and experiment on quantum computers},\ }\bibfield  {journal}
  {\bibinfo  {journal} {arXiv:1910.04735}\ }\href
  {https://doi.org/10.48550/arXiv.1910.04735} {10.48550/arXiv.1910.04735}
  (\bibinfo {year} {2019})\BibitemShut {NoStop}%
\bibitem [{\citenamefont {Selisko}\ \emph {et~al.}(2025)\citenamefont
  {Selisko}, \citenamefont {Amsler}, \citenamefont {Wever}, \citenamefont
  {Kawashima}, \citenamefont {Samsonidze}, \citenamefont {Ul~Haq},
  \citenamefont {Tacchino}, \citenamefont {Tavernelli},\ and\ \citenamefont
  {Eckl}}]{selisko2024DMFT}%
  \BibitemOpen
  \bibfield  {author} {\bibinfo {author} {\bibfnamefont {J.}~\bibnamefont
  {Selisko}}, \bibinfo {author} {\bibfnamefont {M.}~\bibnamefont {Amsler}},
  \bibinfo {author} {\bibfnamefont {C.}~\bibnamefont {Wever}}, \bibinfo
  {author} {\bibfnamefont {Y.}~\bibnamefont {Kawashima}}, \bibinfo {author}
  {\bibfnamefont {G.}~\bibnamefont {Samsonidze}}, \bibinfo {author}
  {\bibfnamefont {R.}~\bibnamefont {Ul~Haq}}, \bibinfo {author} {\bibfnamefont
  {F.}~\bibnamefont {Tacchino}}, \bibinfo {author} {\bibfnamefont
  {I.}~\bibnamefont {Tavernelli}},\ and\ \bibinfo {author} {\bibfnamefont
  {T.}~\bibnamefont {Eckl}},\ }\bibfield  {title} {\bibinfo {title} {Dynamical
  mean field theory for real materials on a quantum computer},\ }\href
  {https://doi.org/10.1038/s41524-025-01772-6} {\bibfield  {journal} {\bibinfo
  {journal} {npj Comput. Mater.}\ }\textbf {\bibinfo {volume} {11}},\ \bibinfo
  {pages} {325} (\bibinfo {year} {2025})}\BibitemShut {NoStop}%
\bibitem [{\citenamefont {Hogan}\ \emph {et~al.}(2026)\citenamefont {Hogan},
  \citenamefont {K\"okc\"u}, \citenamefont {Steckmann}, \citenamefont {Doak},
  \citenamefont {Mejuto-Zaera}, \citenamefont {Camps}, \citenamefont
  {Van~Beeumen}, \citenamefont {de~Jong},\ and\ \citenamefont
  {Kemper}}]{Hogan2026efficient}%
  \BibitemOpen
  \bibfield  {author} {\bibinfo {author} {\bibfnamefont {N.}~\bibnamefont
  {Hogan}}, \bibinfo {author} {\bibfnamefont {E.}~\bibnamefont {K\"okc\"u}},
  \bibinfo {author} {\bibfnamefont {T.}~\bibnamefont {Steckmann}}, \bibinfo
  {author} {\bibfnamefont {L.~P.}\ \bibnamefont {Doak}}, \bibinfo {author}
  {\bibfnamefont {C.}~\bibnamefont {Mejuto-Zaera}}, \bibinfo {author}
  {\bibfnamefont {D.}~\bibnamefont {Camps}}, \bibinfo {author} {\bibfnamefont
  {R.}~\bibnamefont {Van~Beeumen}}, \bibinfo {author} {\bibfnamefont {W.~A.}\
  \bibnamefont {de~Jong}},\ and\ \bibinfo {author} {\bibfnamefont {A.~F.}\
  \bibnamefont {Kemper}},\ }\bibfield  {title} {\bibinfo {title} {Efficient
  quantum implementation of dynamical mean field theory for correlated
  materials},\ }\bibfield  {journal} {\bibinfo  {journal} {npj Comput. Mater.}\
  }\textbf {\bibinfo {volume} {12}},\ \href
  {https://doi.org/10.1038/s41524-026-02289-2} {10.1038/s41524-026-02289-2}
  (\bibinfo {year} {2026})\BibitemShut {NoStop}%
\bibitem [{\citenamefont {Mineh}\ and\ \citenamefont
  {Montanaro}(2022)}]{Mineh2022DMET}%
  \BibitemOpen
  \bibfield  {author} {\bibinfo {author} {\bibfnamefont {L.}~\bibnamefont
  {Mineh}}\ and\ \bibinfo {author} {\bibfnamefont {A.}~\bibnamefont
  {Montanaro}},\ }\bibfield  {title} {\bibinfo {title} {Solving the {Hubbard}
  model using density matrix embedding theory and the variational quantum
  eigensolver},\ }\href {https://doi.org/10.1103/PhysRevB.105.125117}
  {\bibfield  {journal} {\bibinfo  {journal} {Phys. Rev. B}\ }\textbf {\bibinfo
  {volume} {105}},\ \bibinfo {pages} {125117} (\bibinfo {year}
  {2022})}\BibitemShut {NoStop}%
\bibitem [{\citenamefont {Yao}\ \emph {et~al.}(2021{\natexlab{a}})\citenamefont
  {Yao}, \citenamefont {Zhang}, \citenamefont {Wang}, \citenamefont {Ho},\ and\
  \citenamefont {Orth}}]{gqce}%
  \BibitemOpen
  \bibfield  {author} {\bibinfo {author} {\bibfnamefont {Y.-X.}\ \bibnamefont
  {Yao}}, \bibinfo {author} {\bibfnamefont {F.}~\bibnamefont {Zhang}}, \bibinfo
  {author} {\bibfnamefont {C.-Z.}\ \bibnamefont {Wang}}, \bibinfo {author}
  {\bibfnamefont {K.-M.}\ \bibnamefont {Ho}},\ and\ \bibinfo {author}
  {\bibfnamefont {P.~P.}\ \bibnamefont {Orth}},\ }\bibfield  {title} {\bibinfo
  {title} {{Gutzwiller} hybrid quantum-classical computing approach for
  correlated materials},\ }\href
  {https://doi.org/10.1103/PhysRevResearch.3.013184} {\bibfield  {journal}
  {\bibinfo  {journal} {Phys. Rev. Res.}\ }\textbf {\bibinfo {volume} {3}},\
  \bibinfo {pages} {013184} (\bibinfo {year} {2021}{\natexlab{a}})}\BibitemShut
  {NoStop}%
\bibitem [{\citenamefont {Mukherjee}\ \emph {et~al.}(2023)\citenamefont
  {Mukherjee}, \citenamefont {Berthusen}, \citenamefont {Getelina},
  \citenamefont {Orth},\ and\ \citenamefont {Yao}}]{mukherjee2023comparative}%
  \BibitemOpen
  \bibfield  {author} {\bibinfo {author} {\bibfnamefont {A.}~\bibnamefont
  {Mukherjee}}, \bibinfo {author} {\bibfnamefont {N.~F.}\ \bibnamefont
  {Berthusen}}, \bibinfo {author} {\bibfnamefont {J.~C.}\ \bibnamefont
  {Getelina}}, \bibinfo {author} {\bibfnamefont {P.~P.}\ \bibnamefont {Orth}},\
  and\ \bibinfo {author} {\bibfnamefont {Y.-X.}\ \bibnamefont {Yao}},\
  }\bibfield  {title} {\bibinfo {title} {Comparative study of adaptive
  variational quantum eigensolvers for multi-orbital impurity models},\ }\href
  {https://doi.org/10.1038/s42005-022-01089-6} {\bibfield  {journal} {\bibinfo
  {journal} {Commun. Phys.}\ }\textbf {\bibinfo {volume} {6}},\ \bibinfo
  {pages} {4} (\bibinfo {year} {2023})}\BibitemShut {NoStop}%
\bibitem [{\citenamefont {Chen}\ \emph {et~al.}(2026)\citenamefont {Chen},
  \citenamefont {Khindanov}, \citenamefont {Salazar}, \citenamefont {Barona},
  \citenamefont {Harrabi}, \citenamefont {Zhang}, \citenamefont {Wang},
  \citenamefont {Iadecola}, \citenamefont {Lanat{\`a}},\ and\ \citenamefont
  {Yao}}]{chen2025quantum}%
  \BibitemOpen
  \bibfield  {author} {\bibinfo {author} {\bibfnamefont {I.-C.}\ \bibnamefont
  {Chen}}, \bibinfo {author} {\bibfnamefont {A.}~\bibnamefont {Khindanov}},
  \bibinfo {author} {\bibfnamefont {C.~M.}\ \bibnamefont {Salazar}}, \bibinfo
  {author} {\bibfnamefont {H.~M.}\ \bibnamefont {Barona}}, \bibinfo {author}
  {\bibfnamefont {G.}~\bibnamefont {Harrabi}}, \bibinfo {author} {\bibfnamefont
  {F.}~\bibnamefont {Zhang}}, \bibinfo {author} {\bibfnamefont {C.-Z.}\
  \bibnamefont {Wang}}, \bibinfo {author} {\bibfnamefont {T.}~\bibnamefont
  {Iadecola}}, \bibinfo {author} {\bibfnamefont {N.}~\bibnamefont
  {Lanat{\`a}}},\ and\ \bibinfo {author} {\bibfnamefont {Y.-X.}\ \bibnamefont
  {Yao}},\ }\bibfield  {title} {\bibinfo {title} {Quantum-classical embedding
  via ghost {Gutzwiller} approximation for enhanced simulations of correlated
  electron systems},\ }\href {https://doi.org/10.1038/s41534-026-01244-9}
  {\bibfield  {journal} {\bibinfo  {journal} {npj Quantum Inf.}\ }\textbf
  {\bibinfo {volume} {12}},\ \bibinfo {pages} {102} (\bibinfo {year}
  {2026})}\BibitemShut {NoStop}%
\bibitem [{\citenamefont {Sriluckshmy}\ \emph {et~al.}(2026)\citenamefont
  {Sriluckshmy}, \citenamefont {Jamet},\ and\ \citenamefont
  {\v{S}imkovic}}]{Sriluckshmy2025QAgGA}%
  \BibitemOpen
  \bibfield  {author} {\bibinfo {author} {\bibfnamefont {P.~V.}\ \bibnamefont
  {Sriluckshmy}}, \bibinfo {author} {\bibfnamefont {F.}~\bibnamefont {Jamet}},\
  and\ \bibinfo {author} {\bibfnamefont {F.}~\bibnamefont {\v{S}imkovic},
  \bibfnamefont {IV}},\ }\bibfield  {title} {\bibinfo {title} {Quantum assisted
  ghost {Gutzwiller} ansatz},\ }\bibfield  {journal} {\bibinfo  {journal}
  {Phys. Rev. Research}\ }\href {https://doi.org/10.1103/bk4j-kfkq}
  {10.1103/bk4j-kfkq} (\bibinfo {year} {2026}),\ \bibinfo {note} {in
  press}\BibitemShut {NoStop}%
\bibitem [{\citenamefont {{Google AI Quantum and Collaborators}}\ \emph
  {et~al.}(2020)\citenamefont {{Google AI Quantum and Collaborators}},
  \citenamefont {Arute}, \citenamefont {Arya}, \citenamefont {Babbush} \emph
  {et~al.}}]{Arute2020HartreeFock}%
  \BibitemOpen
  \bibfield  {author} {\bibinfo {author} {\bibnamefont {{Google AI Quantum and
  Collaborators}}}, \bibinfo {author} {\bibfnamefont {F.}~\bibnamefont
  {Arute}}, \bibinfo {author} {\bibfnamefont {K.}~\bibnamefont {Arya}},
  \bibinfo {author} {\bibfnamefont {R.}~\bibnamefont {Babbush}}, \emph
  {et~al.},\ }\bibfield  {title} {\bibinfo {title} {Hartree-fock on a
  superconducting qubit quantum computer},\ }\href
  {https://doi.org/10.1126/science.abb9811} {\bibfield  {journal} {\bibinfo
  {journal} {Science}\ }\textbf {\bibinfo {volume} {369}},\ \bibinfo {pages}
  {1084} (\bibinfo {year} {2020})}\BibitemShut {NoStop}%
\bibitem [{\citenamefont {Bonet-Monroig}\ \emph {et~al.}(2020)\citenamefont
  {Bonet-Monroig}, \citenamefont {Babbush},\ and\ \citenamefont
  {O'Brien}}]{BonetMonroig2020}%
  \BibitemOpen
  \bibfield  {author} {\bibinfo {author} {\bibfnamefont {X.}~\bibnamefont
  {Bonet-Monroig}}, \bibinfo {author} {\bibfnamefont {R.}~\bibnamefont
  {Babbush}},\ and\ \bibinfo {author} {\bibfnamefont {T.~E.}\ \bibnamefont
  {O'Brien}},\ }\bibfield  {title} {\bibinfo {title} {Nearly optimal
  measurement scheduling for partial tomography of quantum states},\ }\href
  {https://doi.org/10.1103/PhysRevX.10.031064} {\bibfield  {journal} {\bibinfo
  {journal} {Phys. Rev. X}\ }\textbf {\bibinfo {volume} {10}},\ \bibinfo
  {pages} {031064} (\bibinfo {year} {2020})}\BibitemShut {NoStop}%
\bibitem [{\citenamefont {Zhao}\ \emph {et~al.}(2021)\citenamefont {Zhao},
  \citenamefont {Rubin},\ and\ \citenamefont {Miyake}}]{zhao2021fermionic}%
  \BibitemOpen
  \bibfield  {author} {\bibinfo {author} {\bibfnamefont {A.}~\bibnamefont
  {Zhao}}, \bibinfo {author} {\bibfnamefont {N.~C.}\ \bibnamefont {Rubin}},\
  and\ \bibinfo {author} {\bibfnamefont {A.}~\bibnamefont {Miyake}},\
  }\bibfield  {title} {\bibinfo {title} {Fermionic partial tomography via
  classical shadows},\ }\href@noop {} {\bibfield  {journal} {\bibinfo
  {journal} {Phys. Rev. Lett.}\ }\textbf {\bibinfo {volume} {127}},\ \bibinfo
  {pages} {110504} (\bibinfo {year} {2021})}\BibitemShut {NoStop}%
\bibitem [{\citenamefont {Low}(2022)}]{low2022classical}%
  \BibitemOpen
  \bibfield  {author} {\bibinfo {author} {\bibfnamefont {G.~H.}\ \bibnamefont
  {Low}},\ }\bibfield  {title} {\bibinfo {title} {Classical shadows of fermions
  with particle number symmetry},\ }\href@noop {} {\bibfield  {journal}
  {\bibinfo  {journal} {arXiv preprint arXiv:2208.08964}\ } (\bibinfo {year}
  {2022})}\BibitemShut {NoStop}%
\bibitem [{\citenamefont {Berry}\ \emph {et~al.}(2020)\citenamefont {Berry},
  \citenamefont {Childs}, \citenamefont {Su}, \citenamefont {Wang},\ and\
  \citenamefont {Wiebe}}]{Berry2020TimeDependent}%
  \BibitemOpen
  \bibfield  {author} {\bibinfo {author} {\bibfnamefont {D.~W.}\ \bibnamefont
  {Berry}}, \bibinfo {author} {\bibfnamefont {A.~M.}\ \bibnamefont {Childs}},
  \bibinfo {author} {\bibfnamefont {Y.}~\bibnamefont {Su}}, \bibinfo {author}
  {\bibfnamefont {X.}~\bibnamefont {Wang}},\ and\ \bibinfo {author}
  {\bibfnamefont {N.}~\bibnamefont {Wiebe}},\ }\bibfield  {title} {\bibinfo
  {title} {Time-dependent {H}amiltonian simulation with ${L}^1$-norm scaling},\
  }\href {https://doi.org/10.22331/q-2020-04-20-254} {\bibfield  {journal}
  {\bibinfo  {journal} {Quantum}\ }\textbf {\bibinfo {volume} {4}},\ \bibinfo
  {pages} {254} (\bibinfo {year} {2020})}\BibitemShut {NoStop}%
\bibitem [{\citenamefont {Childs}\ \emph {et~al.}(2021)\citenamefont {Childs},
  \citenamefont {Su}, \citenamefont {Tran}, \citenamefont {Wiebe},\ and\
  \citenamefont {Zhu}}]{Childs2021TrotterError}%
  \BibitemOpen
  \bibfield  {author} {\bibinfo {author} {\bibfnamefont {A.~M.}\ \bibnamefont
  {Childs}}, \bibinfo {author} {\bibfnamefont {Y.}~\bibnamefont {Su}}, \bibinfo
  {author} {\bibfnamefont {M.~C.}\ \bibnamefont {Tran}}, \bibinfo {author}
  {\bibfnamefont {N.}~\bibnamefont {Wiebe}},\ and\ \bibinfo {author}
  {\bibfnamefont {S.}~\bibnamefont {Zhu}},\ }\bibfield  {title} {\bibinfo
  {title} {Theory of {Trotter} error with commutator scaling},\ }\href
  {https://doi.org/10.1103/PhysRevX.11.011020} {\bibfield  {journal} {\bibinfo
  {journal} {Phys. Rev. X}\ }\textbf {\bibinfo {volume} {11}},\ \bibinfo
  {pages} {011020} (\bibinfo {year} {2021})}\BibitemShut {NoStop}%
\bibitem [{\citenamefont {Childs}\ and\ \citenamefont
  {Wiebe}(2012)}]{Childs:2012}%
  \BibitemOpen
  \bibfield  {author} {\bibinfo {author} {\bibfnamefont {A.~M.}\ \bibnamefont
  {Childs}}\ and\ \bibinfo {author} {\bibfnamefont {N.}~\bibnamefont {Wiebe}},\
  }\bibfield  {title} {\bibinfo {title} {Hamiltonian simulation using linear
  combinations of unitary operations},\ }\href
  {https://doi.org/10.26421/QIC12.11-12-1} {\bibfield  {journal} {\bibinfo
  {journal} {Quantum Inf. Comput.}\ }\textbf {\bibinfo {volume} {12}},\
  \bibinfo {pages} {901} (\bibinfo {year} {2012})}\BibitemShut {NoStop}%
\bibitem [{\citenamefont {Berry}\ \emph {et~al.}(2015)\citenamefont {Berry},
  \citenamefont {Childs}, \citenamefont {Cleve}, \citenamefont {Kothari},\ and\
  \citenamefont {Somma}}]{Berry:2015}%
  \BibitemOpen
  \bibfield  {author} {\bibinfo {author} {\bibfnamefont {D.~W.}\ \bibnamefont
  {Berry}}, \bibinfo {author} {\bibfnamefont {A.~M.}\ \bibnamefont {Childs}},
  \bibinfo {author} {\bibfnamefont {R.}~\bibnamefont {Cleve}}, \bibinfo
  {author} {\bibfnamefont {R.}~\bibnamefont {Kothari}},\ and\ \bibinfo {author}
  {\bibfnamefont {R.~D.}\ \bibnamefont {Somma}},\ }\bibfield  {title} {\bibinfo
  {title} {Simulating hamiltonian dynamics with a truncated taylor series},\
  }\href {https://doi.org/10.1103/PhysRevLett.114.090502} {\bibfield  {journal}
  {\bibinfo  {journal} {Phys. Rev. Lett.}\ }\textbf {\bibinfo {volume} {114}},\
  \bibinfo {pages} {090502} (\bibinfo {year} {2015})}\BibitemShut {NoStop}%
\bibitem [{\citenamefont {Low}\ and\ \citenamefont
  {Chuang}(2017)}]{Low2017QSP}%
  \BibitemOpen
  \bibfield  {author} {\bibinfo {author} {\bibfnamefont {G.~H.}\ \bibnamefont
  {Low}}\ and\ \bibinfo {author} {\bibfnamefont {I.~L.}\ \bibnamefont
  {Chuang}},\ }\bibfield  {title} {\bibinfo {title} {Optimal hamiltonian
  simulation by quantum signal processing},\ }\href
  {https://doi.org/10.1103/PhysRevLett.118.010501} {\bibfield  {journal}
  {\bibinfo  {journal} {Phys. Rev. Lett.}\ }\textbf {\bibinfo {volume} {118}},\
  \bibinfo {pages} {010501} (\bibinfo {year} {2017})}\BibitemShut {NoStop}%
\bibitem [{\citenamefont {Low}\ and\ \citenamefont
  {Chuang}(2019)}]{Low2019hamiltonian}%
  \BibitemOpen
  \bibfield  {author} {\bibinfo {author} {\bibfnamefont {G.~H.}\ \bibnamefont
  {Low}}\ and\ \bibinfo {author} {\bibfnamefont {I.~L.}\ \bibnamefont
  {Chuang}},\ }\bibfield  {title} {\bibinfo {title} {Hamiltonian {S}imulation
  by {Q}ubitization},\ }\href {https://doi.org/10.22331/q-2019-07-12-163}
  {\bibfield  {journal} {\bibinfo  {journal} {Quantum}\ }\textbf {\bibinfo
  {volume} {3}},\ \bibinfo {pages} {163} (\bibinfo {year} {2019})}\BibitemShut
  {NoStop}%
\bibitem [{\citenamefont {Cerezo}\ \emph {et~al.}(2021)\citenamefont {Cerezo},
  \citenamefont {Arrasmith}, \citenamefont {Babbush}, \citenamefont {Benjamin},
  \citenamefont {Endo}, \citenamefont {Fujii}, \citenamefont {McClean},
  \citenamefont {Mitarai}, \citenamefont {Yuan}, \citenamefont {Cincio} \emph
  {et~al.}}]{cerezo2021variational}%
  \BibitemOpen
  \bibfield  {author} {\bibinfo {author} {\bibfnamefont {M.}~\bibnamefont
  {Cerezo}}, \bibinfo {author} {\bibfnamefont {A.}~\bibnamefont {Arrasmith}},
  \bibinfo {author} {\bibfnamefont {R.}~\bibnamefont {Babbush}}, \bibinfo
  {author} {\bibfnamefont {S.~C.}\ \bibnamefont {Benjamin}}, \bibinfo {author}
  {\bibfnamefont {S.}~\bibnamefont {Endo}}, \bibinfo {author} {\bibfnamefont
  {K.}~\bibnamefont {Fujii}}, \bibinfo {author} {\bibfnamefont {J.~R.}\
  \bibnamefont {McClean}}, \bibinfo {author} {\bibfnamefont {K.}~\bibnamefont
  {Mitarai}}, \bibinfo {author} {\bibfnamefont {X.}~\bibnamefont {Yuan}},
  \bibinfo {author} {\bibfnamefont {L.}~\bibnamefont {Cincio}}, \emph
  {et~al.},\ }\bibfield  {title} {\bibinfo {title} {Variational quantum
  algorithms},\ }\href {https://doi.org/10.1038/s42254-021-00348-9} {\bibfield
  {journal} {\bibinfo  {journal} {Nat. Rev. Phys.}\ }\textbf {\bibinfo {volume}
  {3}},\ \bibinfo {pages} {625} (\bibinfo {year} {2021})}\BibitemShut {NoStop}%
\bibitem [{\citenamefont {Bharti}\ \emph {et~al.}(2022)\citenamefont {Bharti},
  \citenamefont {Cervera-Lierta}, \citenamefont {Kyaw}, \citenamefont {Haug},
  \citenamefont {Alperin-Lea}, \citenamefont {Anand}, \citenamefont {Degroote},
  \citenamefont {Heimonen}, \citenamefont {Kottmann}, \citenamefont {Menke},
  \citenamefont {Mok}, \citenamefont {Sim}, \citenamefont {Kwek},\ and\
  \citenamefont {Aspuru-Guzik}}]{Bharti2022noisyiq}%
  \BibitemOpen
  \bibfield  {author} {\bibinfo {author} {\bibfnamefont {K.}~\bibnamefont
  {Bharti}}, \bibinfo {author} {\bibfnamefont {A.}~\bibnamefont
  {Cervera-Lierta}}, \bibinfo {author} {\bibfnamefont {T.~H.}\ \bibnamefont
  {Kyaw}}, \bibinfo {author} {\bibfnamefont {T.}~\bibnamefont {Haug}}, \bibinfo
  {author} {\bibfnamefont {S.}~\bibnamefont {Alperin-Lea}}, \bibinfo {author}
  {\bibfnamefont {A.}~\bibnamefont {Anand}}, \bibinfo {author} {\bibfnamefont
  {M.}~\bibnamefont {Degroote}}, \bibinfo {author} {\bibfnamefont
  {H.}~\bibnamefont {Heimonen}}, \bibinfo {author} {\bibfnamefont {J.~S.}\
  \bibnamefont {Kottmann}}, \bibinfo {author} {\bibfnamefont {T.}~\bibnamefont
  {Menke}}, \bibinfo {author} {\bibfnamefont {W.-K.}\ \bibnamefont {Mok}},
  \bibinfo {author} {\bibfnamefont {S.}~\bibnamefont {Sim}}, \bibinfo {author}
  {\bibfnamefont {L.-C.}\ \bibnamefont {Kwek}},\ and\ \bibinfo {author}
  {\bibfnamefont {A.}~\bibnamefont {Aspuru-Guzik}},\ }\bibfield  {title}
  {\bibinfo {title} {Noisy intermediate-scale quantum algorithms},\ }\href
  {https://doi.org/10.1103/RevModPhys.94.015004} {\bibfield  {journal}
  {\bibinfo  {journal} {Rev. Mod. Phys.}\ }\textbf {\bibinfo {volume} {94}},\
  \bibinfo {pages} {015004} (\bibinfo {year} {2022})}\BibitemShut {NoStop}%
\bibitem [{\citenamefont {Li}\ and\ \citenamefont
  {Benjamin}(2017)}]{Li_Benjamin-PRX-2017}%
  \BibitemOpen
  \bibfield  {author} {\bibinfo {author} {\bibfnamefont {Y.}~\bibnamefont
  {Li}}\ and\ \bibinfo {author} {\bibfnamefont {S.~C.}\ \bibnamefont
  {Benjamin}},\ }\bibfield  {title} {\bibinfo {title} {Efficient variational
  quantum simulator incorporating active error minimization},\ }\href
  {https://doi.org/10.1103/PhysRevX.7.021050} {\bibfield  {journal} {\bibinfo
  {journal} {Phys. Rev. X}\ }\textbf {\bibinfo {volume} {7}},\ \bibinfo {pages}
  {021050} (\bibinfo {year} {2017})}\BibitemShut {NoStop}%
\bibitem [{\citenamefont {Yuan}\ \emph {et~al.}(2019)\citenamefont {Yuan},
  \citenamefont {Endo}, \citenamefont {Zhao}, \citenamefont {Li},\ and\
  \citenamefont {Benjamin}}]{theory_vqs}%
  \BibitemOpen
  \bibfield  {author} {\bibinfo {author} {\bibfnamefont {X.}~\bibnamefont
  {Yuan}}, \bibinfo {author} {\bibfnamefont {S.}~\bibnamefont {Endo}}, \bibinfo
  {author} {\bibfnamefont {Q.}~\bibnamefont {Zhao}}, \bibinfo {author}
  {\bibfnamefont {Y.}~\bibnamefont {Li}},\ and\ \bibinfo {author}
  {\bibfnamefont {S.~C.}\ \bibnamefont {Benjamin}},\ }\bibfield  {title}
  {\bibinfo {title} {Theory of variational quantum simulation},\ }\href
  {https://doi.org/10.22331/q-2019-10-07-191} {\bibfield  {journal} {\bibinfo
  {journal} {Quantum}\ }\textbf {\bibinfo {volume} {3}},\ \bibinfo {pages}
  {191} (\bibinfo {year} {2019})}\BibitemShut {NoStop}%
\bibitem [{\citenamefont {Barison}\ \emph {et~al.}(2021)\citenamefont
  {Barison}, \citenamefont {Vicentini},\ and\ \citenamefont
  {Carleo}}]{Barison2021efficientquantum}%
  \BibitemOpen
  \bibfield  {author} {\bibinfo {author} {\bibfnamefont {S.}~\bibnamefont
  {Barison}}, \bibinfo {author} {\bibfnamefont {F.}~\bibnamefont {Vicentini}},\
  and\ \bibinfo {author} {\bibfnamefont {G.}~\bibnamefont {Carleo}},\
  }\bibfield  {title} {\bibinfo {title} {An efficient quantum algorithm for the
  time evolution of parameterized circuits},\ }\href
  {https://doi.org/10.22331/q-2021-07-28-512} {\bibfield  {journal} {\bibinfo
  {journal} {{Quantum}}\ }\textbf {\bibinfo {volume} {5}},\ \bibinfo {pages}
  {512} (\bibinfo {year} {2021})}\BibitemShut {NoStop}%
\bibitem [{\citenamefont {Yao}\ \emph {et~al.}(2021{\natexlab{b}})\citenamefont
  {Yao}, \citenamefont {Gomes}, \citenamefont {Zhang}, \citenamefont {Wang},
  \citenamefont {Ho}, \citenamefont {Iadecola},\ and\ \citenamefont
  {Orth}}]{AVQDS}%
  \BibitemOpen
  \bibfield  {author} {\bibinfo {author} {\bibfnamefont {Y.-X.}\ \bibnamefont
  {Yao}}, \bibinfo {author} {\bibfnamefont {N.}~\bibnamefont {Gomes}}, \bibinfo
  {author} {\bibfnamefont {F.}~\bibnamefont {Zhang}}, \bibinfo {author}
  {\bibfnamefont {C.-Z.}\ \bibnamefont {Wang}}, \bibinfo {author}
  {\bibfnamefont {K.-M.}\ \bibnamefont {Ho}}, \bibinfo {author} {\bibfnamefont
  {T.}~\bibnamefont {Iadecola}},\ and\ \bibinfo {author} {\bibfnamefont
  {P.~P.}\ \bibnamefont {Orth}},\ }\bibfield  {title} {\bibinfo {title}
  {Adaptive variational quantum dynamics simulations},\ }\href
  {https://doi.org/10.1103/PRXQuantum.2.030307} {\bibfield  {journal} {\bibinfo
   {journal} {PRX Quantum}\ }\textbf {\bibinfo {volume} {2}},\ \bibinfo {pages}
  {030307} (\bibinfo {year} {2021}{\natexlab{b}})}\BibitemShut {NoStop}%
\bibitem [{\citenamefont {Zhang}\ \emph {et~al.}(2025)\citenamefont {Zhang},
  \citenamefont {Wang}, \citenamefont {Iadecola}, \citenamefont {Orth},\ and\
  \citenamefont {Yao}}]{Zhang2025AdaptiveVQ}%
  \BibitemOpen
  \bibfield  {author} {\bibinfo {author} {\bibfnamefont {F.}~\bibnamefont
  {Zhang}}, \bibinfo {author} {\bibfnamefont {C.-Z.}\ \bibnamefont {Wang}},
  \bibinfo {author} {\bibfnamefont {T.}~\bibnamefont {Iadecola}}, \bibinfo
  {author} {\bibfnamefont {P.~P.}\ \bibnamefont {Orth}},\ and\ \bibinfo
  {author} {\bibfnamefont {Y.-X.}\ \bibnamefont {Yao}},\ }\bibfield  {title}
  {\bibinfo {title} {Adaptive variational quantum dynamics simulations with
  compressed circuits and fewer measurements},\ }\href
  {https://doi.org/10.1103/PhysRevB.111.094310} {\bibfield  {journal} {\bibinfo
   {journal} {Phys. Rev. B}\ }\textbf {\bibinfo {volume} {111}},\ \bibinfo
  {pages} {094310} (\bibinfo {year} {2025})}\BibitemShut {NoStop}%
\bibitem [{\citenamefont {Linteau}\ \emph {et~al.}(2024)\citenamefont
  {Linteau}, \citenamefont {Barison}, \citenamefont {Lindner},\ and\
  \citenamefont {Carleo}}]{Linteau2024adaptiveprojectedVQD}%
  \BibitemOpen
  \bibfield  {author} {\bibinfo {author} {\bibfnamefont {D.}~\bibnamefont
  {Linteau}}, \bibinfo {author} {\bibfnamefont {S.}~\bibnamefont {Barison}},
  \bibinfo {author} {\bibfnamefont {N.~H.}\ \bibnamefont {Lindner}},\ and\
  \bibinfo {author} {\bibfnamefont {G.}~\bibnamefont {Carleo}},\ }\bibfield
  {title} {\bibinfo {title} {Adaptive projected variational quantum dynamics},\
  }\href {https://doi.org/10.1103/PhysRevResearch.6.023130} {\bibfield
  {journal} {\bibinfo  {journal} {Phys. Rev. Res.}\ }\textbf {\bibinfo {volume}
  {6}},\ \bibinfo {pages} {023130} (\bibinfo {year} {2024})}\BibitemShut
  {NoStop}%
\bibitem [{\citenamefont {Gomes}\ \emph {et~al.}(2023)\citenamefont {Gomes},
  \citenamefont {Williams-Young},\ and\ \citenamefont {de~Jong}}]{Gomes2023}%
  \BibitemOpen
  \bibfield  {author} {\bibinfo {author} {\bibfnamefont {N.}~\bibnamefont
  {Gomes}}, \bibinfo {author} {\bibfnamefont {D.~B.}\ \bibnamefont
  {Williams-Young}},\ and\ \bibinfo {author} {\bibfnamefont {W.~A.}\
  \bibnamefont {de~Jong}},\ }\bibfield  {title} {\bibinfo {title} {Computing
  the many-body green’s function with adaptive variational quantum
  dynamics},\ }\href {https://doi.org/10.1021/acs.jctc.3c00150} {\bibfield
  {journal} {\bibinfo  {journal} {J. Chem. Theory Comput.}\ }\textbf {\bibinfo
  {volume} {19}},\ \bibinfo {pages} {3313} (\bibinfo {year}
  {2023})}\BibitemShut {NoStop}%
\bibitem [{\citenamefont {Mootz}\ \emph {et~al.}(2024)\citenamefont {Mootz},
  \citenamefont {Iadecola},\ and\ \citenamefont {Yao}}]{mootz2024adaptive}%
  \BibitemOpen
  \bibfield  {author} {\bibinfo {author} {\bibfnamefont {M.}~\bibnamefont
  {Mootz}}, \bibinfo {author} {\bibfnamefont {T.}~\bibnamefont {Iadecola}},\
  and\ \bibinfo {author} {\bibfnamefont {Y.-X.}\ \bibnamefont {Yao}},\
  }\bibfield  {title} {\bibinfo {title} {Adaptive variational quantum computing
  approaches for green’s functions and nonlinear susceptibilities},\ }\href
  {https://doi.org/10.1021/acs.jctc.4c00874} {\bibfield  {journal} {\bibinfo
  {journal} {Journal of Chemical Theory and Computation}\ }\textbf {\bibinfo
  {volume} {20}},\ \bibinfo {pages} {8689} (\bibinfo {year}
  {2024})}\BibitemShut {NoStop}%
\bibitem [{\citenamefont {Getelina}\ \emph {et~al.}(2023)\citenamefont
  {Getelina}, \citenamefont {Gomes}, \citenamefont {Iadecola}, \citenamefont
  {Orth},\ and\ \citenamefont {Yao}}]{getelina2023adaptive}%
  \BibitemOpen
  \bibfield  {author} {\bibinfo {author} {\bibfnamefont {J.~C.}\ \bibnamefont
  {Getelina}}, \bibinfo {author} {\bibfnamefont {N.}~\bibnamefont {Gomes}},
  \bibinfo {author} {\bibfnamefont {T.}~\bibnamefont {Iadecola}}, \bibinfo
  {author} {\bibfnamefont {P.~P.}\ \bibnamefont {Orth}},\ and\ \bibinfo
  {author} {\bibfnamefont {Y.-X.}\ \bibnamefont {Yao}},\ }\bibfield  {title}
  {\bibinfo {title} {{Adaptive variational quantum minimally entangled typical
  thermal states for finite temperature simulations}},\ }\href
  {https://doi.org/10.21468/SciPostPhys.15.3.102} {\bibfield  {journal}
  {\bibinfo  {journal} {SciPost Phys.}\ }\textbf {\bibinfo {volume} {15}},\
  \bibinfo {pages} {102} (\bibinfo {year} {2023})}\BibitemShut {NoStop}%
\bibitem [{\citenamefont {Alipanah}\ \emph {et~al.}(2025)\citenamefont
  {Alipanah}, \citenamefont {Zhang}, \citenamefont {Yao}, \citenamefont
  {Thompson}, \citenamefont {Nguyen}, \citenamefont {Liu}, \citenamefont
  {Givi}, \citenamefont {McDermott},\ and\ \citenamefont
  {Mendoza-Arenas}}]{Alipanah2025advection}%
  \BibitemOpen
  \bibfield  {author} {\bibinfo {author} {\bibfnamefont {H.}~\bibnamefont
  {Alipanah}}, \bibinfo {author} {\bibfnamefont {F.}~\bibnamefont {Zhang}},
  \bibinfo {author} {\bibfnamefont {Y.}~\bibnamefont {Yao}}, \bibinfo {author}
  {\bibfnamefont {R.}~\bibnamefont {Thompson}}, \bibinfo {author}
  {\bibfnamefont {N.}~\bibnamefont {Nguyen}}, \bibinfo {author} {\bibfnamefont
  {J.}~\bibnamefont {Liu}}, \bibinfo {author} {\bibfnamefont {P.}~\bibnamefont
  {Givi}}, \bibinfo {author} {\bibfnamefont {B.~J.}\ \bibnamefont
  {McDermott}},\ and\ \bibinfo {author} {\bibfnamefont {J.~J.}\ \bibnamefont
  {Mendoza-Arenas}},\ }\bibfield  {title} {\bibinfo {title} {Quantum dynamics
  simulation of the advection-diffusion equation},\ }\href
  {https://doi.org/10.1103/ndc3-bdwt} {\bibfield  {journal} {\bibinfo
  {journal} {Phys. Rev. Research}\ }\textbf {\bibinfo {volume} {7}},\ \bibinfo
  {pages} {043318} (\bibinfo {year} {2025})}\BibitemShut {NoStop}%
\bibitem [{\citenamefont {Yao}\ \emph {et~al.}(2024)\citenamefont {Yao},
  \citenamefont {Zhang}, \citenamefont {Mootz}, \citenamefont {Getelina},
  \citenamefont {Mukherjee}, \citenamefont {Gomes} \emph {et~al.}}]{CyQC}%
  \BibitemOpen
  \bibfield  {author} {\bibinfo {author} {\bibfnamefont {Y.-X.}\ \bibnamefont
  {Yao}}, \bibinfo {author} {\bibfnamefont {F.}~\bibnamefont {Zhang}}, \bibinfo
  {author} {\bibfnamefont {M.}~\bibnamefont {Mootz}}, \bibinfo {author}
  {\bibfnamefont {J.~C.}\ \bibnamefont {Getelina}}, \bibinfo {author}
  {\bibfnamefont {A.}~\bibnamefont {Mukherjee}}, \bibinfo {author}
  {\bibfnamefont {N.}~\bibnamefont {Gomes}}, \emph {et~al.},\ }\href
  {https://doi.org/10.6084/m9.figshare.26298763} {\bibinfo {title} {{CyQC:
  Quantum computing toolset for correlated materials simulations}}},\ \bibinfo
  {howpublished} {figshare, software} (\bibinfo {year} {2024})\BibitemShut
  {NoStop}%
\bibitem [{\citenamefont {Gomes}\ \emph {et~al.}(2021)\citenamefont {Gomes},
  \citenamefont {Mukherjee}, \citenamefont {Zhang}, \citenamefont {Iadecola},
  \citenamefont {Wang}, \citenamefont {Ho}, \citenamefont {Orth},\ and\
  \citenamefont {Yao}}]{AVQITE}%
  \BibitemOpen
  \bibfield  {author} {\bibinfo {author} {\bibfnamefont {N.}~\bibnamefont
  {Gomes}}, \bibinfo {author} {\bibfnamefont {A.}~\bibnamefont {Mukherjee}},
  \bibinfo {author} {\bibfnamefont {F.}~\bibnamefont {Zhang}}, \bibinfo
  {author} {\bibfnamefont {T.}~\bibnamefont {Iadecola}}, \bibinfo {author}
  {\bibfnamefont {C.-Z.}\ \bibnamefont {Wang}}, \bibinfo {author}
  {\bibfnamefont {K.-M.}\ \bibnamefont {Ho}}, \bibinfo {author} {\bibfnamefont
  {P.~P.}\ \bibnamefont {Orth}},\ and\ \bibinfo {author} {\bibfnamefont
  {Y.-X.}\ \bibnamefont {Yao}},\ }\bibfield  {title} {\bibinfo {title}
  {Adaptive variational quantum imaginary time evolution approach for ground
  state preparation},\ }\href {https://doi.org/10.1002/qute.202100114}
  {\bibfield  {journal} {\bibinfo  {journal} {Adv. Quantum Technol.}\ }\textbf
  {\bibinfo {volume} {4}},\ \bibinfo {pages} {2100114} (\bibinfo {year}
  {2021})}\BibitemShut {NoStop}%
\bibitem [{\citenamefont {Moeckel}\ and\ \citenamefont
  {Kehrein}(2008)}]{Moeckel08InteractionQH}%
  \BibitemOpen
  \bibfield  {author} {\bibinfo {author} {\bibfnamefont {M.}~\bibnamefont
  {Moeckel}}\ and\ \bibinfo {author} {\bibfnamefont {S.}~\bibnamefont
  {Kehrein}},\ }\bibfield  {title} {\bibinfo {title} {Interaction quench in the
  {Hubbard} model},\ }\href {https://doi.org/10.1103/PhysRevLett.100.175702}
  {\bibfield  {journal} {\bibinfo  {journal} {Phys. Rev. Lett.}\ }\textbf
  {\bibinfo {volume} {100}},\ \bibinfo {pages} {175702} (\bibinfo {year}
  {2008})}\BibitemShut {NoStop}%
\bibitem [{\citenamefont {Vatan}\ and\ \citenamefont
  {Williams}(2004)}]{Vatan2004optimal}%
  \BibitemOpen
  \bibfield  {author} {\bibinfo {author} {\bibfnamefont {F.}~\bibnamefont
  {Vatan}}\ and\ \bibinfo {author} {\bibfnamefont {C.}~\bibnamefont
  {Williams}},\ }\bibfield  {title} {\bibinfo {title} {Optimal quantum circuits
  for general two-qubit gates},\ }\href
  {https://doi.org/10.1103/PhysRevA.69.032315} {\bibfield  {journal} {\bibinfo
  {journal} {Phys. Rev. A}\ }\textbf {\bibinfo {volume} {69}},\ \bibinfo
  {pages} {032315} (\bibinfo {year} {2004})}\BibitemShut {NoStop}%
\bibitem [{\citenamefont {Shende}\ \emph {et~al.}(2004)\citenamefont {Shende},
  \citenamefont {Markov},\ and\ \citenamefont {Bullock}}]{Shende2004minimal}%
  \BibitemOpen
  \bibfield  {author} {\bibinfo {author} {\bibfnamefont {V.~V.}\ \bibnamefont
  {Shende}}, \bibinfo {author} {\bibfnamefont {I.~L.}\ \bibnamefont {Markov}},\
  and\ \bibinfo {author} {\bibfnamefont {S.~S.}\ \bibnamefont {Bullock}},\
  }\bibfield  {title} {\bibinfo {title} {Minimal universal two-qubit
  controlled-{NOT}-based circuits},\ }\href
  {https://doi.org/10.1103/PhysRevA.69.062321} {\bibfield  {journal} {\bibinfo
  {journal} {Phys. Rev. A}\ }\textbf {\bibinfo {volume} {69}},\ \bibinfo
  {pages} {062321} (\bibinfo {year} {2004})}\BibitemShut {NoStop}%
\bibitem [{\citenamefont {Lin}\ \emph {et~al.}(2021)\citenamefont {Lin},
  \citenamefont {Dilip}, \citenamefont {Green}, \citenamefont {Smith},\ and\
  \citenamefont {Pollmann}}]{lin2021real}%
  \BibitemOpen
  \bibfield  {author} {\bibinfo {author} {\bibfnamefont {S.-H.}\ \bibnamefont
  {Lin}}, \bibinfo {author} {\bibfnamefont {R.}~\bibnamefont {Dilip}}, \bibinfo
  {author} {\bibfnamefont {A.~G.}\ \bibnamefont {Green}}, \bibinfo {author}
  {\bibfnamefont {A.}~\bibnamefont {Smith}},\ and\ \bibinfo {author}
  {\bibfnamefont {F.}~\bibnamefont {Pollmann}},\ }\bibfield  {title} {\bibinfo
  {title} {Real- and imaginary-time evolution with compressed quantum
  circuits},\ }\href {https://doi.org/10.1103/PRXQuantum.2.010342} {\bibfield
  {journal} {\bibinfo  {journal} {PRX Quantum}\ }\textbf {\bibinfo {volume}
  {2}},\ \bibinfo {pages} {010342} (\bibinfo {year} {2021})}\BibitemShut
  {NoStop}%
\bibitem [{\citenamefont {Barratt}\ \emph {et~al.}(2021)\citenamefont
  {Barratt}, \citenamefont {Dborin}, \citenamefont {Bal}, \citenamefont
  {Stojevic}, \citenamefont {Pollmann},\ and\ \citenamefont
  {Green}}]{Barratt2021}%
  \BibitemOpen
  \bibfield  {author} {\bibinfo {author} {\bibfnamefont {F.}~\bibnamefont
  {Barratt}}, \bibinfo {author} {\bibfnamefont {J.}~\bibnamefont {Dborin}},
  \bibinfo {author} {\bibfnamefont {M.}~\bibnamefont {Bal}}, \bibinfo {author}
  {\bibfnamefont {V.}~\bibnamefont {Stojevic}}, \bibinfo {author}
  {\bibfnamefont {F.}~\bibnamefont {Pollmann}},\ and\ \bibinfo {author}
  {\bibfnamefont {A.~G.}\ \bibnamefont {Green}},\ }\bibfield  {title} {\bibinfo
  {title} {Parallel quantum simulation of large systems on small {NISQ}
  computers},\ }\href {https://doi.org/10.1038/s41534-021-00420-3} {\bibfield
  {journal} {\bibinfo  {journal} {npj Quantum Inf.}\ }\textbf {\bibinfo
  {volume} {7}},\ \bibinfo {pages} {79} (\bibinfo {year} {2021})}\BibitemShut
  {NoStop}%
\bibitem [{\citenamefont {Martin}\ \emph {et~al.}(2024)\citenamefont {Martin},
  \citenamefont {Ayral}, \citenamefont {Jamet}, \citenamefont
  {Ran{\v{c}}i{\'c}},\ and\ \citenamefont {Simon}}]{Martin2024combining}%
  \BibitemOpen
  \bibfield  {author} {\bibinfo {author} {\bibfnamefont {B.~A.}\ \bibnamefont
  {Martin}}, \bibinfo {author} {\bibfnamefont {T.}~\bibnamefont {Ayral}},
  \bibinfo {author} {\bibfnamefont {F.}~\bibnamefont {Jamet}}, \bibinfo
  {author} {\bibfnamefont {M.~J.}\ \bibnamefont {Ran{\v{c}}i{\'c}}},\ and\
  \bibinfo {author} {\bibfnamefont {P.}~\bibnamefont {Simon}},\ }\bibfield
  {title} {\bibinfo {title} {Combining matrix product states and noisy quantum
  computers for quantum simulation},\ }\href
  {https://doi.org/10.1103/PhysRevA.109.062437} {\bibfield  {journal} {\bibinfo
   {journal} {Phys. Rev. A}\ }\textbf {\bibinfo {volume} {109}},\ \bibinfo
  {pages} {062437} (\bibinfo {year} {2024})}\BibitemShut {NoStop}%
\bibitem [{\citenamefont {Self}\ \emph {et~al.}(2024)\citenamefont {Self},
  \citenamefont {Benedetti},\ and\ \citenamefont {Amaro}}]{self2024protecting}%
  \BibitemOpen
  \bibfield  {author} {\bibinfo {author} {\bibfnamefont {C.~N.}\ \bibnamefont
  {Self}}, \bibinfo {author} {\bibfnamefont {M.}~\bibnamefont {Benedetti}},\
  and\ \bibinfo {author} {\bibfnamefont {D.}~\bibnamefont {Amaro}},\ }\bibfield
   {title} {\bibinfo {title} {Protecting expressive circuits with a quantum
  error detection code},\ }\href {https://doi.org/10.1038/s41567-023-02282-2}
  {\bibfield  {journal} {\bibinfo  {journal} {Nature Physics}\ }\textbf
  {\bibinfo {volume} {20}},\ \bibinfo {pages} {219} (\bibinfo {year}
  {2024})}\BibitemShut {NoStop}%
\bibitem [{\citenamefont {Huang}\ \emph {et~al.}(2020)\citenamefont {Huang},
  \citenamefont {Kueng},\ and\ \citenamefont {Preskill}}]{huang2020predicting}%
  \BibitemOpen
  \bibfield  {author} {\bibinfo {author} {\bibfnamefont {H.-Y.}\ \bibnamefont
  {Huang}}, \bibinfo {author} {\bibfnamefont {R.}~\bibnamefont {Kueng}},\ and\
  \bibinfo {author} {\bibfnamefont {J.}~\bibnamefont {Preskill}},\ }\bibfield
  {title} {\bibinfo {title} {Predicting many properties of a quantum system
  from very few measurements},\ }\href
  {https://doi.org/10.1038/s41567-020-0932-7} {\bibfield  {journal} {\bibinfo
  {journal} {Nat. Phys.}\ }\textbf {\bibinfo {volume} {16}},\ \bibinfo {pages}
  {1050} (\bibinfo {year} {2020})}\BibitemShut {NoStop}%
\bibitem [{\citenamefont {Kivlichan}\ \emph {et~al.}(2018)\citenamefont
  {Kivlichan}, \citenamefont {McClean}, \citenamefont {Wiebe}, \citenamefont
  {Gidney}, \citenamefont {Aspuru-Guzik}, \citenamefont {Chan},\ and\
  \citenamefont {Babbush}}]{BabbushLinearDepth}%
  \BibitemOpen
  \bibfield  {author} {\bibinfo {author} {\bibfnamefont {I.~D.}\ \bibnamefont
  {Kivlichan}}, \bibinfo {author} {\bibfnamefont {J.}~\bibnamefont {McClean}},
  \bibinfo {author} {\bibfnamefont {N.}~\bibnamefont {Wiebe}}, \bibinfo
  {author} {\bibfnamefont {C.}~\bibnamefont {Gidney}}, \bibinfo {author}
  {\bibfnamefont {A.}~\bibnamefont {Aspuru-Guzik}}, \bibinfo {author}
  {\bibfnamefont {G.~K.-L.}\ \bibnamefont {Chan}},\ and\ \bibinfo {author}
  {\bibfnamefont {R.}~\bibnamefont {Babbush}},\ }\bibfield  {title} {\bibinfo
  {title} {Quantum simulation of electronic structure with linear depth and
  connectivity},\ }\href {https://doi.org/10.1103/PhysRevLett.120.110501}
  {\bibfield  {journal} {\bibinfo  {journal} {Phys. Rev. Lett.}\ }\textbf
  {\bibinfo {volume} {120}},\ \bibinfo {pages} {110501} (\bibinfo {year}
  {2018})}\BibitemShut {NoStop}%
\bibitem [{\citenamefont {Urb{\'a}nek}\ \emph {et~al.}(2021)\citenamefont
  {Urb{\'a}nek}, \citenamefont {Nachman}, \citenamefont {Pascuzzi},
  \citenamefont {He}, \citenamefont {Bauer},\ and\ \citenamefont
  {de~Jong}}]{Urbnek2021MitigatingDN}%
  \BibitemOpen
  \bibfield  {author} {\bibinfo {author} {\bibfnamefont {M.}~\bibnamefont
  {Urb{\'a}nek}}, \bibinfo {author} {\bibfnamefont {B.}~\bibnamefont
  {Nachman}}, \bibinfo {author} {\bibfnamefont {V.~R.}\ \bibnamefont
  {Pascuzzi}}, \bibinfo {author} {\bibfnamefont {A.}~\bibnamefont {He}},
  \bibinfo {author} {\bibfnamefont {C.~W.}\ \bibnamefont {Bauer}},\ and\
  \bibinfo {author} {\bibfnamefont {W.~A.}\ \bibnamefont {de~Jong}},\
  }\bibfield  {title} {\bibinfo {title} {Mitigating depolarizing noise on
  quantum computers with noise-estimation circuits},\ }\href
  {https://doi.org/10.1103/PhysRevLett.127.270502} {\bibfield  {journal}
  {\bibinfo  {journal} {Phys. Rev. Lett.}\ }\textbf {\bibinfo {volume} {127}},\
  \bibinfo {pages} {270502} (\bibinfo {year} {2021})}\BibitemShut {NoStop}%
\bibitem [{\citenamefont {Zhang}\ \emph {et~al.}(2026)\citenamefont {Zhang},
  \citenamefont {Gomes}, \citenamefont {Aftergood}, \citenamefont {Iadecola},
  \citenamefont {Yao},\ and\ \citenamefont {Orth}}]{Zhang2026sampling}%
  \BibitemOpen
  \bibfield  {author} {\bibinfo {author} {\bibfnamefont {F.}~\bibnamefont
  {Zhang}}, \bibinfo {author} {\bibfnamefont {N.}~\bibnamefont {Gomes}},
  \bibinfo {author} {\bibfnamefont {J.}~\bibnamefont {Aftergood}}, \bibinfo
  {author} {\bibfnamefont {T.}~\bibnamefont {Iadecola}}, \bibinfo {author}
  {\bibfnamefont {Y.-X.}\ \bibnamefont {Yao}},\ and\ \bibinfo {author}
  {\bibfnamefont {P.~P.}\ \bibnamefont {Orth}},\ }\bibfield  {title} {\bibinfo
  {title} {Sampling noise and optimized measurement distribution in
  imaginary-time quantum dynamics simulations},\ }\href
  {https://doi.org/10.1002/qute.70418} {\bibfield  {journal} {\bibinfo
  {journal} {Adv. Quantum Technol.}\ }\textbf {\bibinfo {volume} {9}},\
  \bibinfo {pages} {e70418} (\bibinfo {year} {2026})}\BibitemShut {NoStop}%
\end{thebibliography}%

\end{document}